\documentclass[aps,physrev,reprint,superscriptaddress, amsmath, amssymb]{revtex4-2}
\usepackage{graphicx}
\usepackage{orcidlink}
\usepackage{placeins}
\usepackage[normalem]{ulem}

\usepackage{hyperref}
\hypersetup{
  colorlinks=true,
  linkcolor=blue,
  filecolor=magenta,
  urlcolor=blue,
  citecolor=red,
}
\begin{document}

\title{Heterogeneity-enhanced stochastic resonance improves liquid-state computing in delayed spiking neural networks}

\author{Sandipan Nath\hspace{0.5mm}\orcidlink{0009-0008-4830-6451}}
\email{sandipan.nath@fau.de}
\affiliation{%
 Department of Data Science, Friedrich-Alexander-Universit\"at Erlangen-N\"urnberg, N\"urnberger Str. 74, 91052 Erlangen, Germany
}
\affiliation{Institute of Machine Learning and Neural Computation, Graz University of Technology, Inffeldgasse 16b/I, 8010 Graz, Austria}

\author{Marius E. Yamakou\hspace{0.5mm}\orcidlink{0000-0002-2809-1739}}%
 \email{marius.yamakou@fau.de}
\affiliation{%
 Department of Data Science, Friedrich-Alexander-Universit\"at Erlangen-N\"urnberg, N\"urnberger Str. 74, 91052 Erlangen, Germany
}%
\affiliation{Department of Mathematics, Friedrich-Alexander-Universit\"at Erlangen-N\"urnberg,\\ Cauer Str. 11, 91058 Erlangen, Germany}

\date{\today}

\begin{abstract}
We investigate the joint effects of stochastic forcing and quenched structural heterogeneity on stochastic resonance (SR) and liquid-state computation in a small-world network of excitable FitzHugh--Nagumo neurons. Heterogeneity is introduced separately through coupling strengths and time delays drawn from Gaussian, bimodal, and shifted-exponential distributions. Weak periodic forcing resolves the stochastic-resonance landscape, whereas weak aperiodic driving probes noise-assisted signal encoding and Liquid State Machine (LSM) forecasting. Heterogeneity is not generically beneficial, but reorganizes the resonance landscape and computational performance in a distribution-dependent manner. Gaussian coupling disorder broadens the strong-response region, whereas bimodal disorder produces a more pronounced enhancement and shifts resonance toward weaker noise. Shifted-exponential coupling disorder behaves differently: increasing its scale broadens and shifts the coupling distribution toward less responsive regions. Time-delay heterogeneity can enhance or suppress SR depending on how the delay distribution samples the structured delay-response landscape. Under aperiodic forcing, stronger input--output coherence accompanies lower LSM prediction error, showing that SR can improve LSM performance and that suitable coupling heterogeneity can further enhance this computational benefit; Gaussian and bimodal coupling disorder shift the optimum toward weaker noise and reduce the noise-optimized root-mean-square prediction error, with the largest reduction obtained for the bimodal case. These results identify stochastic forcing and quenched heterogeneity as coupled control parameters and show that the computational enhancement of LSMs depends on disorder structure rather than magnitude alone.
\end{abstract}

\maketitle

\section{\label{sec:introduction}Introduction}
The interplay between stochastic fluctuations and nonlinear dynamics is a longstanding theme in statistical and nonlinear physics. Although noise is often regarded as disorder that degrades deterministic dynamics, its coupling to nonlinearity, threshold structure, and intrinsic timescale separation can instead generate fluctuation-induced order and enhance sensitivity to weak perturbations.

In excitable systems, coherence resonance occurs when intermediate noise optimally regularizes noise-induced oscillations in the absence of periodic forcing~\cite{pikovskyCoherenceResonanceNoisedriven1997}. With an additional weak signal, stochastic fluctuations can instead amplify the response at an optimal noise level, producing stochastic resonance (SR)~\cite{Benzi-1981a}. Noise, however, is not generically constructive. In some dynamical regimes it suppresses sustained activity, giving rise to inverse stochastic resonance and showing that the macroscopic effect of fluctuations depends on the underlying nonlinear phase-space structure~\cite{tuckwellAnalysisInverseStochastic2012}.

In computational neuroscience, SR was first observed in an excitable FitzHugh--Nagumo (FHN) neuron driven by additive periodic and stochastic forcing~\cite{longtin1993stochastic}. Evidence of SR has also been reported in information-processing regions of the human brain, suggesting that noise-assisted amplification of weak signals may support higher-order functions such as perception and cognition~\cite{mori2004stochastic}. At the network level, the dynamics become richer because each excitable element evolves within a collective environment generated by recurrent interactions.

More recently, SR has been shown to persist in small-world networks of FHN motifs, where higher-order interactions and stronger coupling enhance weak-signal transmission by increasing output-neuron firing~\cite{li2024stochastic}. In delayed multilayer excitable networks, interaction order, resonance mismatch, and delay were further identified as key control parameters of the SR amplitude and optimal noise level~\cite{schlabritz2026interactionordercontrolsstochasticresonance}. These results motivate the question of how structural variability in the interaction network reshapes the noise-assisted collective response.

A key source of such variability is heterogeneity. In interacting excitable populations, quenched disorder arises in neuronal excitability, synaptic strengths, and signal-propagation times. Physiological neural networks are intrinsically heterogeneous, with substantial variability in neuronal properties, synaptic strengths, and time delays across pathways~\cite{marder2006variability}. Such diversity can do more than perturb a homogeneous collective state: intrinsic neuronal heterogeneity can enhance information encoding by reducing pairwise spike-train correlations~\cite{padmanabhan2010intrinsic}.
 
More generally, disorder and stochastic forcing can reorganize collective excitable dynamics in nontrivial ways. Parameter heterogeneity combined with spatially independent noise modifies coherence resonance in heterogeneous FHN arrays~\cite{zhouArrayenhancedCoherenceResonance2001} and enhances spatial coherence in map-based neural models~\cite{wang2012combined}; heterogeneous noise amplitudes can induce strong stochastic resonance in coupled particles~\cite{kawai2010effective}; and diversity-induced resonance occurs across multiple parameter distributions~\cite{liu2022diversity} and in coupled FHN networks where population diversity generates hubs~\cite{sciallaHubsDiversitySynchronization2021}.

Importantly, the symmetry of the diversity distribution can govern the emergence of global oscillations, with symmetric diversity enhancing resonant behavior and asymmetric diversity suppressing it~\cite{scialla2025effect}. Conversely, heterogeneity can also reduce oscillator coherence by modulating self-induced stochastic resonance, giving rise to diversity-induced decoherence~\cite{yamakou2022diversity}. However, diversity and noise are distinct sources of disorder: their synchronization effects can act independently or synergistically depending on the position of the diversity mean relative to the intrinsic oscillatory range of the network elements~\cite{sciallaInterplayDiversityNoise2022}. More recently, heterogeneous noise in globally coupled FHN oscillators was shown to synchronize extreme events, further demonstrating that stochastic heterogeneity can organize collective dynamics~\cite{hariharan2026heterogeneous}.

To our knowledge, however, these studies have not addressed how quenched heterogeneity with different distributions in the interaction parameters themselves---specifically, coupling strengths and time delays---reshapes stochastic resonance in excitable networks. This distinction is important because coupling disorder modifies the spatial distribution of interaction scales, whereas time-delay disorder modifies the temporal organization of recurrent interactions; the two may therefore reshape the collective resonance landscape through distinct dynamical mechanisms. It is worth noting that a mean-field approach has been used to investigate heterogeneity in the bifurcation parameters of individual nodes~\cite{liu2022diversity}, rather than in the interaction parameters of the network, as considered in the present study.

Another important question is whether heterogeneity-induced changes in stochastic resonance extend beyond the dynamical response to alter the network's computational capability. We address this within the Liquid State Machine (LSM) framework, where a fixed recurrent spiking network maps the input onto a high-dimensional transient state decoded by a trained linear readout~\cite{maassRealtimeComputingStable2002}.

From a dynamical-systems perspective, the liquid acts as a nonlinear temporal feature map whose utility depends on a sensitive yet reproducible response to the driving signal. This makes the link between stochastic resonance and reservoir computation natural: moderate stochastic forcing can enhance weak-input representation, whereas excessive noise destroys temporal correlations needed by the readout. Related noise-assisted computational effects have been reported in Echo State Networks (ESNs). In one model, a bistable stochastic-resonance potential served as the activation mechanism, providing nonlinearity and short-term memory while improving robustness to input noise~\cite{liao2021echo}. Furthermore, noise injection into ESN training data can produce a resonance-like computational optimum, where an intermediate noise amplitude improves short-term prediction accuracy, stability, and horizon, as well as long-term reconstruction of chaotic attractors~\cite{zhai2023emergence}. In contrast, how stochastic resonance affects the computational performance of a spiking-neuron LSM, and how quenched heterogeneity modifies this relation, remains largely unexplored.

In the present work, we address these two connected problems in a noisy small-world network of FHN neurons. First, we examine how quenched heterogeneity in coupling strengths and time delays reshapes the collective stochastic-resonance response. Rather than characterizing heterogeneity only by its magnitude, we consider Gaussian, bimodal, and shifted-exponential distributions to determine how disorder structure reorganizes the resonance landscape. Periodic subthreshold forcing provides a controlled probe of this landscape, revealing how the location and amplitude of the resonance depend jointly on noise intensity and heterogeneous interaction parameters.

Second, we ask whether these dynamical changes translate into computational consequences for an LSM built from the same recurrent-network model. A weak aperiodic signal drives the liquid, and its noise-assisted encoding is evaluated through the performance of a trained linear readout in a short-horizon forecasting task. The central objective is therefore not only to determine whether heterogeneity enhances or suppresses stochastic resonance, but to establish how the distributional structure of quenched disorder reorganizes the noise-dependent collective dynamics and whether this yields more informative reservoir states for computation.

The remainder of the paper is organized as follows. Section~\ref{sec:math_model_dynamics} introduces the FHN model, small-world topology, and heterogeneity distributions for coupling strengths and time delays. Section~\ref{sec:numerics} describes the integration scheme, SR measures, simulation protocols, and LSM framework. Section~\ref{sec:results_discussion} presents homogeneous-network benchmarks, the effects of coupling-strength and time-delay heterogeneity on SR, and the computational consequences of coupling heterogeneity for LSM performance. Finally, Section~\ref{sec:summary_conclusions} summarizes the main findings and their implications.

\section{\label{sec:math_model_dynamics}Mathematical model}

\subsection{\label{sec:neuron}Spiking neural network model}

As a representative model of excitable dynamics with well-established biological relevance, we consider the FHN neuron model~\cite{fitzhughMathematicalModelsThreshold1955,fitzhughImpulsesPhysiologicalStates1961,nagumoActivePulseTransmission1962}, described by
\begin{equation}\label{eq:1}
\left\{
\begin{aligned}
\dot v_i
&=
v_i-\frac{v_i^3}{3}-w_i
+I_{\mathrm{ext}}(t)+I_{\mathrm{syn},i}(t)+\xi_i(t),\\[0.3em]
\dot w_i
&=
\varepsilon\bigl(v_i+a_0\bigr).
\end{aligned}
\right.
\end{equation}
For each neuron $i=1,\ldots,N$, the state variables
$v_i=v_i(t)\in\mathbb{R}$ and $w_i=w_i(t)\in\mathbb{R}$ denote the fast membrane-potential variable and the slow recovery variable, respectively. The small parameter $0<\varepsilon\ll1$ establishes a separation between their characteristic timescales. Throughout this paper we fix $\varepsilon=0.03$. The parameter $a_0$ controls the excitability properties and, consequently, the qualitative dynamical behavior of the neuron.

Moreover, $I_{\mathrm{ext}}(t)$ represents a weak subthreshold input signal. For periodic stochastic resonance (PSR), we take
$
I_{\mathrm{ext}}(t)=A\cos(\omega t),
$
where $A$ and $\omega$ denote the forcing amplitude and angular frequency, respectively. For aperiodic stochastic resonance (ASR), $I_{\mathrm{ext}}(t)$ may instead be an aperiodic signal whose precise form is not prescribed, provided that its variations occur on a timescale slower than the characteristic timescales of the FHN dynamics~\cite{collins1995aperiodic}. 

The term $I_{\mathrm{syn},i}(t)$ accounts for the synaptic input received by neuron $i$. The stochastic processes $\xi_i(t)$, $i=1,\ldots,N$, are modeled as mutually independent Gaussian white noises with zero mean and amplitude $D$. Their covariance is given by
$
\left\langle \xi_i(t)\xi_j(t')\right\rangle
=
D^2\delta_{ij}\delta(t-t'),
$
where $\delta_{ij}$ is the Kronecker delta and $\delta(\cdot)$ denotes the Dirac delta distribution.

For an isolated neuron, corresponding to $N=1$, the neuron index is omitted, and the synaptic contribution is set to zero, namely,
$I_{\mathrm{syn}}(t)=0$.

To account for synaptic interactions within the neural network, we augment Eq.~\eqref{eq:1} with the synaptic input current $I_{\mathrm{syn},i}(t)$, representing delayed diffusive (electrical-like) coupling from presynaptic neurons to neuron $i$. It is defined by \begin{equation}\label{eq:2} I_{\mathrm{syn},i}(t) = \sum_{\substack{j=1\\j\neq i}}^{N} W_{ij}g_{ij} \left[v_j(t-\tau_{ij})-v_i(t)\right], \end{equation} where $W=(W_{ij})$ is the network connectivity matrix, with $W_{ij}=1$ if neuron $j$ projects to neuron $i$ and $W_{ij}=0$ otherwise. The parameters $g_{ij}\geq 0$ and $\tau_{ij}\geq 0$ denote, respectively, the diffusive coupling strength and time delay of the connection $j\to i$. Thus, $I_{\mathrm{syn},i}(t)$ collects the contributions from all presynaptic neighbors of neuron $i$. The network topology is chosen as a Watts--Strogatz small-world network~\cite{wattsCollectiveDynamicsSmallworldNetworks1998,muldoonSmallworldPropensityWeighted2016}. Autaptic connections are excluded by imposing $W_{ii}=0$ for all $i=1,\ldots,N$.

\subsection{\label{sec:g_tau_dist}Distributions of neural coupling strengths and time delays}

Biological neural networks exhibit substantial variability in synaptic
strengths and time delays. To account for distinct forms of such
heterogeneity, we consider Gaussian, bimodal, and shifted-exponential
distributions.

For each directed edge $j\to i$ with $W_{ij}=1$, the coupling strength
$g_{ij}$ and time delay $\tau_{ij}$ are sampled separately:
when $g_{ij}$ is heterogeneous, the delay is fixed uniformly, whereas when $\tau_{ij}$
is heterogeneous, the coupling strength is fixed uniformly at
$g_{ij}=g=0.0025$. The interval
$[g_{\min},g_{\max}]=[0.005,0.05]$ applies only to the heterogeneous
coupling-strength distributions and therefore does not constrain this
homogeneous reference value.

For all three distribution families, $\mu_g$ and $\mu_\tau$ denote the
location-control parameters, whereas $\sigma_g$ and $\sigma_\tau$ denote
the heterogeneity-control parameters. Their interpretation depends on the
distribution family: standard deviation for the Gaussian distribution,
component separation $\Delta\mu$ for the bimodal distribution, and
exponential scale $1/\lambda$ for the shifted-exponential distribution.

Let $\widetilde g_{ij}$ and $\widetilde\tau_{ij}$ denote the corresponding
unprojected samples. The admissible values are obtained as
\begin{equation}
g_{ij}=\Pi_{[g_{\min},g_{\max}]}(\widetilde g_{ij}),
\quad
\tau_{ij}=\Pi_{[\tau_{\min},\tau_{\max}]}(\widetilde\tau_{ij}),
\end{equation}
where
\begin{equation}
\Pi_{[\alpha,\beta]}(x)=\min\{\beta,\max\{\alpha,x\}\}.
\end{equation}
For all coupling distributions,
$[g_{\min},g_{\max}]=[0.005,0.05]$. For the Gaussian and bimodal delay
distributions, $[\tau_{\min},\tau_{\max}]=[0,2.5]$, whereas for the
shifted-exponential distribution it is $[0,11]$.

Table~\ref{tab:distribution_parameters} summarizes the distribution-specific
parameters and their ranges. The corresponding sampling laws and
representative PDFs are given in the Appendix.

\begin{table*}
\caption{\label{tab:distribution_parameters}
Distribution families and parameter ranges used to generate heterogeneous
coupling strengths and time delays. The symbols $\mu_g$ and $\mu_\tau$
denote the respective location-control parameters, while $\sigma_g$ and
$\sigma_\tau$ denote the corresponding heterogeneity-control parameters.}
\begin{ruledtabular}
\begin{tabular}{lccclcc}
Distribution
& \multicolumn{2}{c}{$\mu_g$}
& \multicolumn{2}{c}{$\sigma_g$}
& $g_{\min}$
& $g_{\max}$ \\
& Parameter & Range & Parameter & Range & & \\ \colrule
Gaussian
& Mean $\mu$ & $[0.01, 0.05]$ & $\sigma$ & $[0, 0.017]$ & $0.005$ & $0.05$ \\
Bimodal
& Mean $\mu$ & $[0.01, 0.045]$ & $\Delta\mu$ & $[0, 0.04]$ & $0.005$ & $0.05$ \\
Shifted-Exp
& lower shift $g_0$ & $[0.0, 0.03]$ & $1/\lambda$ & $[0.0005, 0.1]$ & $0.005$ & $0.05$ \\
\colrule
Distribution
& \multicolumn{2}{c}{$\mu_\tau$}
& \multicolumn{2}{c}{$\sigma_\tau$}
& $\tau_{\min}$
& $\tau_{\max}$ \\
& Parameter & Range & Parameter & Range & & \\ \colrule
Gaussian
& Mean $\mu$ & $[0, 2.5]$ & $\sigma$ & $[0, 0.833]$ & $0$ & $2.5$ \\
Bimodal
& Mean $\mu$ & $[0, 2.5]$ & $\Delta\mu$ & $[0, 2.0]$ & $0$ & $2.5$ \\
Shifted-Exp
& lower shift \ $\tau_0$ & $[0, 5.0]$ & $1/\lambda$ & $[0.01, 2.0]$ & $0$ & $11.0$ \\
\end{tabular}
\end{ruledtabular}
\end{table*}

\section{\label{sec:numerics}Numerical methods and LSM training protocol}

\subsection{Numerical methods}

\subsubsection{\label{subsec:numerical_integration}Time discretization and stochastic integration}

The stochastic network in Eq.~\eqref{eq:1} is integrated by the
Euler--Maruyama method \cite{highamAlgorithmicIntroductionNumerical2001}
with time step $\Delta t=0.01$:
\begin{equation}
\left\{
\begin{aligned}
v_i^{n+1}
&=
v_i^n+\Delta t\,F_i^n
+D\sqrt{\Delta t}\,\xi_i^n,
\\
w_i^{n+1}
&=
w_i^n+\Delta t\,\varepsilon(v_i^n+a_0),
\end{aligned}
\right.
\end{equation}
where $F_i^n$ contains the intrinsic, forcing, and delayed coupling
terms, and $\xi_i^n\sim\mathcal{N}(0,1)$ are independent over neurons
and time steps. Delayed states are retrieved from the stored numerical
history. We set $v_i(0)\sim\mathcal{U}(-1,1)$ and $w_i(0)=0$.

\subsubsection{\label{subsec:sr_measure_periodic}Quantification of periodic stochastic resonance}

When the periodic stimulus $I_{\mathrm{ext}}(t)=A\cos(\omega t)$ drives
the neural network in Eq.~\eqref{eq:1}, the response at the forcing
frequency $\omega$ is quantified over the post-transient observation
interval $[T_{\mathrm{start}},T_{\max}]$, of duration
$T=T_{\max}-T_{\mathrm{start}}$. For each neuron, we define the finite-time
Fourier coefficients
\begin{equation}
\left\{
\begin{aligned}
R_i
&=
\frac{2}{T}
\int_{T_{\mathrm{start}}}^{T_{\max}}
v_i(t)\cos(\omega t)\,\mathrm{d}t,
\\
S_i
&=
\frac{2}{T}
\int_{T_{\mathrm{start}}}^{T_{\max}}
v_i(t)\sin(\omega t)\,\mathrm{d}t.
\end{aligned}
\right.
\end{equation}

Adapting the stochastic-resonance measure of
Ref.~\cite{ozerStochasticResonanceNewman2009}, the network-level response
and its maximum over the sampled set of noise amplitudes $\mathcal{D}$
are defined by
\begin{equation}
\left\{
\begin{aligned}
Q(\mu,\sigma,D)
&=
\frac{1}{N}\sum_{i=1}^{N}\sqrt{R_i^2+S_i^2},
\\
Q_{\max}(\mu,\sigma)
&=
\max_{D\in\mathcal{D}}Q(\mu,\sigma,D).
\end{aligned}
\right.
\end{equation}
Here, $\mu$ and $\sigma$ denote, respectively, the location- and
heterogeneity-control parameters of the coupling-strength or
time-delay distribution under consideration. A bell-shaped
dependence of $Q$ on $D$ is taken as the numerical signature of periodic
stochastic resonance. A larger value of $Q_{\max}(\mu,\sigma)$ indicates
a stronger maximal network response at the forcing frequency and,
therefore, a more pronounced stochastic-resonance effect.
\newline
\subsubsection{\label{subsec:asr_measure}Quantification of aperiodic stochastic resonance}

For any discrete time series $X^n$, let $\langle X^n\rangle_T$ denote
its temporal average over the post-transient observation interval. The
aperiodic input is first centered according to
\begin{equation}
S^n
=
I_{\mathrm{ext}}(t_n)
-
\left\langle I_{\mathrm{ext}}(t_n)\right\rangle_T .
\end{equation}

For each neuron, the binary spike indicator is defined by
\begin{equation}
z_i^n
=
\begin{cases}
1,
& v_i(t_{n-1})<v_{\mathrm{th}}
\ \text{and}\
v_i(t_n)\geq v_{\mathrm{th}},\\
0,
& \text{otherwise},
\end{cases}
\qquad
v_{\mathrm{th}}=0,
\end{equation}
so that a spike is recorded whenever the membrane potential crosses
the threshold from below. The binned population spike activity is then
defined by
\begin{equation}
R^n
=
\frac{1}{N}
\sum_{i=1}^{N}z_i^n,
\end{equation}
where each bin has width equal to the integration time step,
$\Delta t=0.01$. Thus, $R^n$ represents the fraction of neurons that
spike during the interval $(t_{n-1},t_n]$.

Following Ref.~\cite{collins1995aperiodic}, the zero-lag power norm
and its normalized form are defined by
\begin{widetext}
\begin{equation}
\left\{
\begin{aligned}
C_0(\mu,\sigma,D)
&=
\left\langle S^nR^n\right\rangle_T,
\\[0.3em]
\bar{Q}(\mu,\sigma,D)
&=
\left\{
\begin{array}{c@{\quad}l}
\frac{C_0(\mu,\sigma,D)}
{\sqrt{
\left\langle(S^n)^2\right\rangle_T
\left\langle
\bigl(R^n-\langle R^n\rangle_T\bigr)^2
\right\rangle_T
}}
& \text{if } \mathrm{Var}(R^n)>0
\\[1em]
0
& \text{if } \mathrm{Var}(R^n)=0
\end{array}
\right.
.
\end{aligned}
\right.
\label{eq:asr_measure}
\end{equation}
\end{widetext}

Since $\langle S^n\rangle_T=0$, $\bar{Q}$ is the normalized zero-lag
correlation between the aperiodic input and the binned population spike
activity. A nonmonotone, bell-shaped dependence of $\bar{Q}$ on $D$, with a
maximum at an intermediate noise amplitude, indicates aperiodic
stochastic resonance.

\subsubsection{\label{subsec:long_time_averaging}Simulation protocol}

We consider a Watts--Strogatz network of $N=50$ neurons with mean
degree $\langle K\rangle=4$ and rewiring probability $\beta=0.3$.
For each parameter set, the network topology and the sampled coupling
strengths and time delays are held fixed throughout the
simulation.

A single stochastic realization is integrated up to
$T_{\max}=5.0\times10^{4}$ time units. After discarding the transient
interval $T_{\mathrm{start}}=5000$, the periodic response $Q$ and the
aperiodic response $\bar{Q}$ are evaluated over the remaining interval
of length $T=T_{\max}-T_{\mathrm{start}}=4.5\times10^{4}$.
No optimization over temporal lags is performed; accordingly,
$\bar{Q}$ quantifies zero-lag input--output coherence.

All observables are evaluated from the post-transient portion of each trajectory. For the experiments on heterogeneity-enhanced aperiodic input--output coherence and LSM prediction shown in Figs.~\ref{fig:min_rmse}, \ref{fig:gaussian_asr_lsm}, \ref{fig:2gmm_lsm}, and \ref{fig:exponential_lsm}, the reported quantities are additionally ensemble-averaged over 20 independent realizations. In each realization, the stochastic forcing, Watts--Strogatz network topology, initial conditions, and quenched coupling-strength disorder are generated independently; the sampled topology and coupling strengths are then held fixed throughout that realization. The aperiodic input sequence is kept fixed across ensemble realizations. For Fig.~\ref{fig:min_rmse}, the RMSE is first ensemble-averaged at each sampled noise amplitude \(D\), after which \(\mathrm{RMSE}_{\min}\) is obtained by minimizing the ensemble-averaged prediction error over \(D\).

\subsection{\label{subsec:lsm_protocol}Liquid state machine architecture and training protocol}

\begin{figure*}
\centering
\includegraphics{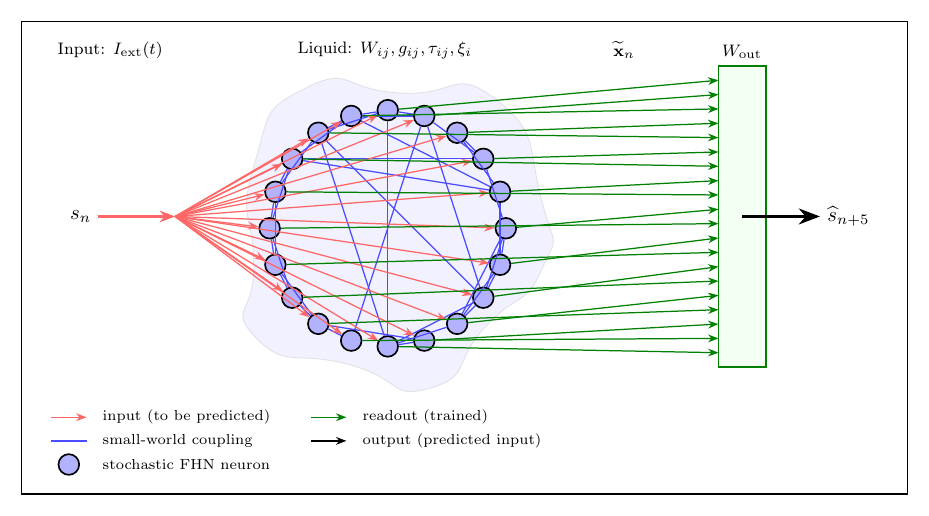}
\caption{\label{fig:lsm}Liquid state machine architecture used for five-step-ahead
forecasting. The discrete weak aperiodic input $s_n$ is applied as the
external drive to a noisy heterogeneous small-world reservoir with
connectivity matrix $W_{ij}$, coupling strengths $g_{ij}$, and
time delays $\tau_{ij}$. The resulting reservoir activity is
preprocessed into the feature vector
$\widetilde{\mathbf{x}}_n$ and mapped by the trained linear readout
$\mathbf{W}_{\mathrm{out}}$ to the prediction
$\widehat{s}_{n+5}$. Only the readout weights are trained; the network
topology, coupling parameters, time delays, and intrinsic liquid
dynamics remain fixed throughout training.}
\end{figure*}

We formulate the liquid state machine (LSM) as a reservoir-computing
model for short-horizon forecasting of a weak aperiodic signal. The
architecture consists of three components: an external input signal, a
fixed stochastic recurrent neural network acting as the liquid, and a
trainable linear readout, as illustrated in Fig.~\ref{fig:lsm}. In
contrast to a fully trained recurrent neural network, the internal
parameters of the liquid are not optimized. The recurrent network
therefore acts as a fixed nonlinear temporal feature map, while learning
is restricted to the output layer.

The computational task is formulated as supervised forecasting of an
aperiodic time series. The use of an aperiodic input is important for
assessing the computational properties of the liquid: unlike a strictly
periodic signal, whose future evolution can in principle be reconstructed
once its period is known, an aperiodic signal contains non-repeating
temporal variations that must be represented by the reservoir dynamics
and exploited by the readout.

\subsubsection{Aperiodic input generation}

For the forecasting experiments, the input sequence is generated from
the membrane-potential trace of a Hodgkin--Huxley (HH) neuron
\cite{hodgkinQuantitativeDescriptionMembrane1952}. To distinguish the
physical time used in the HH simulation from the dimensionless time of
the FHN reservoir, we denote the former by $t_{\mathrm{HH}}$. The HH
membrane potential $V_{\mathrm{HH}}(t_{\mathrm{HH}})$ satisfies
\begin{eqnarray}
C_m\frac{dV_{\mathrm{HH}}}{dt_{\mathrm{HH}}}
=&&
-g_{\mathrm{Na}}m^3h
\left(V_{\mathrm{HH}}-E_{\mathrm{Na}}\right)\nonumber\\-&&g_{\mathrm{K}}n^4
\left(V_{\mathrm{HH}}-E_{\mathrm{K}}\right)\nonumber\\
-&&g_{\mathrm{L}}
\left(V_{\mathrm{HH}}-E_{\mathrm{L}}\right)
+\frac{I_{\mathrm{DC}}}{A_m},
\label{eq:hh_voltage}
\end{eqnarray}
where $A_m$ denotes the membrane area, so that
$I_{\mathrm{DC}}/A_m$ has the units of current density required by the
conductance-density formulation of the HH equations. The gating
variables $x\in\{m,h,n\}$ evolve according to
\begin{equation}
\frac{dx}{dt_{\mathrm{HH}}}
=
\alpha_x(V_{\mathrm{HH}})(1-x)
-
\beta_x(V_{\mathrm{HH}})x,
\label{eq:hh_gating}
\end{equation}
where $\alpha_x$ and $\beta_x$ are the standard voltage-dependent
opening and closing rate functions \cite{yamakou2023combined}.

The HH parameters are $C_m = 1.0\,\mu\mathrm{F}/\mathrm{cm}^{2}$, $A_{m}=10^{-4}\,\mathrm{cm}^{2}$,
$g_{\mathrm{Na}} = 120\,\mathrm{mS}/\mathrm{cm}^{2},
g_{\mathrm{K}} = 36\,\mathrm{mS}/\mathrm{cm}^{2},
g_{\mathrm{L}} = 0.3\,\mathrm{mS}/\mathrm{cm}^{2}$ 
with reversal potentials
$E_{\mathrm{Na}} = 50\,\mathrm{mV},
E_{\mathrm{K}} = -77\,\mathrm{mV},
E_{\mathrm{L}} = -54.4\,\mathrm{mV}.$ 

The HH membrane potential $V_{\mathrm{HH}}$ is perturbed by a piecewise-constant stochastic current,
\begin{equation}
I_{\rm DC}(t_{\mathrm{HH}})
=
I^0_{\mathrm{DC}}
+
\sigma \eta_k,
\quad
\eta_k \overset{\mathrm{i.i.d.}}{\sim} \mathcal{N}(0,1),
\end{equation}
for
\begin{equation}
t_{\mathrm{HH}}
\in
[k\Delta t_{\rm noise},(k+1)\Delta t_{\rm noise}),
\quad
k=0,1,2,\ldots .
\end{equation}
Here, $I^0_{\mathrm{DC}}=8\times10^{-4}\,\mathrm{\mu A}$ is the constant
depolarizing current that places the neuron in the tonic-spiking regime,
$\sigma=10^{-4}\,\mathrm{\mu A}$ is the noise amplitude, and
$\Delta t_{\rm noise}=1\,\mathrm{ms}$ is the noise-update interval.
Thus, the random variable $\eta_k$ is held fixed over each interval of
duration $\Delta t_{\rm noise}$ and is independently resampled from a
standard normal distribution at the beginning of the next interval.
This produces a stochastic, aperiodic HH membrane-potential trace.
The HH system is integrated with time step
$\Delta t_{\mathrm{HH}}=0.01\,\mathrm{ms}$ over a total duration
$T_{\mathrm{HH}}=5\times10^{4}\,\mathrm{ms}$, yielding
$5\times10^{6}$ membrane-potential samples.
 
The raw HH trace is rescaled to the weak-amplitude interval used for
the stochastic-resonance experiments. Let $V_{\min}$ and $V_{\max}$
denote the minimum and maximum of the complete HH membrane-potential
trace. We define
\begin{equation}
u_{\mathrm{HH}}(t_{\mathrm{HH}})
=
\frac{V_{\mathrm{HH}}(t_{\mathrm{HH}})-V_{\min}}
     {V_{\max}-V_{\min}}
\,2A_{\mathrm{in}}
-
A_{\mathrm{in}},
\label{eq:hh_rescaling}
\end{equation}
with $A_{\mathrm{in}}=0.015.$ Consequently,
\begin{equation}
u_{\mathrm{HH}}(t_{\mathrm{HH}})
\in[-0.015,0.015].
\end{equation}

The continuous HH trace is converted into the discrete input sequence
by sampling it on the HH time grid,
\begin{equation}
s_n
=
u_{\mathrm{HH}}\!\left(n\Delta t_{\mathrm{HH}}\right).
\label{eq:hh_input_sequence}
\end{equation}
The sequence $\{s_n\}$ is then supplied sample-by-sample to the FHN
reservoir. Thus, at the dimensionless reservoir time
$t_n=n\Delta t$, the external drive is defined by
\begin{equation}
I_{\mathrm{ext}}(t_n)=s_n.
\label{eq:discrete_external_input}
\end{equation}
This construction identifies the sample index of the two simulations,
rather than identifying the physical HH time $t_{\mathrm{HH}}$ with the
dimensionless FHN time $t$.

The input amplitude is chosen such that the deterministic liquid remains
in the subthreshold regime. Hence, in the absence of noise, the weak
input alone does not generate sustained spiking activity. Instead, its
temporal structure is encoded through the interaction between the
subthreshold forcing, the recurrent reservoir dynamics, and stochastic
fluctuations. In this regime, aperiodic stochastic resonance can increase
the correspondence between the input signal and the activity generated
by the liquid.

\subsubsection{Reservoir representation}

The liquid is the heterogeneous Watts--Strogatz small-world network
introduced in Sec.~\ref{sec:math_model_dynamics}. Its recurrent
connectivity is specified by $W_{ij}$, with edge-dependent coupling
strengths $g_{ij}$ and time delays $\tau_{ij}$. Together with the
stochastic forcing, these fixed recurrent interactions transform the
scalar input into a high-dimensional dynamical representation of its
recent history. From a machine-learning perspective, the liquid therefore
provides a nonlinear temporal feature representation, whereas the readout
constitutes the only trainable component of the model.

At the discrete reservoir times
\begin{equation}
t_n=n\Delta t,
\quad
\Delta t=0.01,
\end{equation}
the external input is the corresponding discrete sample $s_n$ defined
in Eq.~\eqref{eq:hh_input_sequence}. The reservoir state supplied to
the readout is constructed from the membrane potentials of all $N=50$
FHN neurons,
\begin{equation}
\mathbf{x}_n
=
\bigl[
v_1(t_n),\,
v_2(t_n),\,
\ldots,\,
v_{50}(t_n)
\bigr]^{\top}
\in\mathbb{R}^{50}.
\label{eq:reservoir_state}
\end{equation}
Only the membrane-potential variables are exposed to the readout. The
recovery variables $w_i$ are not included in the feature vector, and no
constant bias feature is appended to $\mathbf{x}_n$.

Before fitting the readout, the reservoir states are subjected to the
same element-wise preprocessing used in the numerical implementation.
We define
\begin{equation}
\mathcal{H}(v)
=
\begin{cases}
0, & v\in[-1.2,-0.8],\\[2pt]
v, & \text{otherwise},
\end{cases}
\label{eq:state_threshold}
\end{equation}
and apply $\mathcal{H}$ independently to each component of
$\mathbf{x}_n$. The resulting feature vector is therefore
\begin{equation}
\widetilde{\mathbf{x}}_n
=
\mathcal{H}(\mathbf{x}_n),
\label{eq:thresholded_state}
\end{equation}
where the operation is understood component-wise. The interval $[-1.2, -0.8]$ is centered on the resting potential $v = -1.0$ of the FHN neuron, so that subthreshold fluctuations around it are suppressed while spike-driven excursions are retained as features for the
readout during this preprocessing step.

\subsubsection{Forecasting target and linear readout}

We consider a five-step-ahead forecasting task. The supervised target
associated with the reservoir state at time $t_n$ is
\begin{equation}
y_n^{\mathrm{target}}
=
s_{n+5}.
\label{eq:lsm_target}
\end{equation}
The corresponding prediction is generated by a linear readout,
\begin{equation}
\widehat{s}_{n+5}
=
\widetilde{\mathbf{x}}_n^{\top}
\mathbf{W}_{\mathrm{out}},
\label{eq:lsm_prediction}
\end{equation}
where $\mathbf{W}_{\mathrm{out}}\in\mathbb{R}^{50\times1}$
contains the trainable readout weights.

After removal of the initial transient, suppose that the available
post-transient input--reservoir sequence contains $M$ samples. Because
the forecasting target associated with state $\widetilde{\mathbf{x}}_n$
is $s_{n+5}$, only the first $M-5$ reservoir states have corresponding
five-step-ahead targets. Hence,
\begin{equation}
N_{\mathrm{samp}}=M-5.
\label{eq:number_samples}
\end{equation}
The corresponding preprocessed reservoir states are stacked row-wise to
form the design matrix
\begin{equation}
\widetilde{\mathbf{R}}
=
\begin{bmatrix}
\widetilde{\mathbf{x}}_1^{\top}\\
\widetilde{\mathbf{x}}_2^{\top}\\
\vdots\\
\widetilde{\mathbf{x}}_{N_{\mathrm{samp}}}^{\top}
\end{bmatrix}
\in
\mathbb{R}^{N_{\mathrm{samp}}\times 50}.
\label{eq:design_matrix}
\end{equation}

To preserve the temporal ordering of the time series, no random
shuffling is performed. Instead, the available state--target pairs are
divided chronologically into a training set
$\mathcal{I}_{\mathrm{train}}$, containing the first $75\%$ of the
samples, and a test set $\mathcal{I}_{\mathrm{test}}$, containing the
remaining $25\%$. This chronological hold-out split prevents future
samples from being used to train the readout.

Let
$\widetilde{\mathbf{R}}_{\mathrm{train}}
\in\mathbb{R}^{N_{\mathrm{train}}\times 50}$
denote the training design matrix, where
$N_{\mathrm{train}}=|\mathcal{I}_{\mathrm{train}}|$, and let
$\mathbf{s}_{\mathrm{train}}
\in\mathbb{R}^{N_{\mathrm{train}}}$
contain the corresponding targets $s_{n+5}$.
The readout-weight vector is
$\mathbf{W}_{\mathrm{out}}\in\mathbb{R}^{50}$.
The auxiliary vector
$\mathbf{w}\in\mathbb{R}^{50}$ denotes a candidate readout-weight vector
over which the ridge-regression objective is minimized. The readout
weights are estimated by
\begin{equation}
\mathbf{W}_{\mathrm{out}}
=
\arg\min_{\mathbf{w}\in\mathbb{R}^{50}}
\left\{
\left\|
\widetilde{\mathbf{R}}_{\mathrm{train}}\mathbf{w}
-
\mathbf{s}_{\mathrm{train}}
\right\|_2^2
+
\lambda\|\mathbf{w}\|_2^2
\right\},
\label{eq:ridge_objective}
\end{equation}
with regularization parameter $\lambda=10^{-4}.$ The corresponding closed-form estimator is
\begin{equation}
\mathbf{W}_{\mathrm{out}}
=
\left(
\widetilde{\mathbf{R}}_{\mathrm{train}}^{\top}
\widetilde{\mathbf{R}}_{\mathrm{train}}
+
\lambda\mathbf{I}_{50}
\right)^{-1}
\widetilde{\mathbf{R}}_{\mathrm{train}}^{\top}
\mathbf{s}_{\mathrm{train}},
\label{eq:ridge_solution}
\end{equation}
where $\mathbf{I}_{50}\in\mathbb{R}^{50\times50}$ is the identity matrix.
No normalization or standardization of the reservoir features is applied
before fitting the readout.

\subsubsection{Test-time prediction and performance measure}

For the held-out test interval, the raw predictions are obtained as
\begin{equation}
\widehat{\mathbf{s}}_{\mathrm{test}}^{\,\mathrm{raw}}
=
\widetilde{\mathbf{R}}_{\mathrm{test}}
\mathbf{W}_{\mathrm{out}}.
\label{eq:test_prediction}
\end{equation}

Following the numerical implementation, the raw predictions are
post-processed by an affine transformation so that their mean and
standard deviation match those of the training targets. Let
\begin{equation}
\mu_{\mathrm{train}}
=
\frac{1}{|\mathcal{I}_{\mathrm{train}}|}
\sum_{n\in\mathcal{I}_{\mathrm{train}}}s_{n+5},
\quad
\sigma_{\mathrm{train}}
=
\operatorname{std}\!\left(\mathbf{s}_{\mathrm{train}}\right),
\end{equation}
and let
\begin{equation}
\mu_{\mathrm{raw}}
=
\operatorname{mean}\!\left(
\widehat{\mathbf{s}}_{\mathrm{test}}^{\,\mathrm{raw}}
\right),
\quad
\sigma_{\mathrm{raw}}
=
\operatorname{std}\!\left(
\widehat{\mathbf{s}}_{\mathrm{test}}^{\,\mathrm{raw}}
\right).
\end{equation}

The final test predictions are then
\begin{equation}
\widehat{\mathbf{s}}_{\mathrm{test}}
=
\mu_{\mathrm{train}}\mathbf{1}
+
\frac{\sigma_{\mathrm{train}}}{\sigma_{\mathrm{raw}}}
\left(
\widehat{\mathbf{s}}_{\mathrm{test}}^{\,\mathrm{raw}}
-
\mu_{\mathrm{raw}}\mathbf{1}
\right),
\label{eq:test_rescaling}
\end{equation}
where $\mathbf{1}$ denotes the vector of ones of length
$|\mathcal{I}_{\mathrm{test}}|$. Because $\mu_{\rm raw}$ and
$\sigma_{\rm raw}$ are computed from the complete set of raw predictions
on the held-out test interval, this affine rescaling constitutes an
offline batch post-processing step. It does not use the test targets and
therefore does not affect the fitted readout parameters, reservoir
features, or learned readout weights, but it is not intended to represent
strictly causal online forecasting. In all simulations reported here,
$\sigma_{\rm raw}>0$, so the rescaling in
Eq.~\eqref{eq:test_rescaling} is well defined.

The prediction horizon is five integration steps. Since
$\Delta t=0.01$, this corresponds to
\begin{equation}
T_{\mathrm{pred}}
=
5\Delta t
=
0.05
\label{eq:prediction_horizon}
\end{equation}
in the dimensionless time units of the FHN model.

Generalization performance is evaluated on the held-out test interval
using the root mean square error
\begin{equation}
\mathrm{RMSE}
=
\left[
\frac{1}{|\mathcal{I}_{\mathrm{test}}|}
\sum_{n\in\mathcal{I}_{\mathrm{test}}}
\left(
\widehat{s}_{n+5}-s_{n+5}
\right)^2
\right]^{1/2}.
\label{eq:lsm_rmse}
\end{equation}
Smaller RMSE therefore corresponds to more accurate forecasting of the aperiodic input.


\section{\label{sec:results_discussion}Results and discussion}
We begin by characterizing how quenched heterogeneity in coupling
strengths and time delays reshapes the noise-induced response of
the excitable network. Periodic forcing provides a particularly clean
probe of this effect because the response can be resolved at the imposed
frequency $\omega$ through the spectral measure $Q$ introduced in Sec.~\ref{subsec:sr_measure_periodic}. We therefore use periodic
stochastic resonance (PSR) to map systematically the response landscape
generated by Gaussian, bimodal, and shifted-exponential heterogeneity in
the coupling strengths and time delays.
For coupling-strength heterogeneity, full parameter maps under aperiodic forcing exhibit qualitatively similar trends and are therefore not shown separately; selected noise-dependent sections are presented below together with the LSM prediction error.

The aperiodically driven system is instead used to address the
computational consequences of stochastic resonance. For the LSM
experiments, the weak aperiodic signal introduced in
Sec.~\ref{subsec:lsm_protocol} drives the same recurrent-network model,
and the normalized zero-lag coherence $\bar{Q}$ quantifies the extent to
which the population response retains the temporal structure of the
input. The central question is then whether changes in this
noise-assisted input--output coherence are accompanied by corresponding
changes in the prediction error of the linear readout. Thus, PSR serves
as a controlled dynamical probe of the resonance structure, whereas ASR
is used to determine whether the same noise--heterogeneity interplay
produces a dynamically informative reservoir state and, consequently,
improved LSM performance.

Before introducing heterogeneity, we first identify a deterministic
subthreshold reference regime. This step is essential because the
resonant responses discussed below should originate from
noise-assisted threshold crossings rather than from deterministic
spiking generated directly by the external forcing or by the recurrent
coupling.

\begin{figure*}
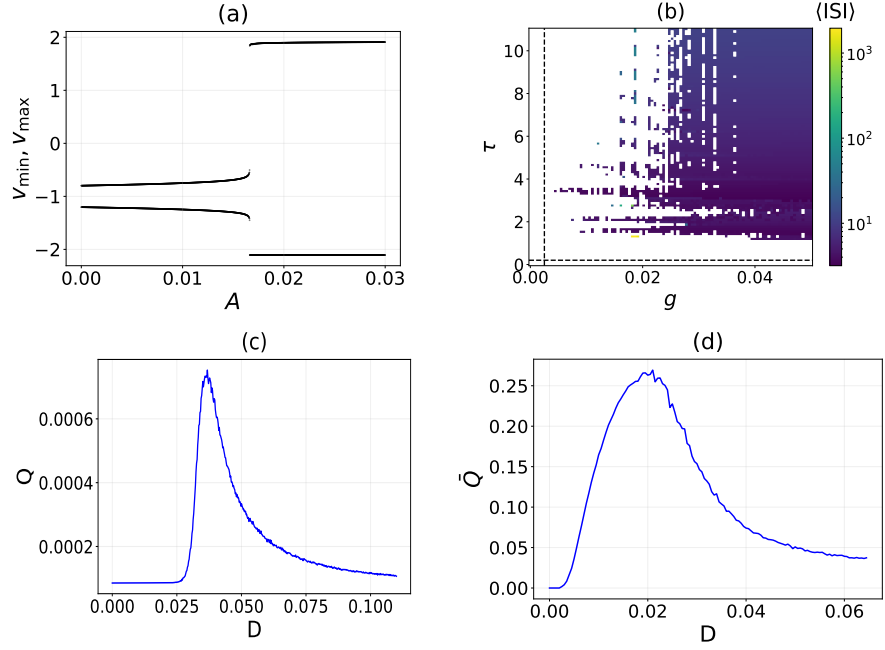

\centering
\includegraphics[width=0.32\textwidth,height=0.24\textwidth]
{bif\_a.pdf}
\hspace{0.5em}
\includegraphics[width=0.32\textwidth,height=0.24\textwidth]
{bif\_g\_tau.pdf}
\\
\includegraphics[width=0.32\textwidth,height=0.24\textwidth]
{psr\_hom\_Q.pdf}
\includegraphics[width=0.32\textwidth,height=0.24\textwidth]
{asr\_hom\_C1.pdf}
\caption{\label{fig:1}Deterministic subthreshold regime and baseline stochastic
resonance. (a) Membrane-potential extrema $v_{\min}$ and $v_{\max}$
versus forcing amplitude $A$ for an isolated FHN neuron with
$\omega=0.08$, $a_0=1.0$, $\varepsilon=0.03$ and $D=0$; repetitive spiking begins at
$A_{\mathrm{th}}\approx0.016$. (b) Mean interspike interval
$\langle\mathrm{ISI}\rangle$ in the $(g,\tau)$ plane for the homogeneous
Watts--Strogatz network with $N=50$, $A=0.015$, and $D=0$; white regions
indicate silent dynamics, and dashed lines mark $\tau=0.3$ and
$g=0.0025$. (c) Periodic response $Q(D)$ at $g = 0.0162$, $\tau = 0.3$ and (d) normalized zero-lag
aperiodic coherence $\bar{Q}(D)$ at $g = 0.01859$, $\tau = 0.3$, for the homogeneous subthreshold network, both exhibiting a maximum at intermediate noise amplitude.}
\end{figure*}

Figure~\ref{fig:1}(a) shows the deterministic response of an isolated
FHN neuron as the forcing amplitude is varied. For $A<A_{\mathrm{th}}\approx0.016$, both membrane-potential extrema remain
on the negative-voltage branch and the oscillation amplitude remains
small, corresponding to subthreshold motion around the excitable rest
state. At $A\approx A_{\mathrm{th}}$, the voltage excursion increases
abruptly and $v_{\max}$ becomes positive, signaling the onset of
repetitive spiking. We therefore choose
$A=0.015<A_{\mathrm{th}}$ for the periodically driven simulations, so
that the external forcing alone remains below the deterministic firing
threshold.

The isolated-neuron condition is not sufficient, however, because
recurrent coupling can itself destabilize the silent state. We therefore
examine in Fig.~\ref{fig:1}(b) the deterministic homogeneous network,
with $g_{ij}=g$, $\tau_{ij}=\tau$, and $D=0$. The mean interspike
interval reveals a structured partition of the $(g,\tau)$ plane into
spiking and silent regions. The white regions correspond to parameter
values for which no spikes are detected over the observation interval.
The cuts $\tau=0.3$ for coupling-strength variations and $g=0.0025$ for
delay variations are chosen so that the corresponding reference dynamics
remain subthreshold in the absence of noise. These parameter paths
therefore provide a controlled background against which the effect of
stochastic forcing and quenched heterogeneity can be isolated.

Having established this deterministic reference state, Figs.~\ref{fig:1}(c) and \ref{fig:1}(d) show the response of the
homogeneous network when noise is introduced. In both the periodically
and aperiodically driven cases, the response is nonmonotone in the noise
amplitude. For weak noise, threshold-crossing events are too rare to
produce a strong coherent response. At intermediate noise amplitude,
the stochastic fluctuations cooperate with the weak forcing and produce
a pronounced maximum in $Q$ or $\bar{Q}$. For still larger $D$, the
dynamics become increasingly dominated by stochastic fluctuations and
the coherence with the external signal decreases. The resulting
bell-shaped response establishes the stochastic-resonance regime that
serves as the homogeneous reference for the heterogeneity-dependent
analysis below.

\begin{figure*}[ht]
\centering
\includegraphics[height=6cm]
{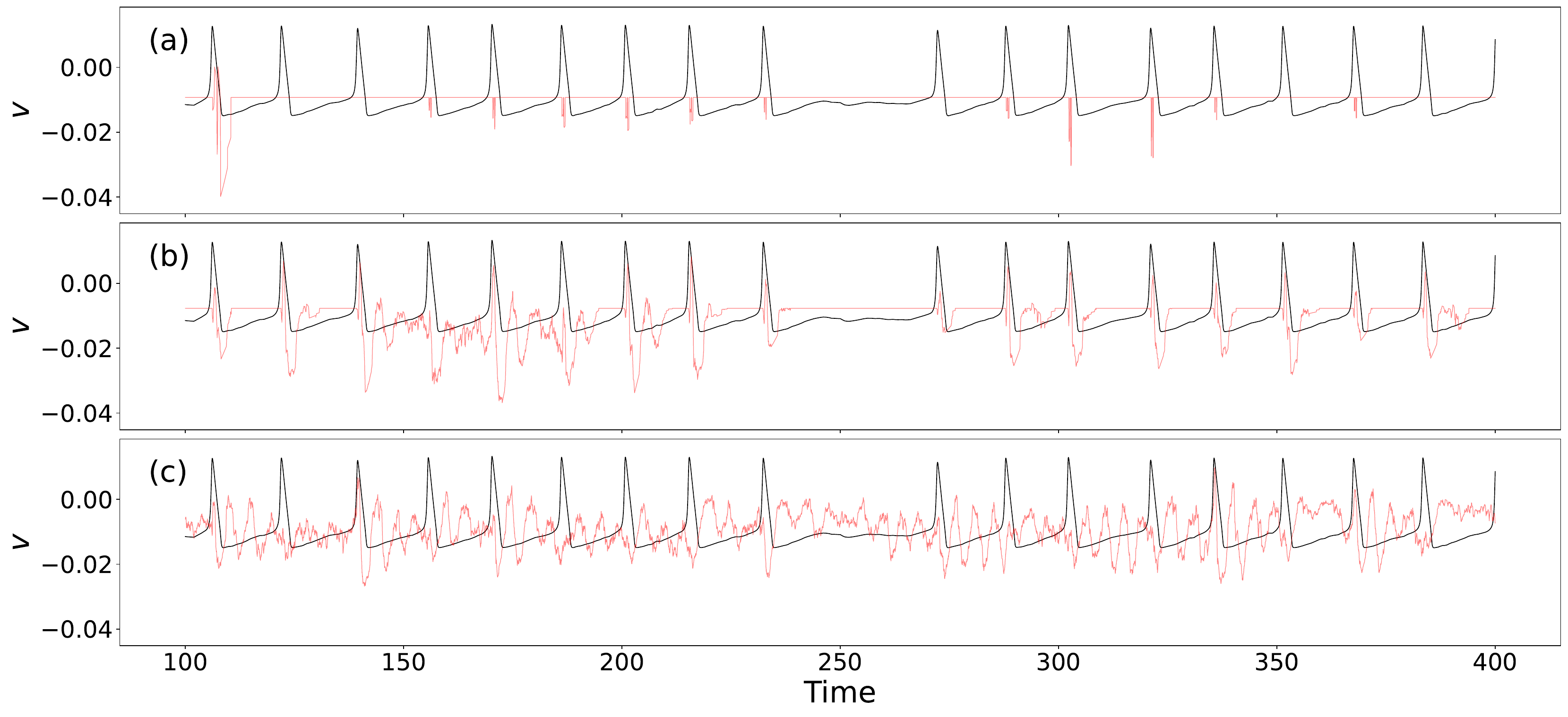}
\caption{\label{fig:ts_rmse}
Noise-dependent LSM forecasting across the stochastic-resonance regime.
Representative segment of the held-out test set showing the target aperiodic signal (black) and the corresponding five-step-ahead LSM prediction (red) for a homogeneous FHN reservoir with $g=0.01859$ and $\tau=0.3$.
The noise amplitudes are (a) $D=0.006$, (b) $D=0.022$, and (c) $D=0.042$, yielding test errors $\mathrm{RMSE}=0.00800$, $0.00721$, and $0.00820$, respectively.
The intermediate-noise regime produces the closest target--prediction agreement and the minimum prediction error, consistent with noise-assisted signal encoding near the stochastic-resonance optimum.
}
\end{figure*}

 We now see that in Fig.~\ref{fig:ts_rmse}, which overlays the target and predicted signals for different noise amplitudes $D$, the prediction tracks the signal markedly better at the intermediate amplitude $D = 0.022$ in Fig.~\ref{fig:ts_rmse}(b) (RMSE $= 0.00721$) than at the low-noise value $D = 0.006$ in Fig.~\ref{fig:ts_rmse}(a) (RMSE $= 0.00800$) or the high-noise value $D = 0.042$ in Fig.~\ref{fig:ts_rmse}(c) (RMSE $= 0.00820$), illustrating the dependence of readout quality on stochastic resonance.

The next question to address is therefore not
whether heterogeneity creates stochastic resonance from an otherwise
nonresonant system, but how quenched disorder in coupling strengths deforms this underlying resonance landscape. In
particular, we examine whether heterogeneity changes the amplitude and
location of the resonance maximum, enlarges or suppresses
high-response regions in parameter space, and, under aperiodic forcing,
whether the corresponding modification of input--output coherence is
reflected in the predictive performance of the liquid state machine. 

It is worth noting that under aperiodic forcing, varying the time delay moves a large fraction of the relevant $(g,\tau)$  parameter space into a deterministically spiking regime. To retain a controlled subthreshold operating state in the LSM experiments, we therefore fix the delay at $\tau=0.3$ and investigate computational effects only for coupling-strength heterogeneity.

\begin{figure*}
\centering
\includegraphics[width=0.32\textwidth]
{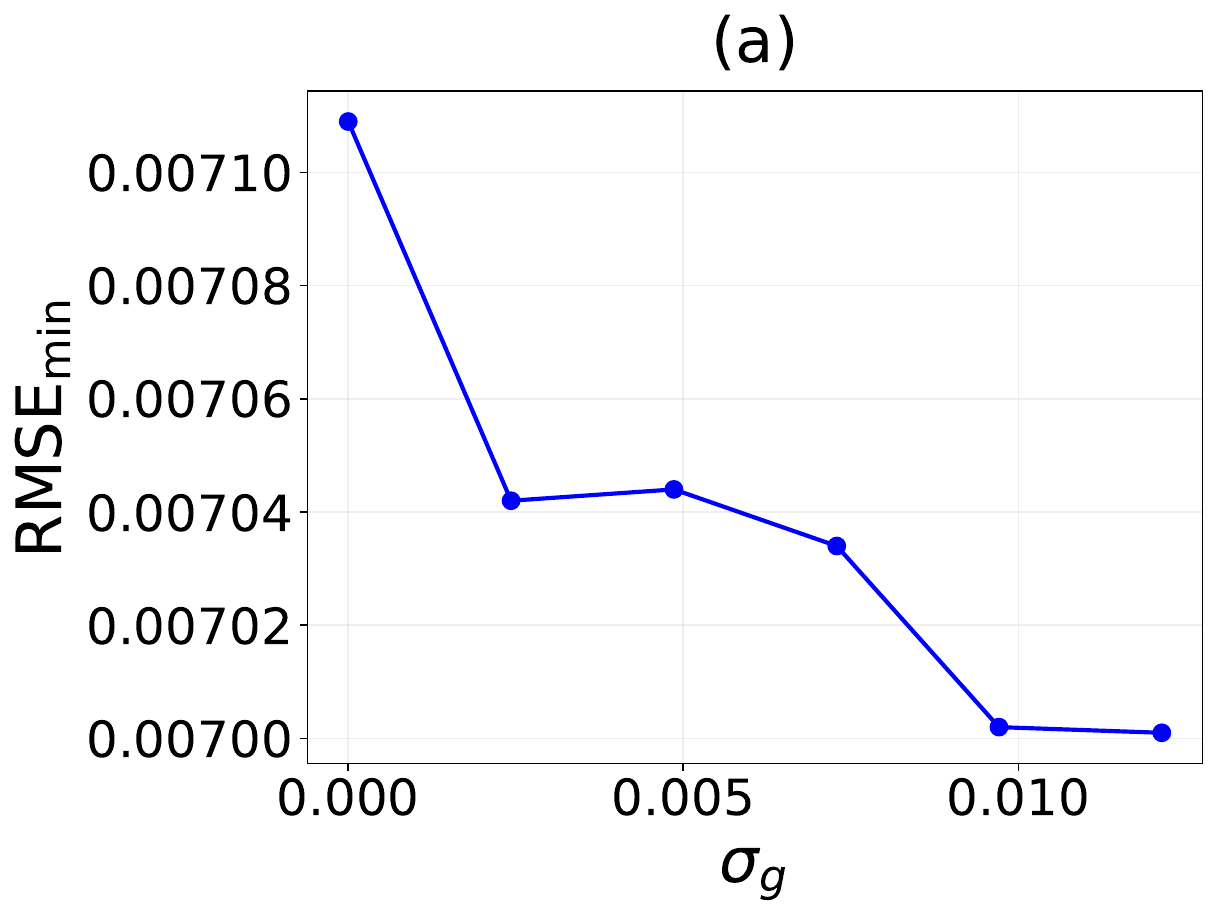}
\includegraphics[width=0.32\textwidth]
{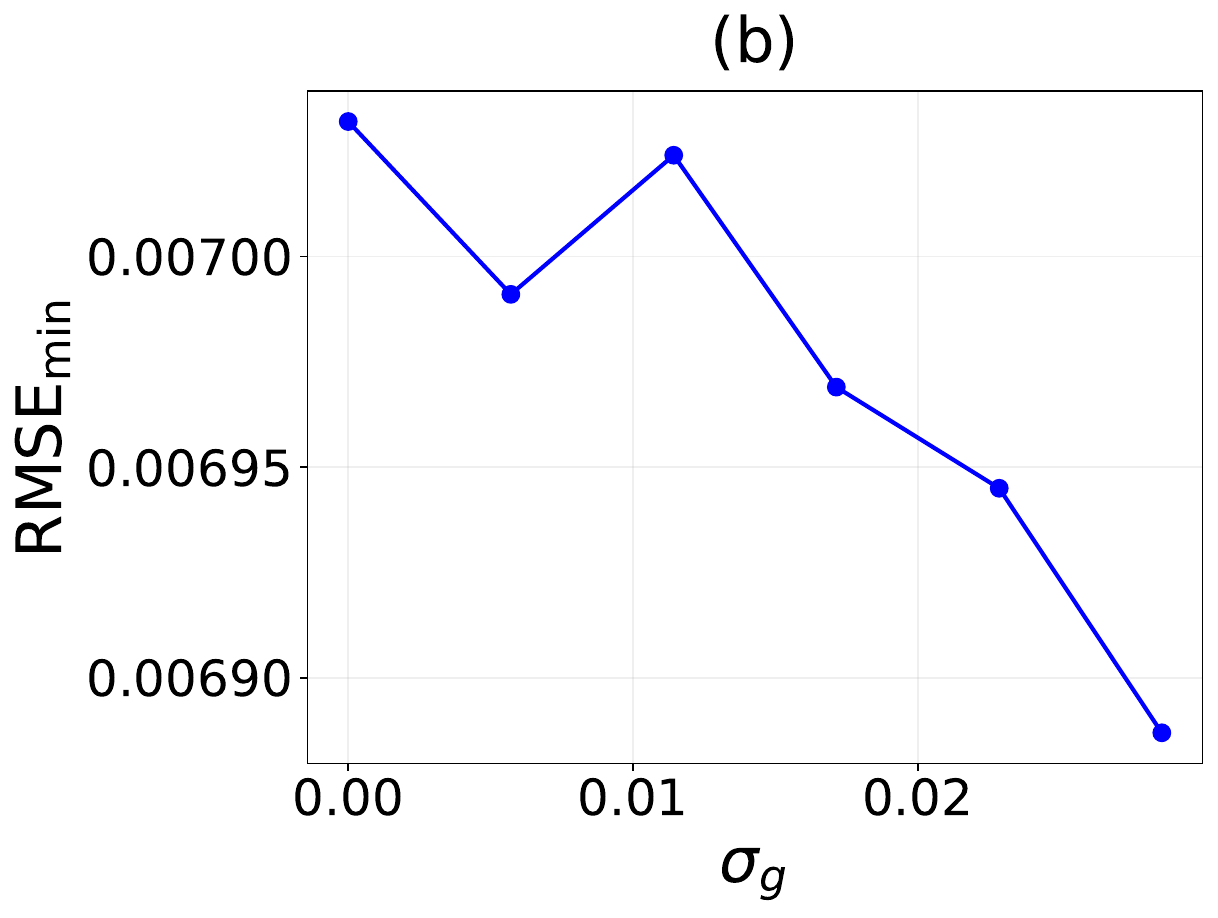}
\includegraphics[width=0.32\textwidth]
{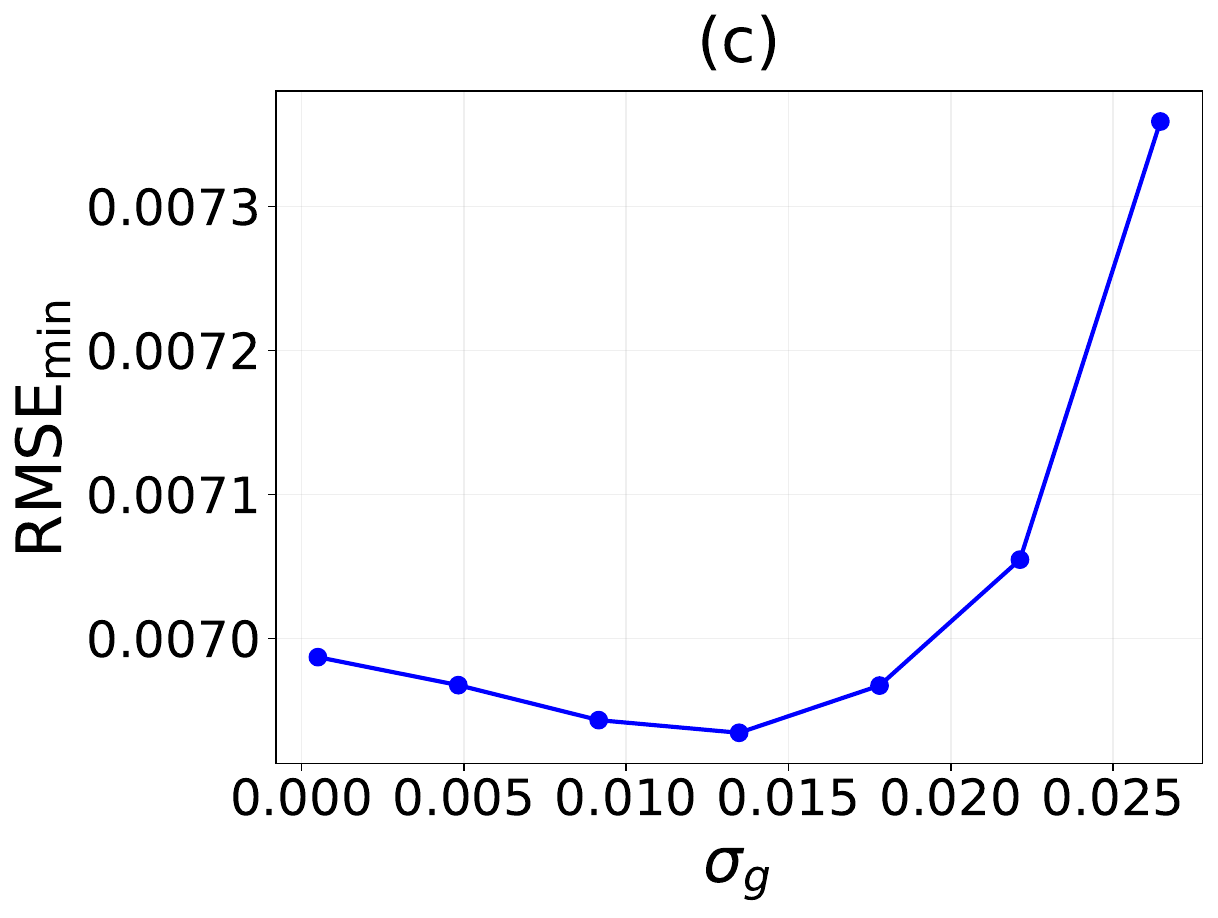}
\caption{Distribution-dependent reorganization of the noise-optimized LSM prediction error by quenched coupling disorder. Here,
$\mathrm{RMSE}_{\min}(\sigma_g)=\min_{D\in\mathcal{D}}\mathrm{RMSE}(\sigma_g,D)$
for prediction horizon $h=5$.
(a) Gaussian disorder at $\tau=0.3$ and $\mu_g=0.01859$: increasing $\sigma_g$ produces an overall modest reduction of $\mathrm{RMSE}_{\min}$.
(b) Bimodal disorder at $\tau=0.3$ and $\mu_g=0.0162$: increasing the mode separation yields a larger overall reduction for the parameter values.
(c) Shifted-exponential disorder at $\tau=0.3$ and $\mu_g=0.005$: $\mathrm{RMSE}_{\min}$ decreases weakly at intermediate scales and rises markedly for larger $\sigma_g=1/\lambda_g$, reflecting the one-sided, non-mean-preserving redistribution of coupling strengths.}\label{fig:min_rmse}
\end{figure*}

In the sequel, we fix $\mu_g$, $\tau$ and define
the noise-optimized prediction error as
\begin{equation}
\mathrm{RMSE}_{\min}(\sigma_g)
=
\min_{D\in\mathcal{D}}
\mathrm{RMSE}(\sigma_g,D).
\label{eq:rmse_min}
\end{equation}
so that $\mathrm{RMSE}_{\min}$ represents the lower envelope of the
noise-dependent prediction-error landscape. In this representation,
changes in $\mathrm{RMSE}_{\min}$ directly quantify whether increasing
coupling heterogeneity improves or degrades the optimal computational
state accessible to the noisy liquid.

For Gaussian coupling disorder, Fig.~\ref{fig:min_rmse}(a), increasing
$\sigma_g$ produces an overall modest reduction of
$\mathrm{RMSE}_{\min}$. A larger overall reduction is observed for the bimodal case shown in  Fig.~\ref{fig:min_rmse}(b), where increasing the separation between the two coupling scales lowers the minimum prediction error.
These trends are consistent
with the dynamical picture developed below: symmetric coupling
disorder can be interpreted as broadening the set of effective local
activation and propagation scales available to the recurrent network,
thereby allowing noise-assisted reservoir states with stronger input
encoding and improved generalization to be reached. In particular, the
coexistence of weak and strong coupling sectors in the bimodal case is consistent with a two-scale organization of effective local
susceptibilities, in which informative collective states can be accessed
with reduced stochastic forcing.

A qualitatively different, nonmonotonic trend is observed for the shifted-exponential distribution in Fig.~\ref{fig:min_rmse}(c). $\mathrm{RMSE}_{\min}$ initially decreases slightly, reaches a shallow minimum at intermediate $\sigma_g$ and then increases markedly at larger exponential scales.
This behavior should not be interpreted as a generic destructive effect
of heterogeneity alone. Unlike the Gaussian and bimodal families, the
shifted-exponential distribution is one-sided and non-mean-preserving:
at fixed lower shift $\mu_g$, increasing $\sigma_g$ simultaneously
broadens the distribution and shifts its unprojected mean, $\mathbb{E}[\widetilde g_{ij}]=\mu_g+\sigma_g$, toward larger coupling
strengths. The control parameter $\sigma_g$ therefore moves the
reservoir through coupling space while increasing the strength of the
quenched disorder. For the present parameter regime, this redistribution
drives statistical weight toward coupling sectors with weaker coherent
response and consequently raises the lowest prediction error attainable
over $D$.

The opposite signs of the trends in Fig.~\ref{fig:min_rmse} therefore
show that computational enhancement is not determined by the magnitude
of heterogeneity alone, but by the manner in which the coupling
distribution reorganizes the effective dynamical response landscape
of the liquid. Symmetric Gaussian and bimodal disorder can lower the
noise-optimized prediction error, whereas the asymmetric,
non-mean-preserving shifted-exponential disorder degrades, at sufficiently large scales, 
the globally optimized performance over the parameter range considered.
This distinction does not exclude a finite-noise computational benefit
at a fixed shifted-exponential scale: sufficiently broad distributions
can still exhibit a noise interval in which prediction improves relative
to their own weak-noise state. Rather, Fig.~\ref{fig:min_rmse} addresses
the complementary question of how the \emph{best} prediction error,
after optimization over the noise amplitude, changes as the coupling
distribution itself is made increasingly heterogeneous.

\subsection{\label{sec:res_g_gauss}Stochastic resonance and LSM performance under Gaussian heterogeneity}

For Gaussian heterogeneity, $\mu_g$ and $\mu_\tau$ denote the means of the
unprojected coupling-strength and time-delay distributions, while
$\sigma_g$ and $\sigma_\tau$ measure their dispersions. Increasing either
standard deviation therefore broadens the corresponding unimodal distribution
about its prescribed mean. Since the samples are subsequently projected onto
their admissible intervals, the realized moments can deviate from those of the
underlying Gaussian distribution when appreciable probability mass reaches a
boundary. See Fig.~\ref{fig:g_gauss_pdfs}(a1)--(b2) in the Appendix. 

Figure~\ref{fig:gaussian_coupling}(a1) identifies three coupling
regimes. For small values of the mean coupling $\mu_g$, the maximal
response $Q_{\max}$ remains comparatively large and depends only weakly
on the coupling dispersion $\sigma_g$. By contrast, sufficiently large
values of $\mu_g$ produce a uniformly weak response. In the intermediate
transition region, approximately
$0.015\lesssim\mu_g\lesssim0.030$, increasing $\sigma_g$ shifts the
high-response region toward larger values of $\mu_g$. Thus, Gaussian
coupling heterogeneity extends the parameter domain over which
pronounced stochastic resonance occurs.

The section at $\mu_g=0.01859$, shown in
Fig.~\ref{fig:gaussian_coupling}(a2), resolves the corresponding
dependence on the noise amplitude. A well-defined resonance ridge is
observed at an intermediate value of $D$. Its amplitude increases with
$\sigma_g$, while the location of the optimal noise amplitude changes
only weakly. In this parameter regime, Gaussian coupling heterogeneity
therefore acts primarily by amplifying the noise-induced response rather
than by substantially shifting the stochastic-resonance condition.

In the homogeneous-delay limit $\sigma_\tau=0$,
Fig.~\ref{fig:gaussian_coupling}(b1) displays a strongly nonmonotonic
dependence of $Q_{\max}$ on the mean delay $\mu_\tau$, characterized by
alternating local maxima and minima. As $\sigma_\tau$ increases, the
contrast between these high- and low-response bands progressively
decreases: local maxima are attenuated, whereas local minima are elevated.
From a dynamical perspective, this behavior is consistent with
distribution-weighted averaging over a broader range of delay-dependent
response states, which smooths the structured dependence on $\mu_\tau$
and makes the network response less sensitive to the mean delay.

The panels in Figs.~\ref{fig:gaussian_coupling}(b2) and
\ref{fig:gaussian_coupling}(b3) illustrate the two possible effects of
delay heterogeneity. At $\mu_\tau=1.207$, which lies near a
low-response region of the homogeneous system, introducing delay
dispersion increases the maximum of the noise-dependent response. At
$\mu_\tau=1.5$, which lies near a high-response region, increasing
$\sigma_\tau$ reduces the resonance maximum. Hence, Gaussian delay
heterogeneity is neither intrinsically constructive nor destructive:
its effect is determined by the position of the mean delay relative to
the high- and low-response regions of the homogeneous network.

\begin{figure*}
\centering
\includegraphics[width=0.32\textwidth]
{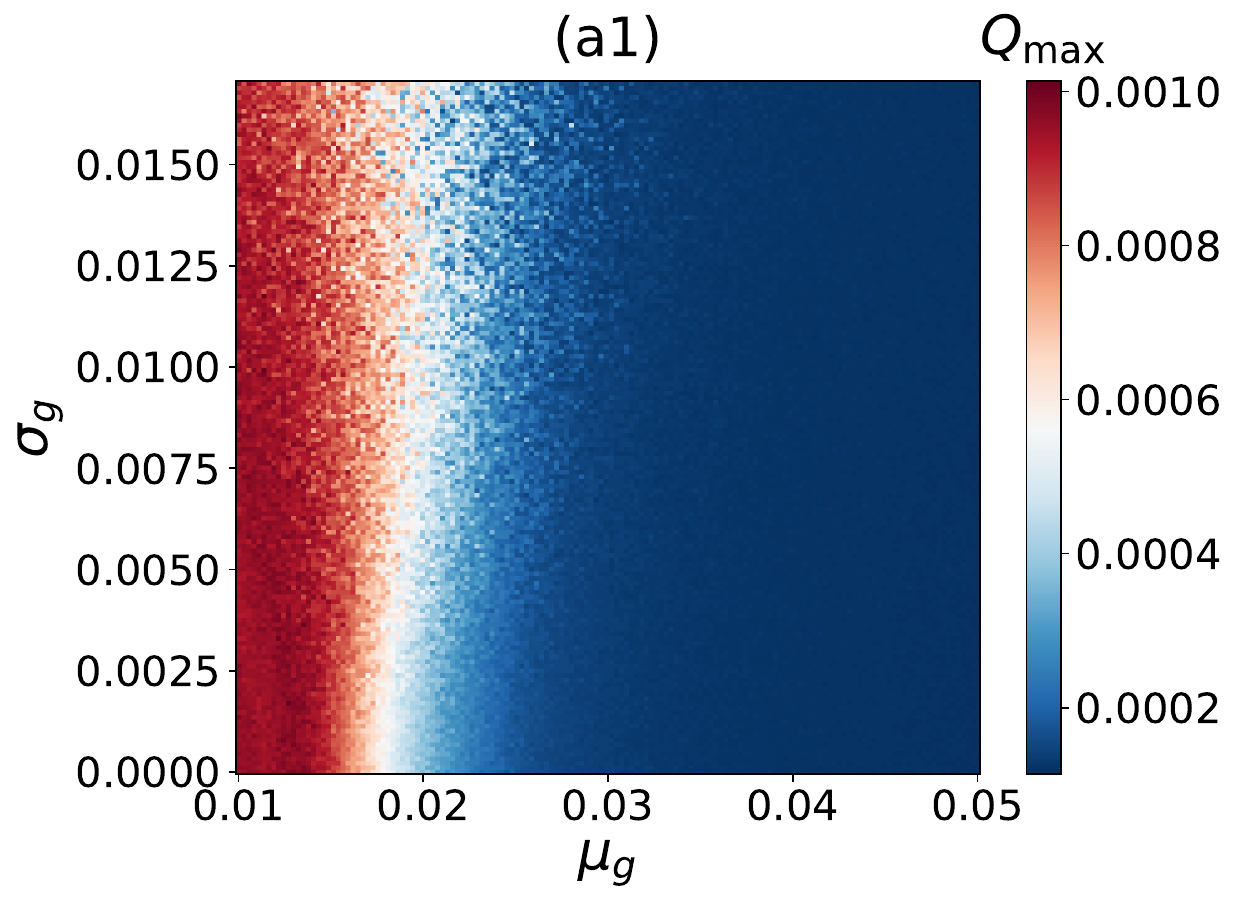}
\includegraphics[width=0.32\textwidth]
{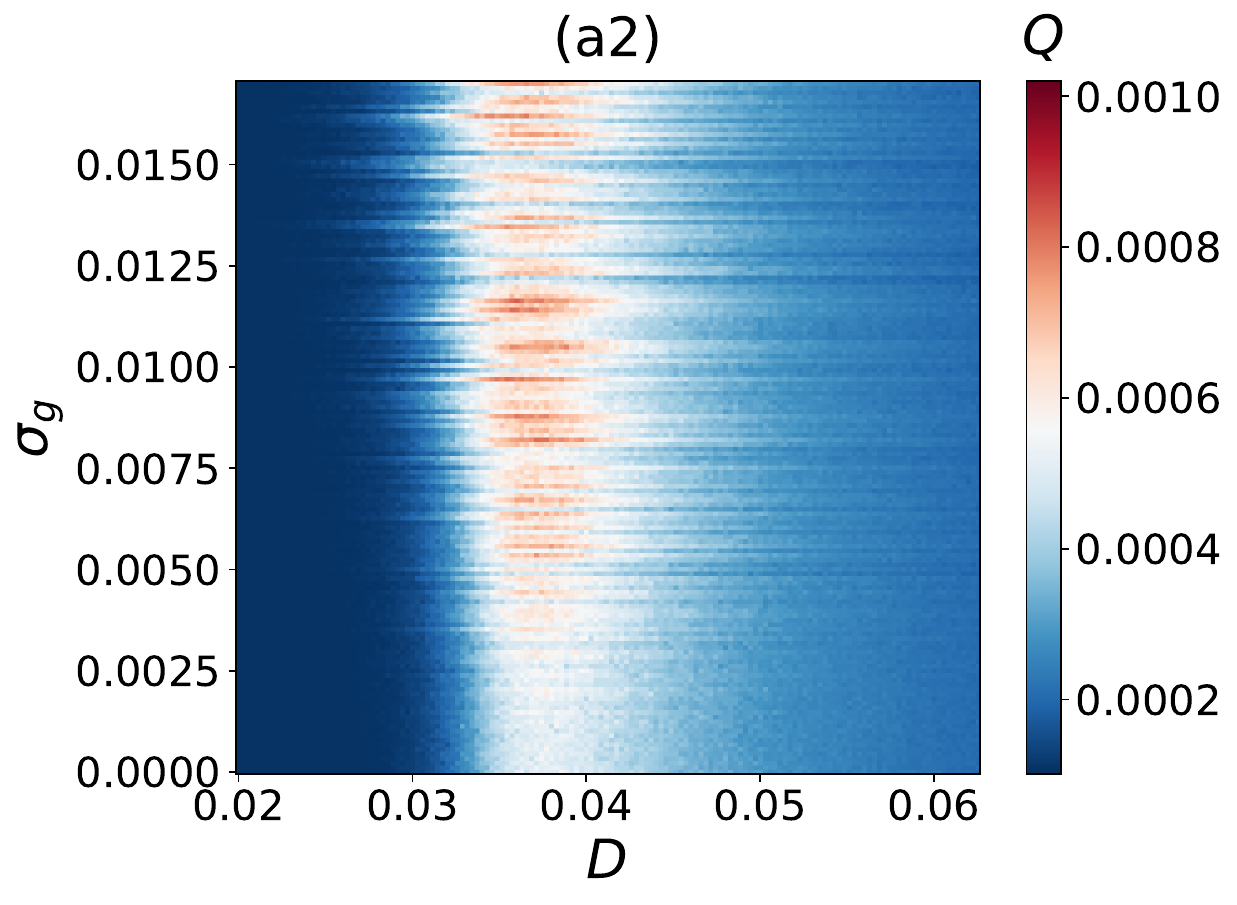}
\\
\includegraphics[width=0.32\textwidth]
{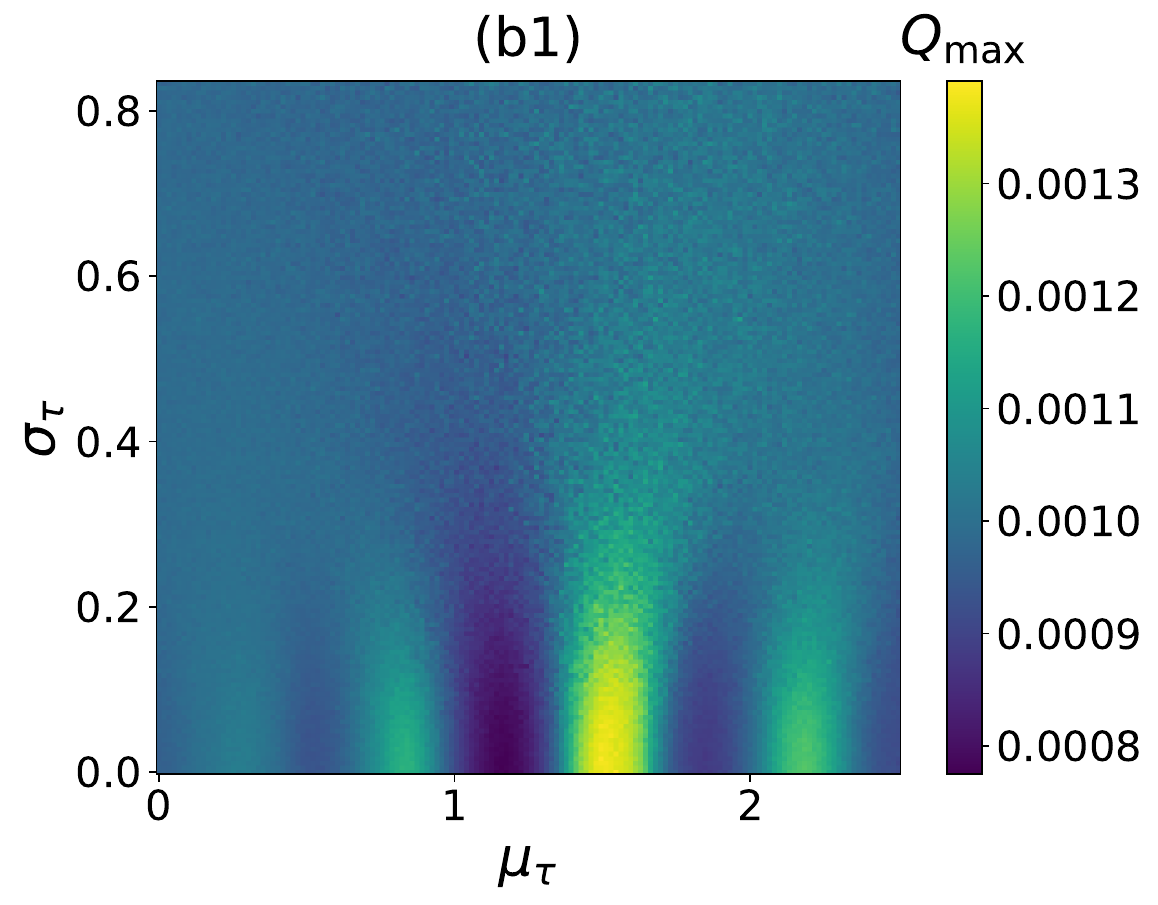}
\hfill
\includegraphics[width=0.32\textwidth]
{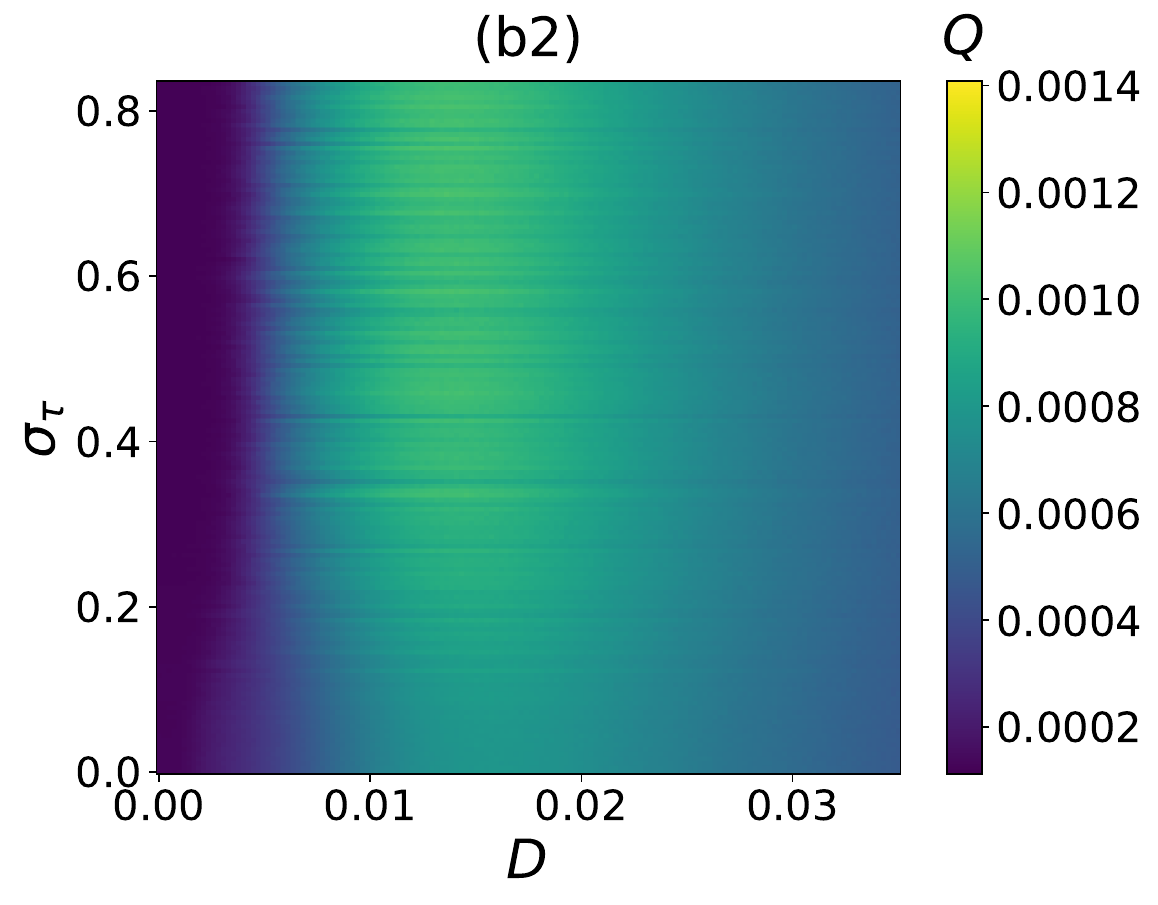}
\hfill
\includegraphics[width=0.32\textwidth]
{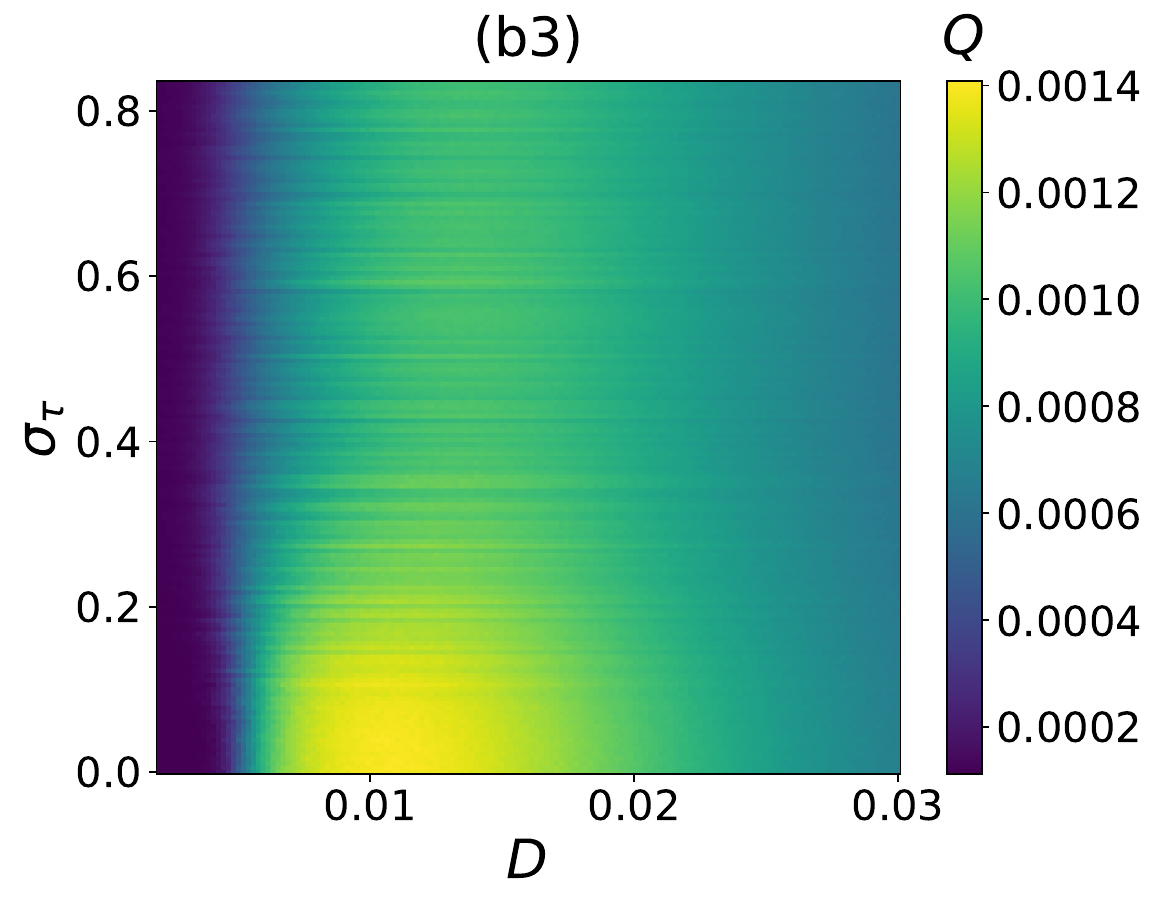}
\caption{Periodic stochastic resonance under Gaussian heterogeneity. For coupling-strength heterogeneity at $\tau=0.3$, (a1) shows $Q_{\max}(\mu_g,\sigma_g)$ and (a2) $Q(D,\sigma_g)$ at $\mu_g=0.01859$. For time-delay heterogeneity at $g=0.0025$, (b1) shows $Q_{\max}(\mu_\tau,\sigma_\tau)$ and (b2)--(b3) $Q(D,\sigma_\tau)$ at $\mu_\tau=1.207$ and $1.5$, respectively. Coupling dispersion extends the strong-response region, whereas time-delay dispersion can enhance or suppress the resonance depending on the mean delay.} \label{fig:gaussian_coupling}
\end{figure*}

\begin{figure*}
\centering
\includegraphics[width=0.32\textwidth]
{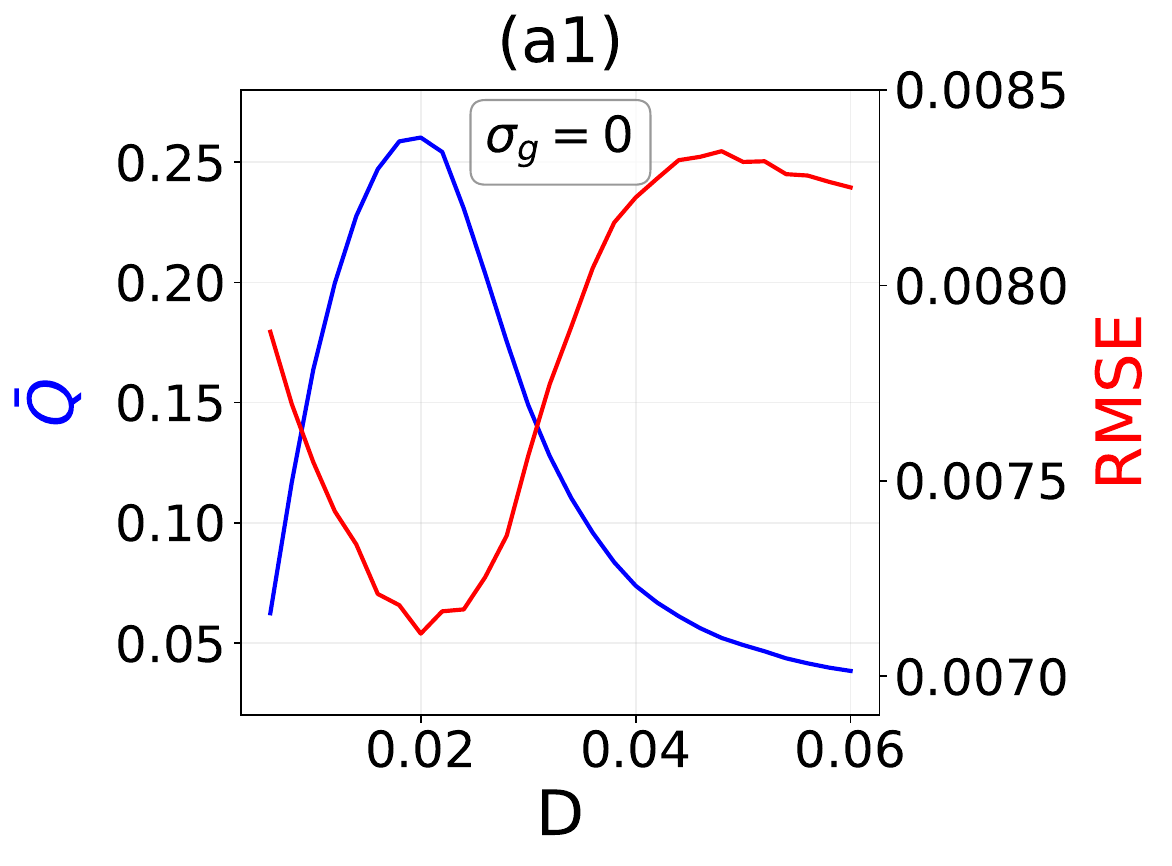}
\includegraphics[width=0.32\textwidth]
{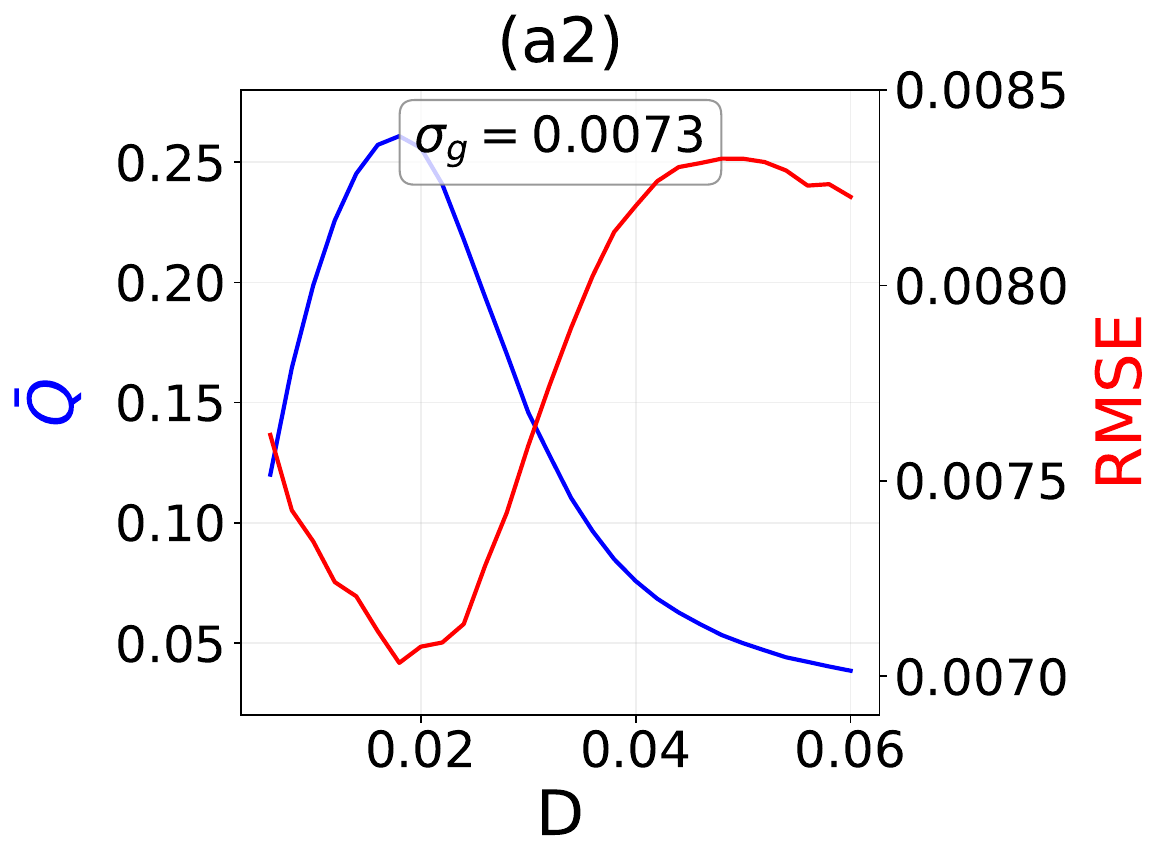}
\includegraphics[width=0.32\textwidth]
{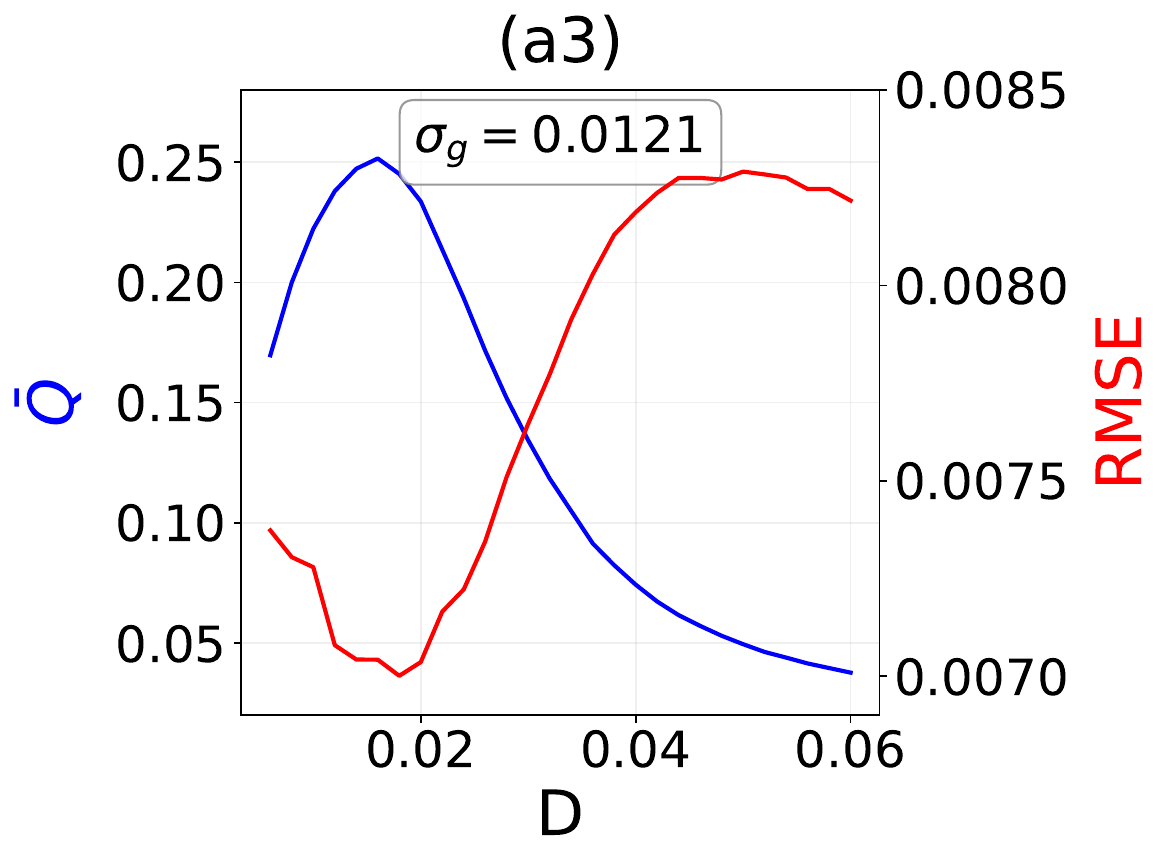}
\caption{Aperiodic input--output coherence and LSM prediction performance under Gaussian coupling heterogeneity. The normalized zero-lag coherence $\bar{Q}$ (blue) and test RMSE (red) are shown versus $D$ at $\tau=0.3$ and $\mu_g=0.01859$ for (a1)--(a3) $\sigma_g=0$, $0.0073$, and $0.0121$, respectively. Increasing the coupling dispersion shifts the finite-noise optimum toward weaker noise.}\label{fig:gaussian_asr_lsm}
\end{figure*}

We next ask whether the modification of stochastic resonance by
Gaussian heterogeneity observed under periodic forcing has a
computational counterpart under aperiodic driving.
Figure~\ref{fig:gaussian_asr_lsm} therefore compares the normalized
zero-lag aperiodic input--output coherence $\bar{Q}$ with the test-set
prediction error as the noise amplitude $D$ is varied. Since the
periodic and aperiodic experiments employ different driving signals and
different response measures, their optimal noise amplitudes need not
coincide. The coupling-strength sections in
Figs.~\ref{fig:gaussian_coupling}(a2) and
\ref{fig:gaussian_asr_lsm}(a1)--(a3) are evaluated at the same mean
coupling strength, $\mu_g=0.01859$. Thus, quantitative differences
between the corresponding PSR and ASR responses arise from the different
driving protocols and response observables rather than from a change in
the mean coupling strength.
Accordingly, the relevant question is not whether PSR and ASR reproduce
identical resonance curves, but whether enhanced input--output coherence
is accompanied by improved readout performance of the liquid state machine.

A pronounced inverse relation between $\bar{Q}$ and the RMSE is
observed, most clearly for coupling-strength heterogeneity: as the
noise-induced zero-lag coherence increases, the prediction error
decreases, whereas the reduction of coherence at larger noise
amplitudes is accompanied by a deterioration of prediction accuracy.
Thus, the computationally useful noise regime is not simply the one
producing frequent threshold crossings, but rather the one in which the
stochastic response remains strongly correlated with the temporal
structure of the weak aperiodic input.

For Gaussian coupling-strength heterogeneity,
Figs.~\ref{fig:gaussian_asr_lsm}(a1)--(a3), the homogeneous network
exhibits the characteristic nonmonotonic dependence of $\bar{Q}$ on
$D$. Starting from weak noise, the coherence increases, reaches a
maximum at an intermediate noise amplitude, and subsequently decreases
as the stochastic forcing becomes too strong. The RMSE exhibits the
complementary behavior: it decreases as the resonance develops, reaches
a minimum within approximately the same intermediate-noise window, and
then increases once the reservoir dynamics become increasingly
noise-dominated. The coincidence of large $\bar{Q}$ with small RMSE
shows that noise-enhanced encoding of the aperiodic input is directly
reflected in the predictive performance of the liquid state machine.

Increasing the coupling dispersion from $\sigma_g=0$ to $0.0073$ and
$0.0121$ shifts both the maximum of $\bar{Q}$ and the minimum of the
prediction error toward smaller values of $D$. The maximal coherence
changes comparatively little, whereas the minimum RMSE is slightly
reduced, most clearly for the largest coupling dispersion. Thus, in the
aperiodically driven liquid, the dominant computational effect of
Gaussian coupling heterogeneity is a displacement of the optimal
noise window toward weaker stochastic forcing.

This shift is more pronounced than in the periodic-response curves of
Fig.~\ref{fig:gaussian_coupling}(a2), where increasing $\sigma_g$
primarily enhances the resonance amplitude and changes the optimal
noise amplitude only weakly. This difference does not constitute an
inconsistency: PSR and ASR probe different temporal structures of the
external forcing and employ different response observables, while the
mean coupling strength $\mu_g=0.01859$ is the same in both cases.
Rather, the two figures show complementary consequences of coupling
heterogeneity. Under periodic forcing, heterogeneity enlarges and
strengthens the high-response region, whereas under aperiodic forcing it
can additionally reduce the stochastic forcing required to reach the
most informative dynamical state of the liquid.

From a dynamical-systems perspective, quenched coupling disorder can be
viewed as broadening the landscape of effective local activation scales
and propagation strengths within the recurrent network. More weakly
coupled sectors can retain sensitivity to local stochastic excursions,
whereas stronger pathways can facilitate recruitment and redistribution
once an excursion has occurred. It is therefore the coexistence of
different coupling scales, rather than a uniform strengthening of the
network, that provides a natural interpretation of the shift in the
optimal noise scale. The collective effect is to move the regime of
maximal input--output coherence, and hence optimal liquid state machine performance,
toward weaker noise.

Figure~\ref{fig:gaussian_asr_lsm} therefore establishes a direct
connection between noise-enhanced zero-lag input--output coherence and
liquid-state computation, most clearly for coupling-strength
heterogeneity. Predictive accuracy is highest when the noisy excitable
reservoir preserves the temporal structure of the weak input and
deteriorates once excessive stochastic forcing reduces this
input--response coherence. Gaussian coupling heterogeneity can shift
this computationally optimal regime toward weaker noise and slightly
reduce the minimum prediction error. 

\subsection{\label{sec:res_g_2gmm}Stochastic resonance and LSM performance under bimodal heterogeneity}

For bimodal heterogeneity, $\mu_g$ and $\mu_\tau$ specify the centers
of the unprojected coupling-strength and time-delay mixtures,
whereas $\sigma_g$ and $\sigma_\tau$ quantify the separation between
the two component means. Specifically, the two modes are centered at
$\mu_g\pm\sigma_g/2$ and $\mu_\tau\pm\sigma_\tau/2$, respectively,
while the widths of the individual Gaussian components remain fixed.
Thus, increasing $\sigma_g$ or $\sigma_\tau$ does not simply broaden a
single distribution, as in the Gaussian case, but progressively
separates the edge-parameter distribution into two distinct coupling
scales or delay modes. The center of the underlying mixture remains fixed,
although projection onto the admissible intervals can modify the
moments of the realized edge parameters near the boundaries. See
Fig.~\ref{fig:g_2gmm_pdfs}(a)--(b) in the Appendix.

Figure~\ref{fig:2gmm_coupling}(a1) shows that the maximal periodic
response is controlled jointly by the mean coupling strength and the
separation of the two coupling modes. For sufficiently weak
$\mu_g$, $Q_{\max}$ remains large over essentially the full range of
$\sigma_g$, whereas sufficiently large values of $\mu_g$ correspond
to a uniformly weak response. In the intermediate coupling regime,
approximately $0.018\lesssim\mu_g\lesssim0.038$, increasing
$\sigma_g$ displaces the boundary between the high- and low-response
regions toward larger values of $\mu_g$. The emergence of two
distinct coupling scales in the edge-parameter distribution therefore
extends the domain in which pronounced stochastic resonance can be
sustained.

The section at $\mu_g=0.02386$, shown in
Fig.~\ref{fig:2gmm_coupling}(a2), resolves the corresponding
noise-dependent dynamics. For small component separation, the
noise-induced response is comparatively weak. As $\sigma_g$ increases,
a pronounced resonance ridge develops and its maximal amplitude grows.
At the same time, the ridge bends systematically toward smaller values
of $D$. Bimodal coupling heterogeneity therefore produces two coupled
effects: it enhances the maximal coherent response while simultaneously
reducing the stochastic forcing required to reach it.

This behavior can be interpreted in terms of a heterogeneous activation
landscape generated by two coupling scales. Increasing $\sigma_g$
creates one population of relatively weak couplings and a second of
relatively strong couplings. The former can preserve locally excitable,
noise-sensitive degrees of freedom, while the latter can promote the
propagation and recruitment of activity once a fluctuation-induced
excursion has occurred. The coexistence of these two scales broadens
the distribution of effective local susceptibilities and propagation
strengths without shifting the center of the underlying mixture.
Consequently, the collective resonance condition can be reached at a
smaller external noise amplitude.

A richer structure is obtained when the time delays are
bimodally distributed. Figure~\ref{fig:2gmm_coupling}(b1) exhibits a
strongly nonmonotone dependence of $Q_{\max}$ on both the mean delay
$\mu_\tau$ and the component separation $\sigma_\tau$. The
high-response regions organize into curved branches in the
$(\mu_\tau,\sigma_\tau)$ plane. Since the two modal delays are $\tau_-=\mu_\tau-\sigma_\tau/2$, and $\tau_+=\mu_\tau+\sigma_\tau/2,$
increasing $\sigma_\tau$ moves the two delay populations in opposite
directions while their center remains fixed. In the phenomenological
response-landscape picture introduced above, the bimodal distribution
therefore samples two separated delay sectors simultaneously. As the
two modes move through high- and low-response intervals of the
homogeneous network, the resulting collective response can alternate between enhanced and
suppressed regimes. The branch structure
in Fig.~\ref{fig:2gmm_coupling}(b1) should therefore be interpreted as
a modulation of the collective response by the two delay populations,
rather than as a direct measurement of inter-neuronal phase
cancellation.

At $\mu_\tau=1.2$,
Fig.~\ref{fig:2gmm_coupling}(b2) shows that increasing the modal
separation initially enhances the resonance response, with a local
maximum near $\sigma_\tau\approx0.6$. The response then decreases to
a local minimum around $\sigma_\tau\approx1.4$ and increases again
toward larger separations. Hence, a mean delay associated with a
relatively weak homogeneous response can be transformed into a
high-response state by appropriately separating the two delay
populations.

A complementary sequence occurs at $\mu_\tau=1.5$, as shown in
Fig.~\ref{fig:2gmm_coupling}(b3). The zero-separation unimodal delay distribution initially lies in a high-response regime. Increasing $\sigma_\tau$ first suppresses this response and produces a low-response region near
$\sigma_\tau\approx0.7$. A second high-response branch emerges around
$\sigma_\tau\approx1.4$, followed by another reduction at larger
separations. Bimodal delay heterogeneity is therefore neither
monotonically constructive nor monotonically destructive. Instead, the
response is governed by how the two modal delays sample the structured
delay-response landscape of the recurrent network.

\begin{figure*}
\centering
\includegraphics[width=0.32\textwidth]
{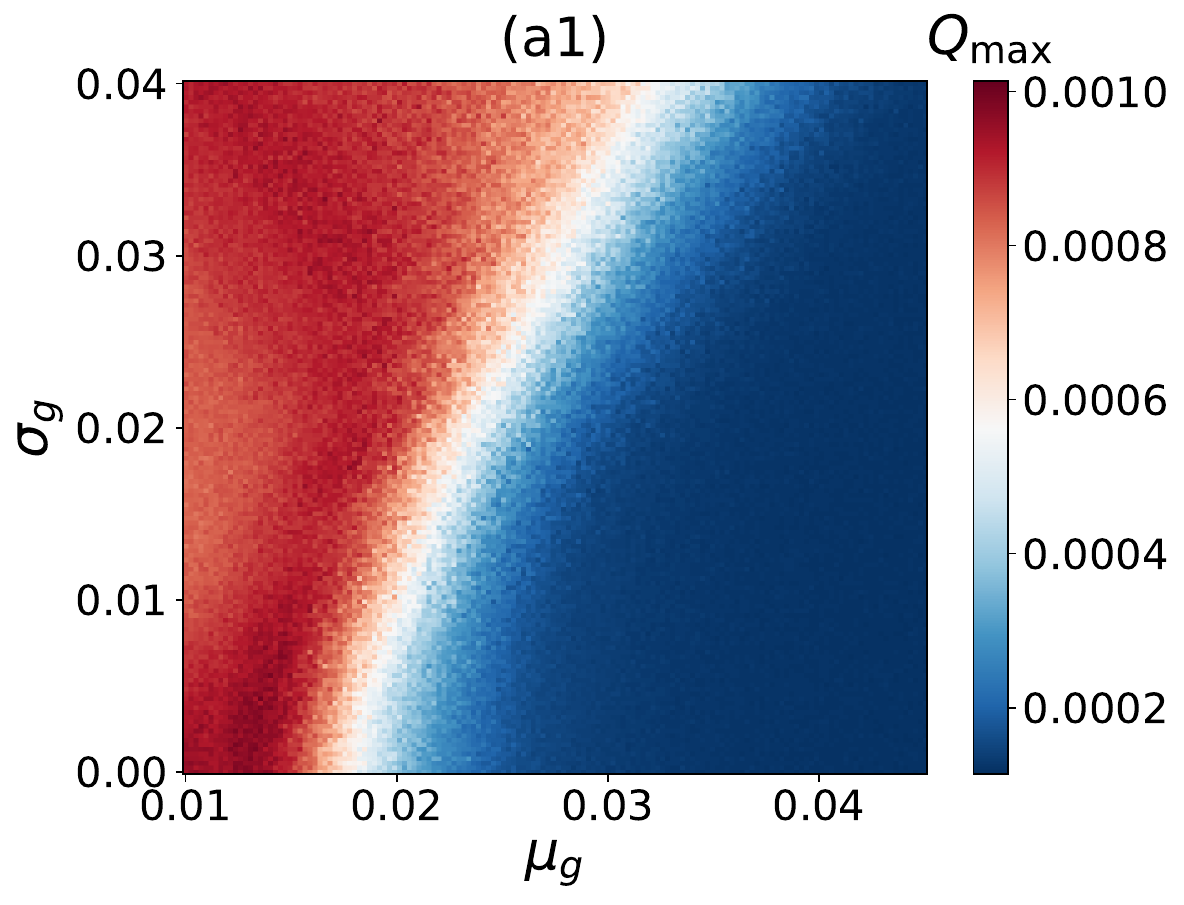}
\includegraphics[width=0.32\textwidth]
{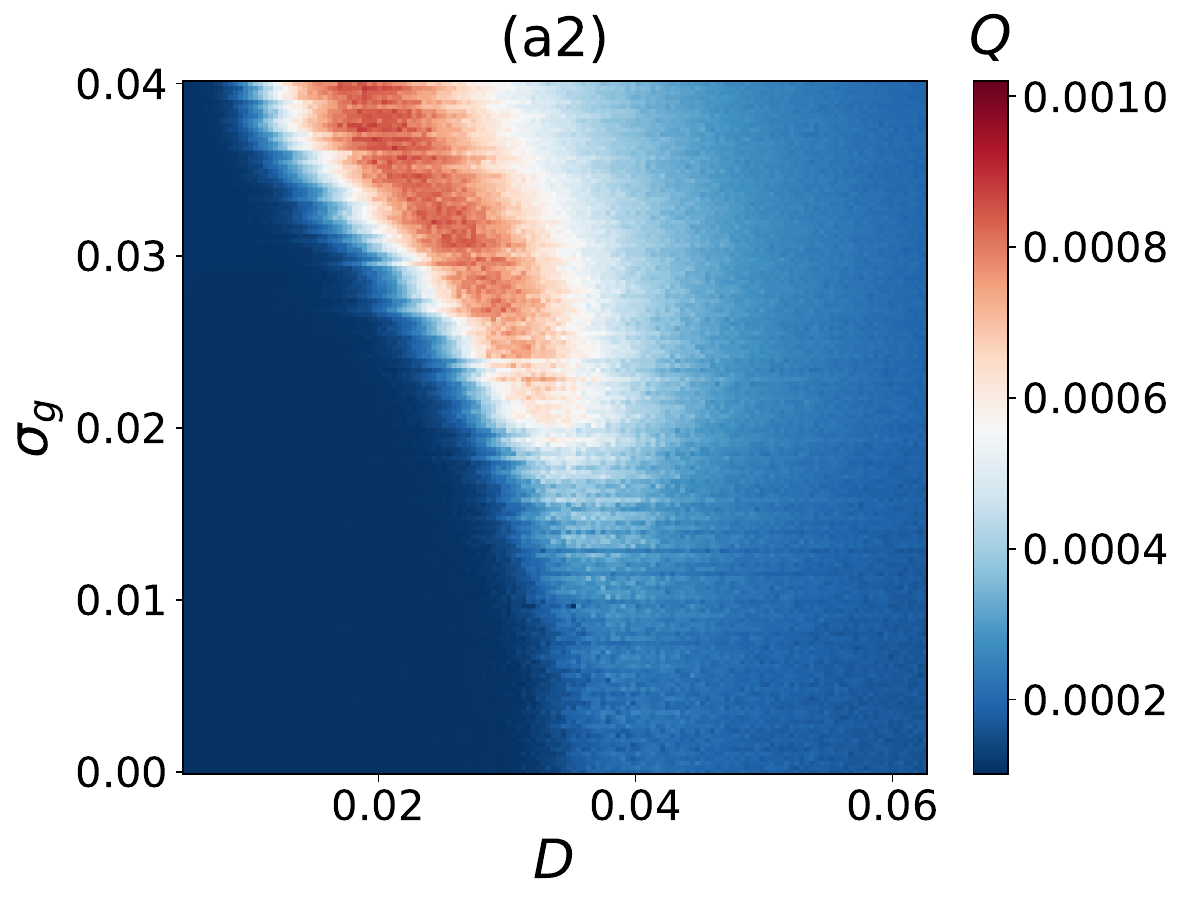}
\\
\includegraphics[width=0.32\textwidth]
{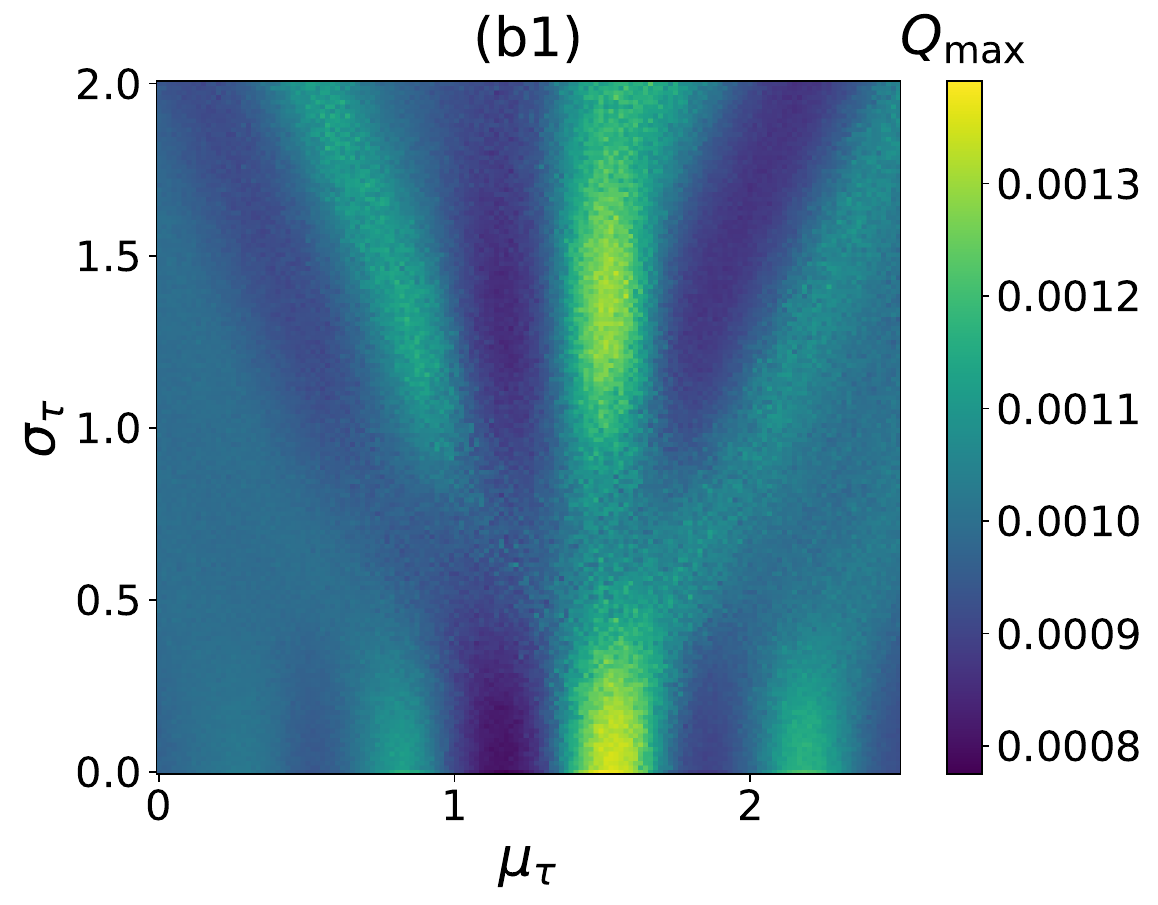}
\hfill
\includegraphics[width=0.32\textwidth]
{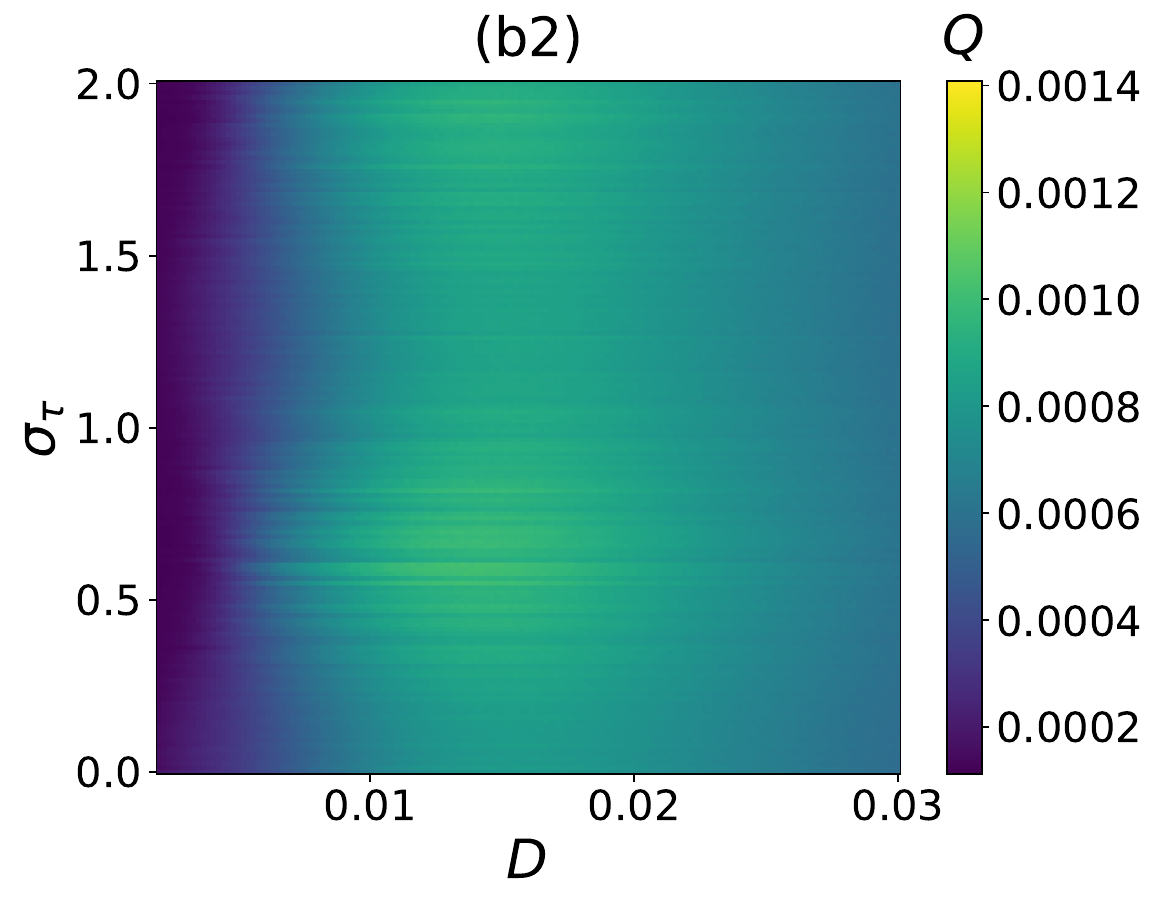}
\hfill
\includegraphics[width=0.32\textwidth]
{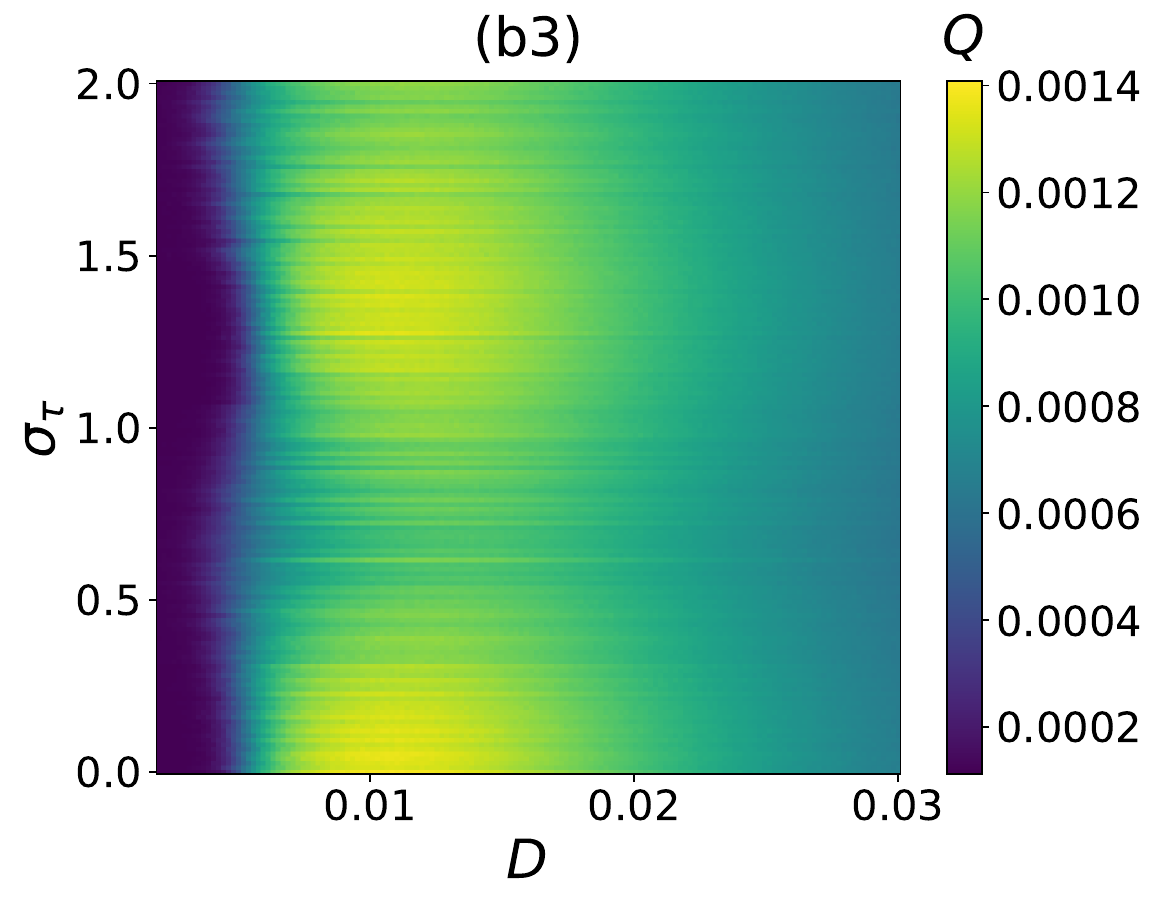}
\caption{\label{fig:2gmm_coupling}
Periodic stochastic resonance under bimodal heterogeneity. For coupling-strength heterogeneity at $\tau=0.3$, (a1) shows $Q_{\max}(\mu_g,\sigma_g)$ and (a2) $Q(D,\sigma_g)$ at $\mu_g=0.02386$. For time-delay heterogeneity at $g=0.0025$, (b1) shows $Q_{\max}(\mu_\tau,\sigma_\tau)$ and (b2)--(b3) $Q(D,\sigma_\tau)$ at $\mu_\tau=1.2$ and $1.5$, respectively. Here, $\sigma_g$ and $\sigma_\tau$ denote the separations of the two component means. Increasing the coupling-mode separation enhances the resonant response and shifts the optimum toward weaker noise, whereas time-delay bimodality generates alternating high- and low-response regimes.}
\end{figure*}

\begin{figure*}
\centering
\includegraphics[width=0.32\textwidth]
{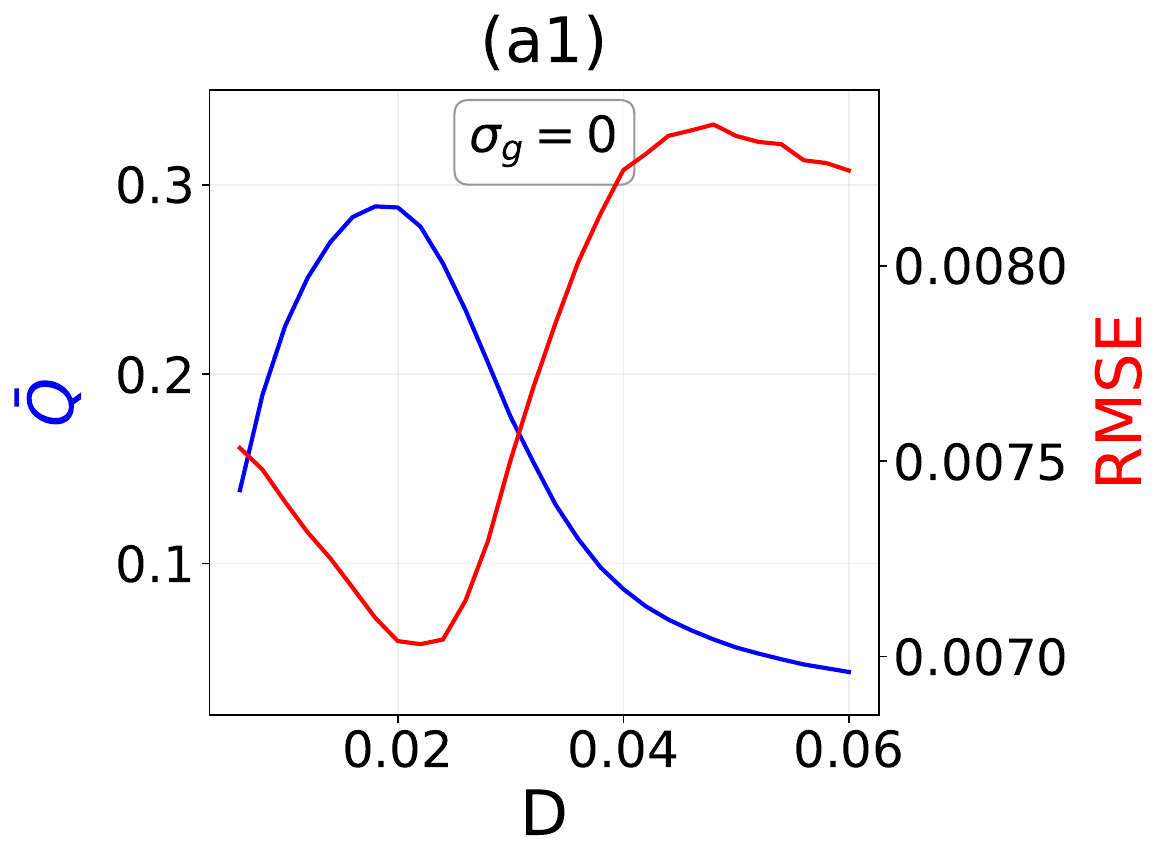}
\includegraphics[width=0.32\textwidth]
{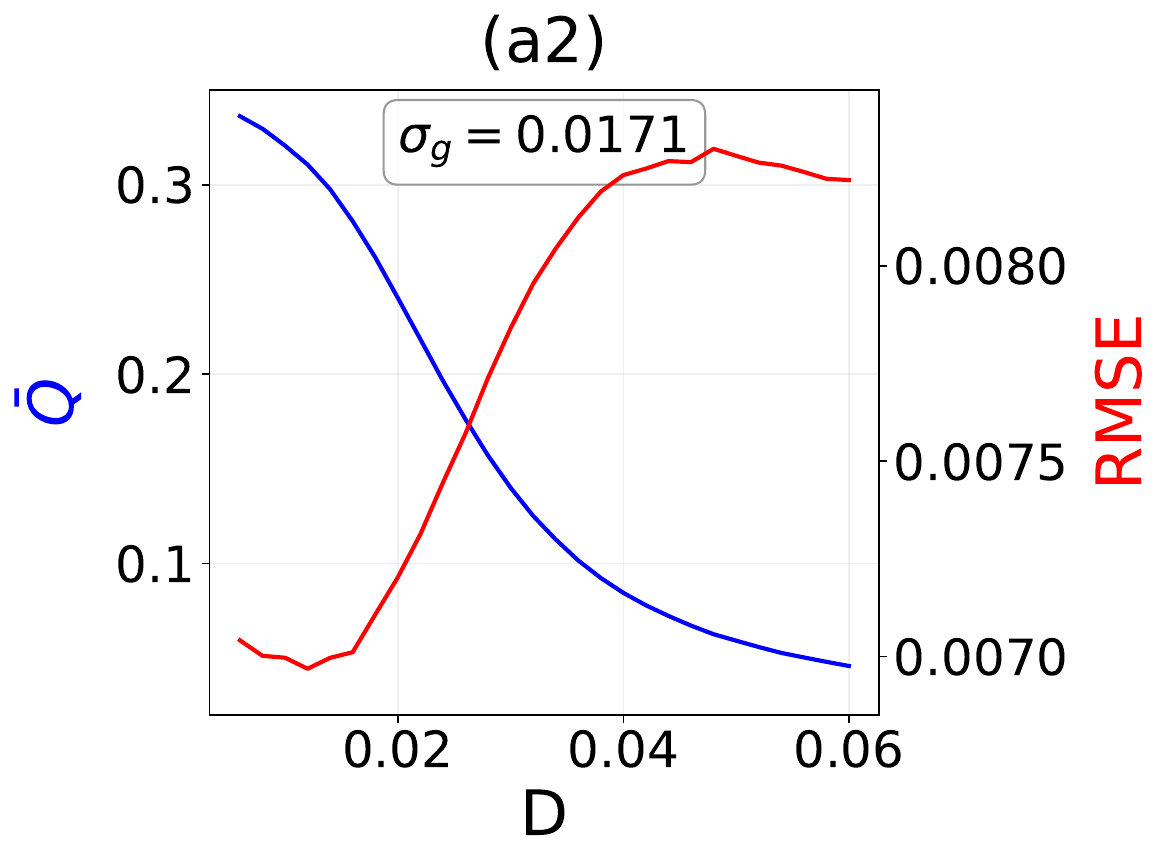}
\includegraphics[width=0.32\textwidth]
{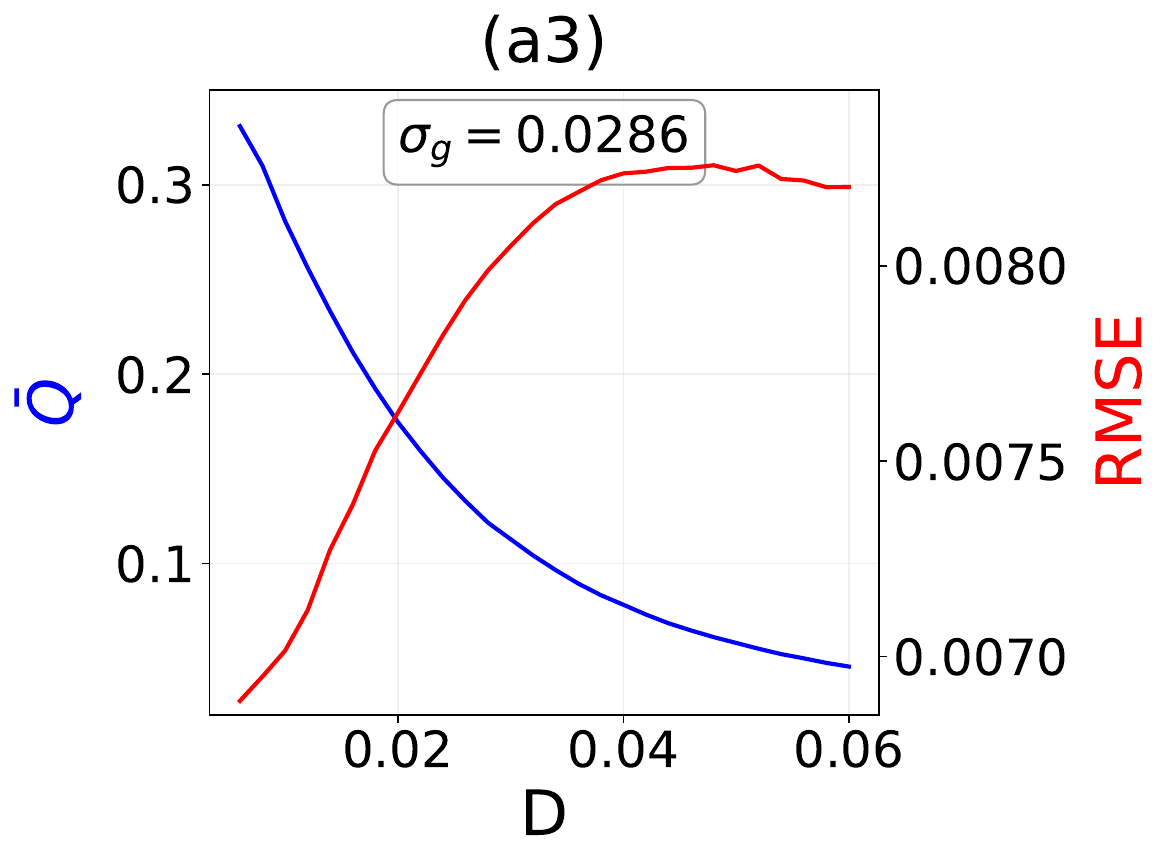}
\caption{Aperiodic input--output coherence and LSM prediction performance under bimodal coupling heterogeneity. The normalized zero-lag coherence $\bar{Q}$ (blue) and test RMSE (red) are shown versus $D$ at $\tau=0.3$ and $\mu_g=0.0162$ for (a1)--(a3) $\sigma_g=0$, $0.0171$, and $0.0286$, respectively. Increasing the coupling-mode separation shifts the optimal regime toward weaker noise, with the largest separation becoming weak-noise dominated.}\label{fig:2gmm_lsm}
\end{figure*}

We next examine whether the two-population organization revealed by
the periodic response has a counterpart under aperiodic driving and
whether it modifies the computational performance of the liquid.
Figure~\ref{fig:2gmm_lsm} compares the normalized zero-lag aperiodic
input--output coherence $\bar{Q}$ with the test-set RMSE as the noise
amplitude $D$ is varied. As in the Gaussian case, the PSR and LSM
experiments employ different forcing signals and different response
observables. Moreover, the coupling-strength sections are evaluated at
different mean couplings: $\mu_g=0.02386$ in
Fig.~\ref{fig:2gmm_coupling}(a2) and $\mu_g=0.0162$ in
Fig.~\ref{fig:2gmm_lsm}(a1)--(a3). Quantitative coincidence of the
resonance curves is therefore not expected. The relevant comparison is
whether the displacement of the noise-dependent coherence caused by
bimodal heterogeneity is accompanied by a corresponding displacement
of the prediction-error minimum.

For coupling-strength heterogeneity,
Figs.~\ref{fig:2gmm_lsm}(a1)--(a3), such a correspondence is clearly
visible. In the unimodal limit $\sigma_g=0$, $\bar{Q}$ exhibits a
well-defined finite-noise maximum, while the RMSE reaches its minimum
within approximately the same noise interval. The liquid therefore
displays the usual ASR balance: weak stochastic forcing is insufficient
to encode the subthreshold input efficiently, whereas excessive noise
reduces the temporal correlation between the input and the population
response.

When the component separation is increased to $\sigma_g=0.0171$, the
coherence is already appreciable at weaker noise, and both the maximum
of $\bar{Q}$ and the minimum of the RMSE are displaced toward smaller
values of $D$. Increasing the separation further to
$\sigma_g=0.0286$ produces a qualitatively different limit: the largest
coherence is attained at the weak-noise edge of the sampled interval
and subsequently decreases as $D$ increases. The prediction error
shows the complementary behavior, being smallest in the same
weak-noise regime and increasing as stochastic forcing becomes
stronger. Thus, for the largest bimodal separation, a distinct
interior finite-noise ASR maximum is no longer resolved; the response
has become weak-noise dominated.

This progression is qualitatively consistent with the PSR ridge in
Fig.~\ref{fig:2gmm_coupling}(a2), which bends toward smaller noise
amplitudes as the coupling modes are separated. The principal effect of
bimodal coupling heterogeneity is therefore not merely an increase of
response amplitude, but a systematic reduction of the noise scale
required for coherent signal transmission. Under aperiodic driving,
this reduction shifts the best readout performance toward weaker
stochastic forcing.

From a dynamical perspective, this behavior is consistent with the
two-scale activation landscape identified in the periodic response.
The simultaneous presence of weaker and stronger couplings generates a
broader set of effective local susceptibilities and propagation
strengths, allowing informative reservoir states to be reached without
requiring a large global noise amplitude. In this sense, bimodal
coupling disorder shifts the computationally useful regime toward
weaker stochastic forcing.

Taken together, Figs.~\ref{fig:2gmm_coupling} and
\ref{fig:2gmm_lsm} demonstrate that bimodality affects coupling
strengths and time delays through physically distinct
mechanisms. Separating the coupling modes creates two effective
activation scales and shifts the regime of optimal signal encoding and
prediction toward weaker noise. Separating the delay modes, by
contrast, causes the network to sample two distinct sectors of the
delay-response landscape and thereby generates reentrant high- and
low-response branches.


\subsection{\label{sec:res_g_exp}Stochastic resonance and LSM performance under shifted-exponential heterogeneity}

For shifted-exponential heterogeneity, the parameters
$\mu_g=g_0$ and $\mu_\tau=\tau_0$ specify the lower shifts of the
unprojected coupling-strength and time-delay distributions,
whereas $\sigma_g=1/\lambda_g$ and $\sigma_\tau=1/\lambda_\tau$
determine their exponential tail scales. In contrast to the Gaussian
and bimodal distributions considered above, increasing
$\sigma_g$ or $\sigma_\tau$ does not preserve the center of the
underlying distribution. Instead, it simultaneously broadens the
positively skewed tail and increases the corresponding unprojected mean,
$\mathbb{E}[\widetilde g_{ij}]=\mu_g+\sigma_g$, $\mathbb{E}[\widetilde\tau_{ij}]=\mu_\tau+\sigma_\tau.$
The resulting samples are subsequently projected onto their admissible
intervals. Shifted-exponential heterogeneity therefore represents a
one-sided, non-mean-preserving form of quenched disorder. See
Fig.~\ref{fig:g_exp_pdfs}(a1)--(b2) in the Appendix.

Figure~\ref{fig:exponential_coupling}(a1) shows that pronounced
periodic stochastic resonance is concentrated at small values of both
$\mu_g$ and $\sigma_g$. For small exponential scales, increasing
$\mu_g$ produces a relatively sharp transition from a high-response
to a low-response regime. As $\sigma_g$ is increased, this boundary
moves toward smaller values of $\mu_g$, and the region of strong
$Q_{\max}$ progressively contracts. This behavior differs
qualitatively from the Gaussian and bimodal cases, for which increasing
heterogeneity can extend the high-response region. The difference is
consistent with the one-sided, non-mean-preserving redistribution of
coupling strengths: increasing $\sigma_g$ simultaneously broadens the
distribution and transfers statistical weight toward larger $g$,
where the response maps of the present network show weaker periodic
resonance.

The section at $\mu_g=0.01490$, shown in
Fig.~\ref{fig:exponential_coupling}(a2), resolves the corresponding
noise-dependent response. At small $\sigma_g$, a well-defined
resonance ridge is centered near $D\simeq0.037$. Increasing the
exponential scale progressively reduces the amplitude of this ridge,
while its position in $D$ changes comparatively little over the range
for which the resonance remains clearly visible. Thus, for periodic
forcing, the dominant effect of increasing the shifted-exponential
coupling scale is a suppression of the coherent response rather than
a substantial displacement of the stochastic-resonance condition.

From a dynamical perspective, the effect should not be interpreted as
a consequence of dispersion alone. Because the shifted-exponential
family is not mean preserving, increasing $\sigma_g$ simultaneously
changes the disorder strength and the characteristic coupling scale.
The recurrent network is therefore displaced through coupling space
while its heterogeneity is increased. The contraction of the
high-response region in Fig.~\ref{fig:exponential_coupling}(a1) is
consistent with this combined shift of the coupling distribution toward
a less responsive sector of parameter space.

A similarly nontrivial reorganization occurs for heterogeneous
time delays. Figure~\ref{fig:exponential_coupling}(b1) displays
a sequence of alternating high- and low-response bands as a function
of the lower delay shift $\mu_\tau$ when $\sigma_\tau$ is small.
Increasing $\sigma_\tau$ broadens the distribution toward larger
delays and, because
$\mathbb{E}[\widetilde\tau_{ij}]=\mu_\tau+\sigma_\tau$, also shifts
its characteristic delay scale. Consequently, the resonance bands
move toward smaller values of $\mu_\tau$ and progressively lose
contrast.

Within the phenomenological response-landscape picture, the
shifted-exponential distribution performs a one-sided weighted average
over the homogeneous delay response while simultaneously translating
probability mass toward larger delays. The resulting deformation
therefore combines smoothing with a systematic displacement of the
effective delay scale. This provides a natural interpretation of the
curved high-response structures in the
$(\mu_\tau,\sigma_\tau)$ plane without requiring a direct measurement
of inter-neuronal phase cancellation.

The sections in Figs.~\ref{fig:exponential_coupling}(b2) and
\ref{fig:exponential_coupling}(b3) demonstrate that the resulting
effect depends strongly on the lower delay shift. At
$\mu_\tau=1.1$, where the small-$\sigma_\tau$ response is relatively
weak, increasing the exponential scale enhances the response over an
extended range of $\sigma_\tau$. Hence, spreading the delays toward
larger values can move appreciable probability mass into delay sectors
with a stronger collective response. At $\mu_\tau=1.5$, by contrast,
the system already lies in a high-response region for small tail
scales. Stronger delay heterogeneity eventually reduces this response
as the distribution samples a broader range of the structured
delay-response landscape. Shifted-exponential delay disorder can
therefore be either constructive or destructive, depending on the
position of its lower shift and tail scale relative to that landscape.

\begin{figure*}
\centering
\includegraphics[width=0.32\textwidth]
{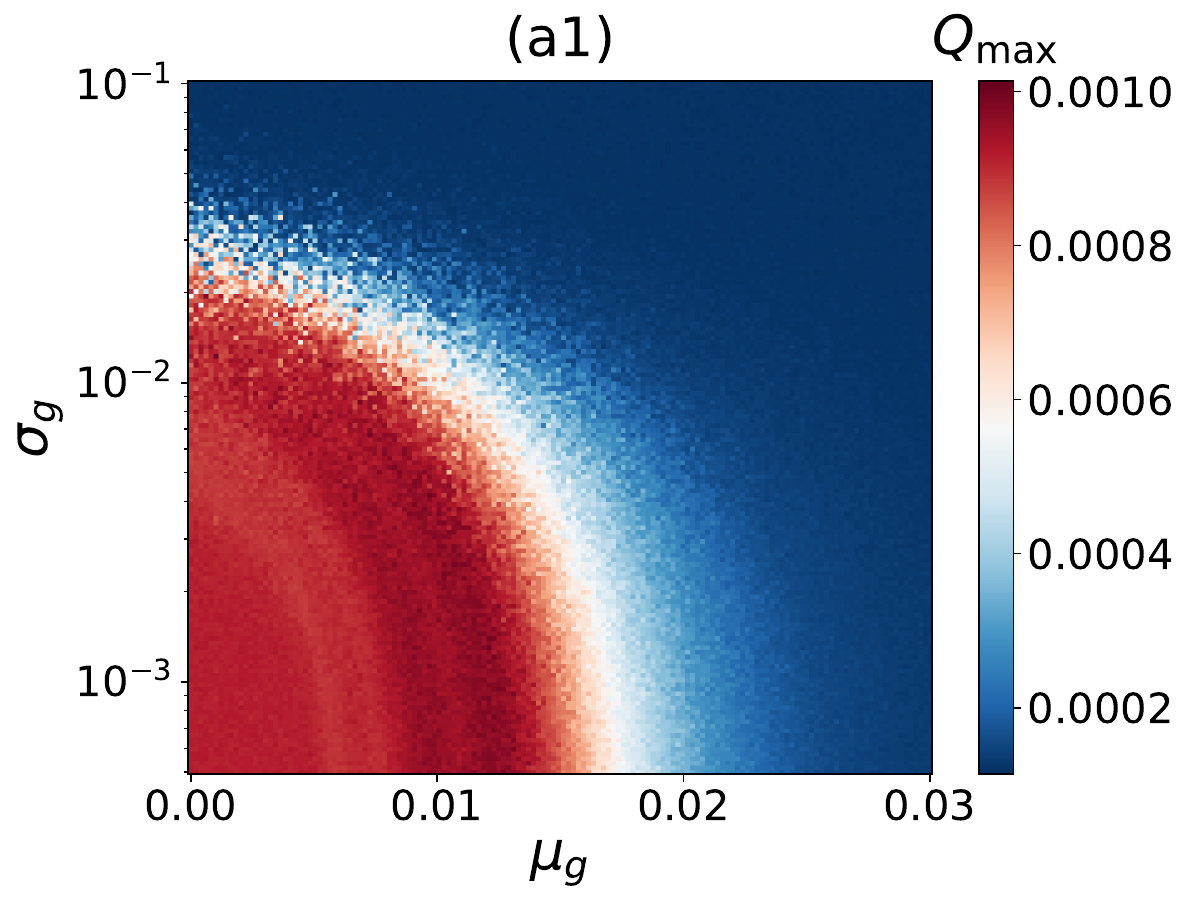}
\includegraphics[width=0.32\textwidth]
{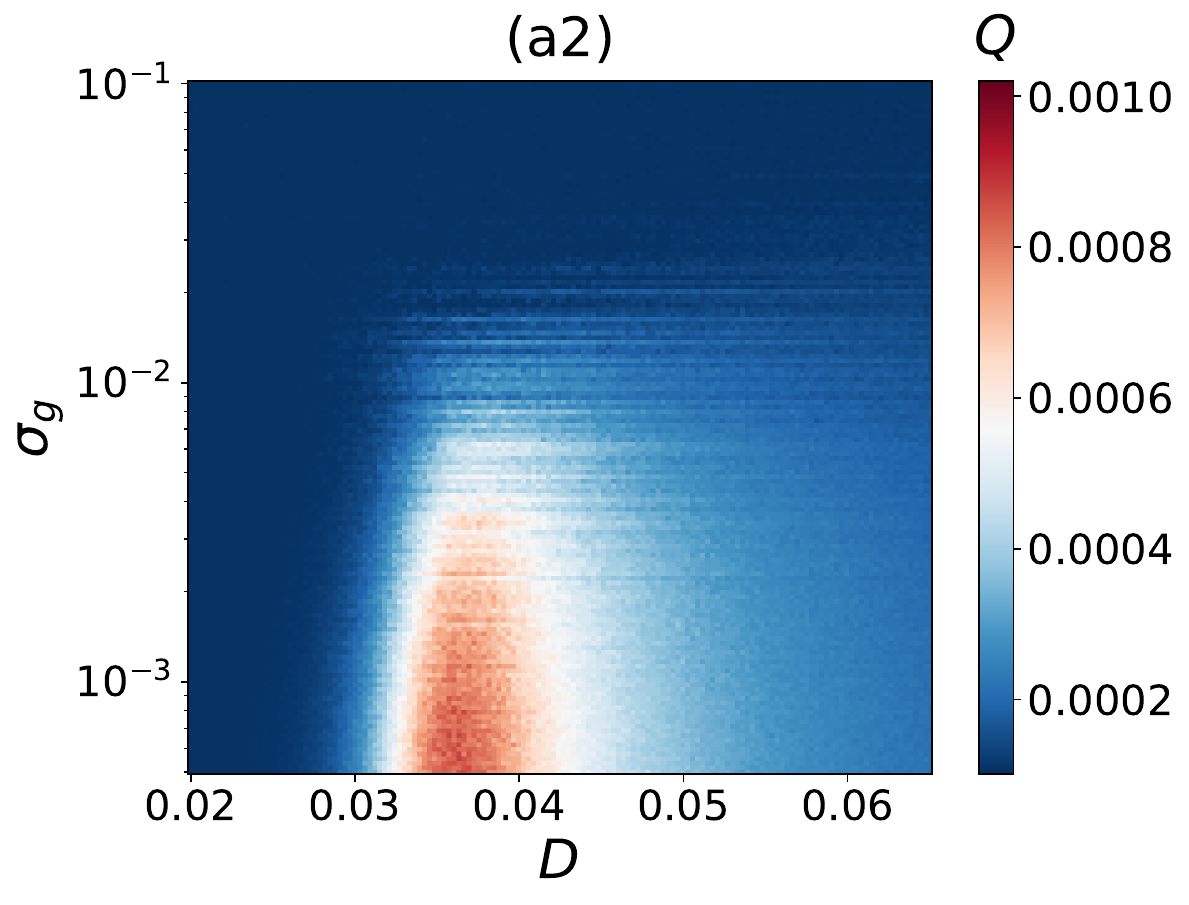}
\\
\includegraphics[width=0.32\textwidth]
{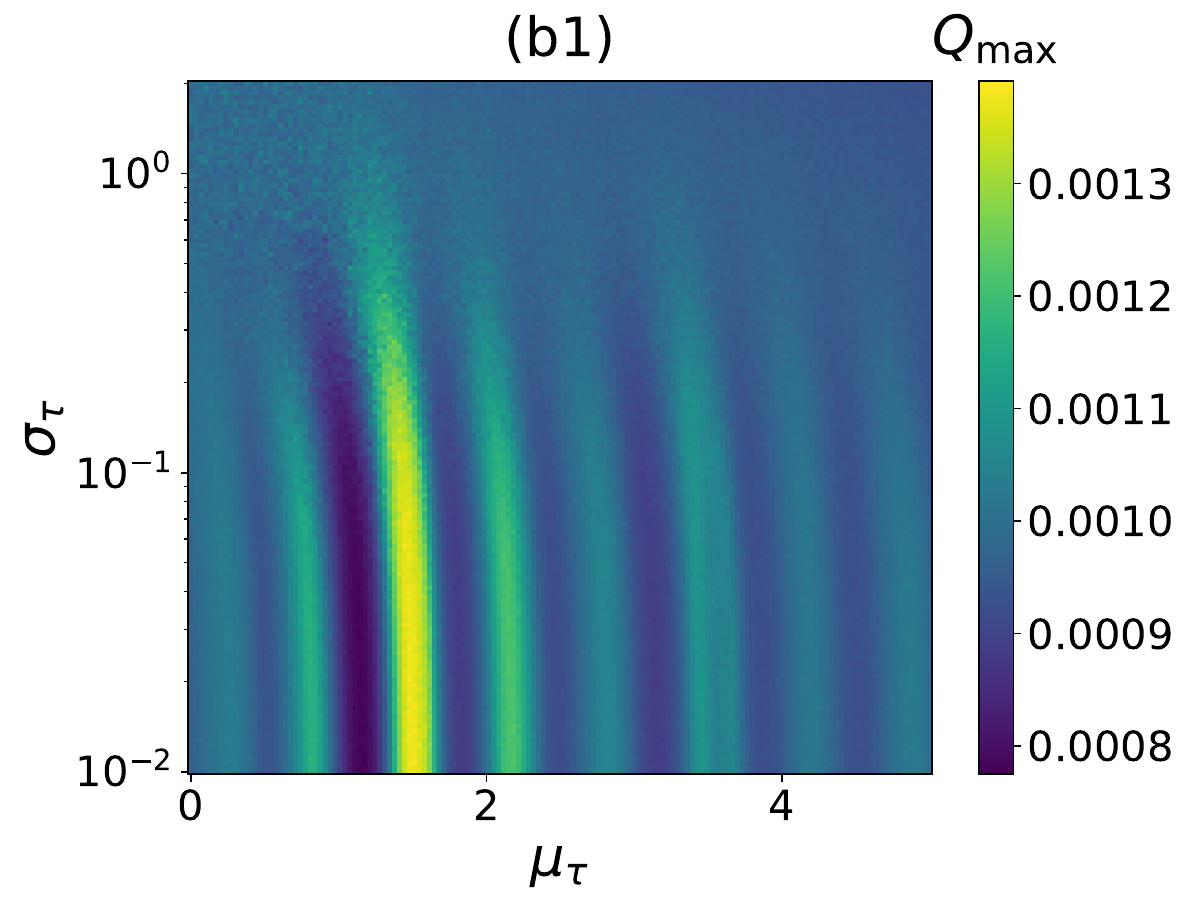}
\includegraphics[width=0.32\textwidth]
{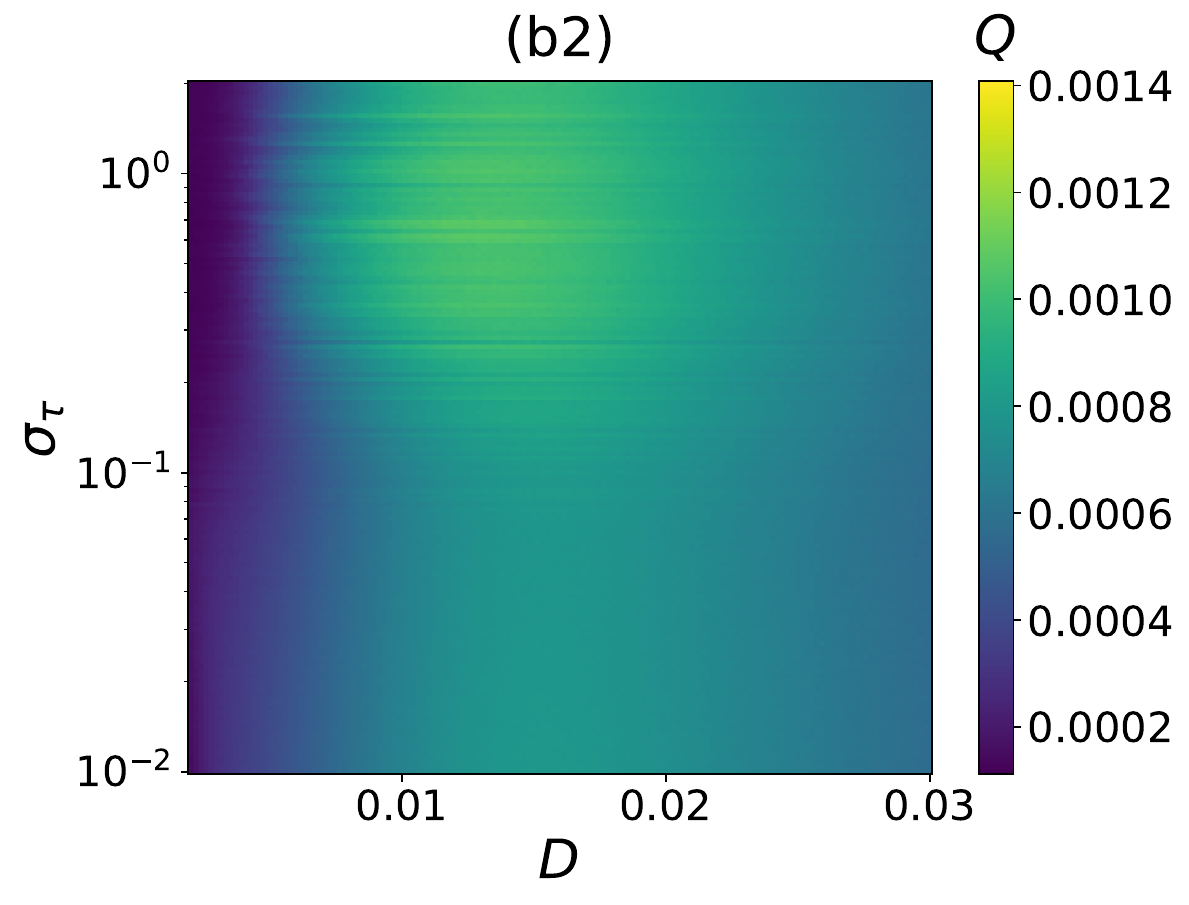}
\includegraphics[width=0.32\textwidth]
{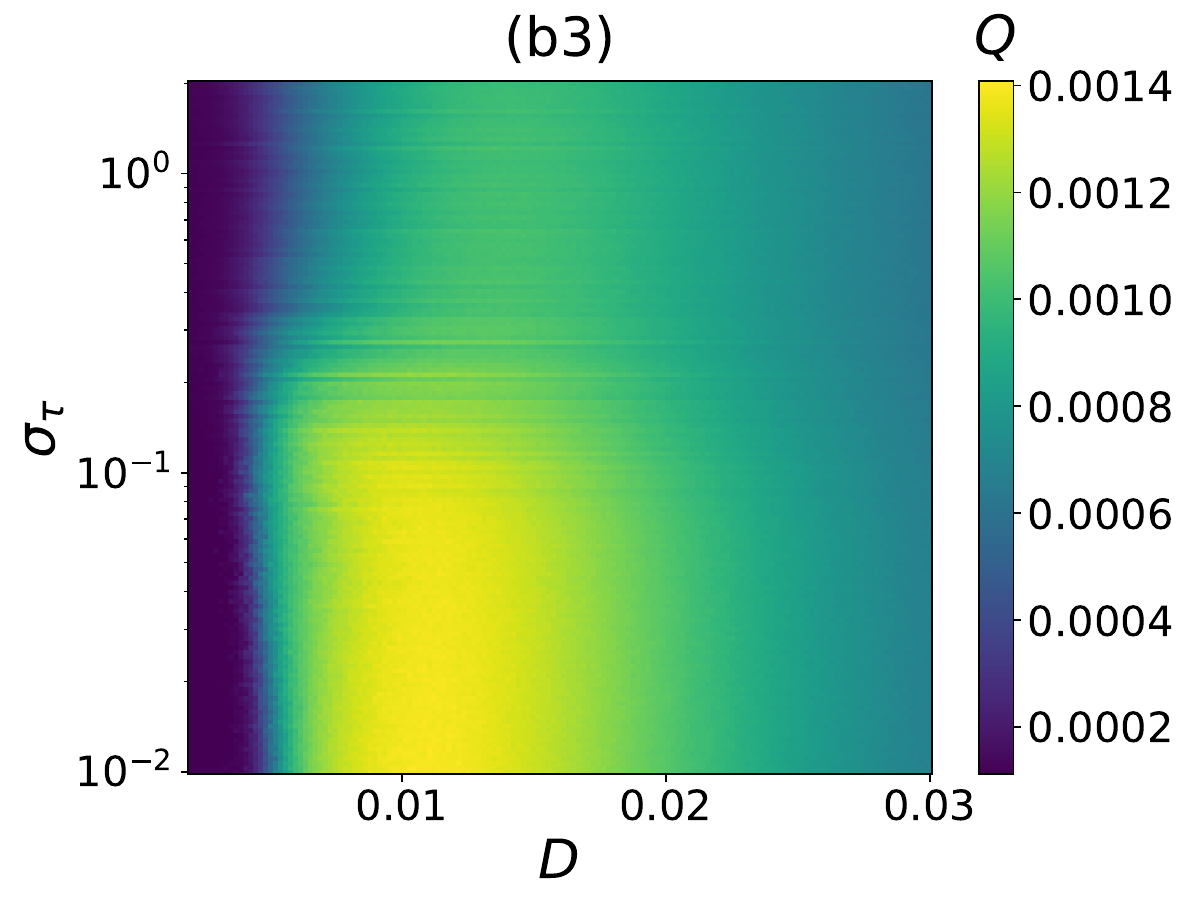}
\caption{\label{fig:exponential_coupling}
Periodic stochastic resonance under shifted-exponential heterogeneity. For coupling-strength heterogeneity at $\tau=0.3$, (a1) shows $Q_{\max}(\mu_g,\sigma_g)$ and (a2) $Q(D,\sigma_g)$ at $\mu_g=0.01490$. For time-delay heterogeneity at $g=0.0025$, (b1) shows $Q_{\max}(\mu_\tau,\sigma_\tau)$ and (b2)--(b3) $Q(D,\sigma_\tau)$ at $\mu_\tau=1.1$ and $1.5$, respectively. Here, $\sigma_g$ and $\sigma_\tau$ denote the exponential tail scales. Increasing the coupling tail scale suppresses the strong-response region, whereas time-delay heterogeneity can enhance or suppress the resonance depending on the lower delay shift.}
\end{figure*}

\begin{figure*}
\centering
\includegraphics[width=0.32\textwidth]
{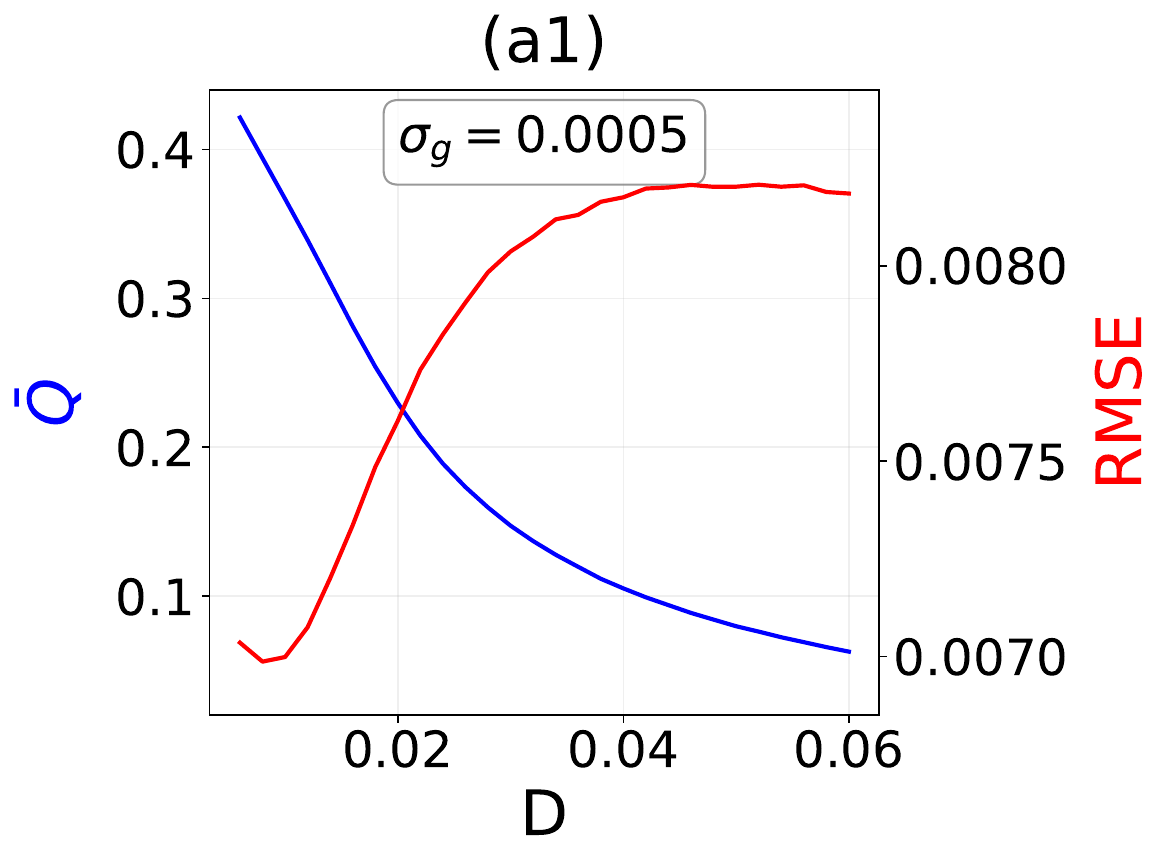}
\includegraphics[width=0.32\textwidth]
{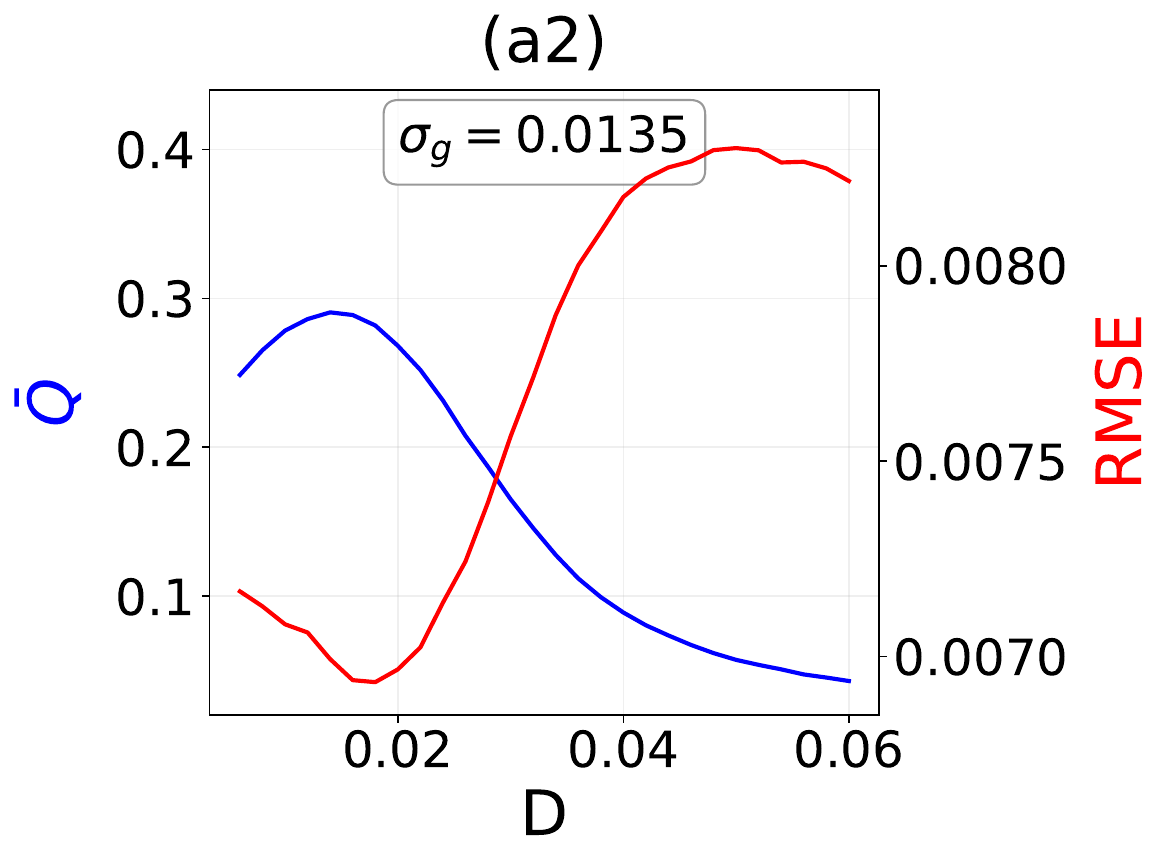}
\includegraphics[width=0.32\textwidth]
{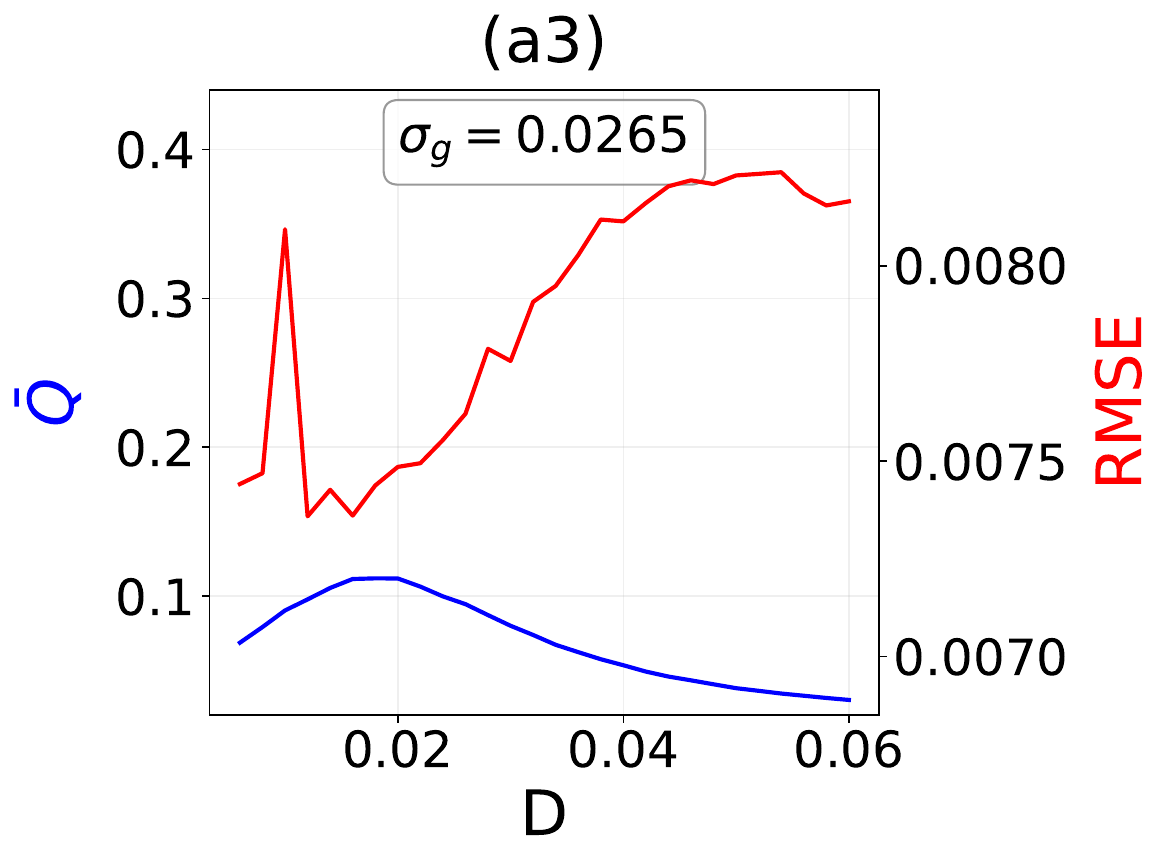}
\caption{Aperiodic input--output coherence and LSM prediction performance under shifted-exponential coupling heterogeneity. The normalized zero-lag coherence $\bar{Q}$ (blue) and test RMSE (red) are shown versus $D$ at $\tau=0.3$ and $\mu_g=0.005$ for (a1)--(a3) $\sigma_g=0.0005$, $0.0135$, and $0.0265$, respectively. Increasing the exponential tail scale shifts the response toward a finite-noise optimum for the broadest distribution.}\label{fig:exponential_lsm}
\end{figure*}

We next examine how this asymmetric form of disorder affects the
aperiodic input--output coherence and the computational performance of
the liquid. Figure~\ref{fig:exponential_lsm} compares the normalized
zero-lag coherence $\bar{Q}$ with the test-set RMSE as the noise
amplitude $D$ is varied. As in the preceding sections, the PSR and LSM
simulations use different forcing signals and different response
observables. Furthermore, the coupling-strength sections are evaluated
at $\mu_g=0.01490$ in Fig.~\ref{fig:exponential_coupling}(a2) and at
$\mu_g=0.005$ in Fig.~\ref{fig:exponential_lsm}(a1)--(a3).
Quantitative coincidence of the two sets of resonance curves is
therefore not expected. The relevant question is whether the
suppression and redistribution of the coherent response produced by
the shifted-exponential coupling distribution is reflected in the
prediction error.

For coupling-strength heterogeneity,
Figs.~\ref{fig:exponential_lsm}(a1)--(a3), increasing the exponential
scale produces a systematic change in both the magnitude and the
location of the noise-dependent coherence. At the smallest scale,
$\sigma_g=0.0005$, $\bar{Q}$ is already large at the weak-noise edge
of the sampled interval and decreases as $D$ is increased. The RMSE
exhibits the corresponding complementary behavior, remaining smallest
at weak noise and increasing as the zero-lag input--response coherence
is reduced. Thus, for this narrow shifted-exponential distribution,
there is no pronounced interior ASR maximum over the sampled range of
$D$.

At $\sigma_g=0.0135$, the coherence develops a broad maximum at weak but
finite noise before decreasing for larger $D$. The RMSE reaches its
minimum over approximately the same low-noise interval. Increasing the
tail scale further to $\sigma_g=0.0265$ produces a more pronounced
finite-noise optimum: $\bar{Q}$ rises from its weak-noise value,
reaches a clear maximum at an intermediate $D$, and then decreases as
the stochastic forcing becomes dominant. The RMSE displays the
opposite variation, with a pronounced reduction from its weak-noise
value and a minimum in approximately the same interval in which
$\bar{Q}$ is maximal.

The three panels therefore reveal an important distinction between
absolute resonance strength and computational benefit. The maximal
value of $\bar{Q}$ decreases as the shifted-exponential coupling scale
becomes large, consistent with the suppression of the periodic response
in Fig.~\ref{fig:exponential_coupling}(a2). Nevertheless, for the
broader distribution the addition of a finite amount of noise can
substantially improve prediction relative to the weak-noise state.
Shifted-exponential coupling heterogeneity therefore does not enhance
the absolute coherent response in the same manner as Gaussian or
bimodal coupling disorder. Instead, it changes the dynamical operating
point of the liquid and can create a finite-noise window in which the
aperiodic signal is encoded and predicted more effectively than in the
corresponding weak-noise state.

The different displacement of the optimal noise scale in
Figs.~\ref{fig:exponential_coupling}(a2) and
\ref{fig:exponential_lsm}(a1)--(a3) is not contradictory. In addition
to the different forcing protocols and response observables, changing
$\sigma_g$ in the shifted-exponential family changes both the
heterogeneity and the mean of the unprojected coupling distribution.
The PSR and ASR measurements therefore sample different sections of a
dynamical landscape in which the characteristic coupling scale and the
noise amplitude vary simultaneously.

Taken together, Figs.~\ref{fig:exponential_coupling} and
\ref{fig:exponential_lsm} show that shifted-exponential heterogeneity
differs fundamentally from the symmetric underlying Gaussian and
bimodal distribution families considered above.
Because the disorder is one-sided and non-mean-preserving, increasing
its scale changes both the breadth and the characteristic magnitude of
the couplings or delays. For coupling strengths, this progressively
suppresses the maximal coherent response, although sufficiently broad
distributions can still create a finite-noise interval in which
prediction is improved relative to the weak-noise state. More generally, the computational consequences
of heterogeneity depend not only on its magnitude, but on how the
underlying distribution redistributes coupling strengths, including its symmetry, multimodality, and any
concomitant shift of the effective mean.


\section{\label{sec:summary_conclusions}Summary and concluding remarks}
We have investigated how quenched heterogeneity in coupling strengths and
time delays modifies stochastic resonance and the computational
performance of a liquid state machine formed by a noisy small-world network
of FitzHugh--Nagumo neurons. Periodic forcing was used as a controlled
dynamical probe of the resonance landscape, while a weak aperiodic signal
was used to assess noise-assisted encoding and five-step-ahead forecasting
with a fixed nonlinear reservoir and a ridge-regularized linear readout.

The homogeneous subthreshold network exhibits the characteristic
stochastic-resonance balance: weak noise produces too few coherent
threshold crossings, intermediate noise maximizes the response to the
external signal, and stronger fluctuations progressively destroy
input--response coherence. Quenched disorder deforms this resonance
landscape in a distribution-dependent manner. Gaussian coupling
heterogeneity extends the region of strong resonant response, whereas
bimodal coupling heterogeneity produces a more pronounced reorganization,
enhancing the coherent response while shifting the resonance ridge toward
weaker stochastic forcing.

By contrast, shifted-exponential coupling
heterogeneity is one-sided and non-mean-preserving; increasing its scale
changes both the disorder strength and the characteristic coupling scale. 
At sufficiently large scales, this redistribution suppresses the globally optimized prediction performance. Time-delay heterogeneity
acts differently: by redistributing the network over the structured
delay-response landscape, it can either enhance or suppress the resonant
response depending on the location and form of the underlying delay
distribution.

The aperiodic simulations reveal an analogous computational organization.
Large zero-lag input--output coherence is generally accompanied by small
held-out prediction error, whereas noise-dominated reservoir dynamics
produce both reduced coherence and poorer forecasts. In particular,
Gaussian and bimodal coupling disorder shift the regime of best prediction
toward weaker noise and reduce the minimum RMSE, with the largest reduction obtained for the bimodal case considered. 
The computational benefit is therefore not controlled by heterogeneity magnitude alone, but
by how the disorder redistributes the effective activation and propagation
scales of the recurrent network.

These results identify stochastic forcing and structural heterogeneity as
coupled control parameters of reservoir dynamics. Appropriate quenched
disorder can reorganize an excitable network so that informative collective
states are reached with weaker stochastic forcing, thereby improving the
temporal representation available to a simple linear readout. Conversely,
heterogeneity need not be beneficial when it displaces the system toward
less responsive regions of parameter space. Thus, the relevant design
principle is not maximal disorder, but a distribution-dependent matching
between noise, coupling statistics, time delays, and the dynamical
operating point of the reservoir. Establishing the robustness of the full PSR response landscapes across ensembles of network topologies and disorder realizations, different input processes, and longer prediction horizons provides a natural direction for further work.

\section*{Acknowledgements} This work was funded by the Department of Data Science (DDS), Friedrich-Alexander-Universit\"at Erlangen-Nürnberg, Germany, and the Deutsche Forschungsgemeinschaft (DFG, German Research Foundation) via the grant YA 764/1-1 to M.E.Y—Project No. 456989199.

\appendix*
\section{}

The continuous density curves shown in 
Figs.~\ref{fig:g_gauss_pdfs}--\ref{fig:g_exp_pdfs} correspond to the 
underlying unprojected sampling distributions of $\widetilde g_{ij}$ and 
$\widetilde\tau_{ij}$. After projection, the realized distributions of 
$g_{ij}$ and $\tau_{ij}$ may additionally contain probability mass at the 
admissible interval boundaries shown in Table \ref{tab:distribution_parameters} and below.

\subsection{\label{sec:g_gauss}Gaussian distribution of coupling strengths and time delays}

For each edge $j\to i$, the coupling strength and time delay are sampled separately, one at a time, as
\begin{equation}
\widetilde g_{ij}\sim\mathcal{N}(\mu_g,\sigma_g^2),
\quad
\widetilde\tau_{ij}\sim\mathcal{N}(\mu_\tau,\sigma_\tau^2),
\end{equation}
and projected onto their admissible intervals,
\begin{equation}
g_{ij}=\Pi_{[0.005,\,0.05]}(\widetilde g_{ij}),
\quad
\tau_{ij}=\Pi_{[0,\,2.5]}(\widetilde\tau_{ij}).
\end{equation}
Here, $\mu_g$ and $\mu_\tau$ control the respective locations, whereas
$\sigma_g$ and $\sigma_\tau$ control the corresponding heterogeneity.

\begin{figure}
\centering
\includegraphics[width=0.25\textwidth]{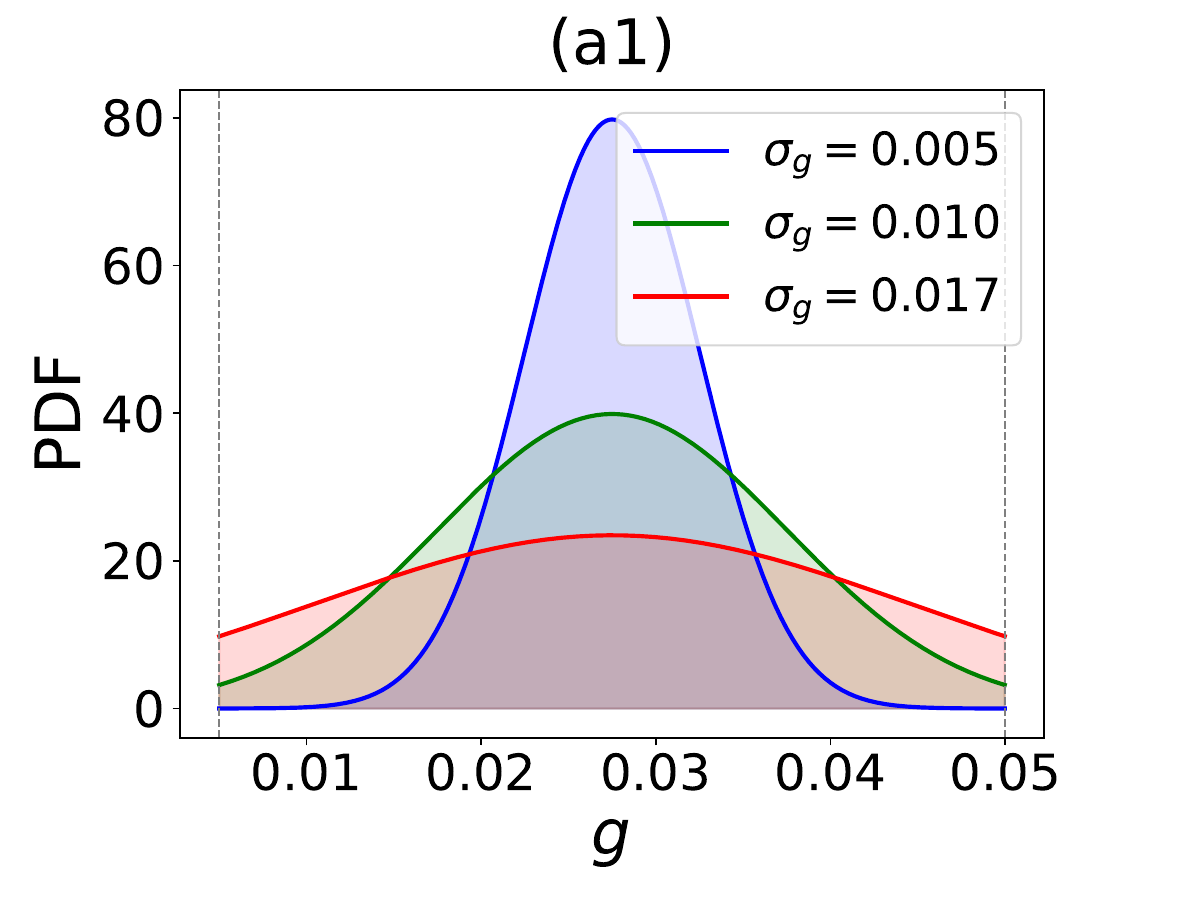}\includegraphics[width=0.25\textwidth]{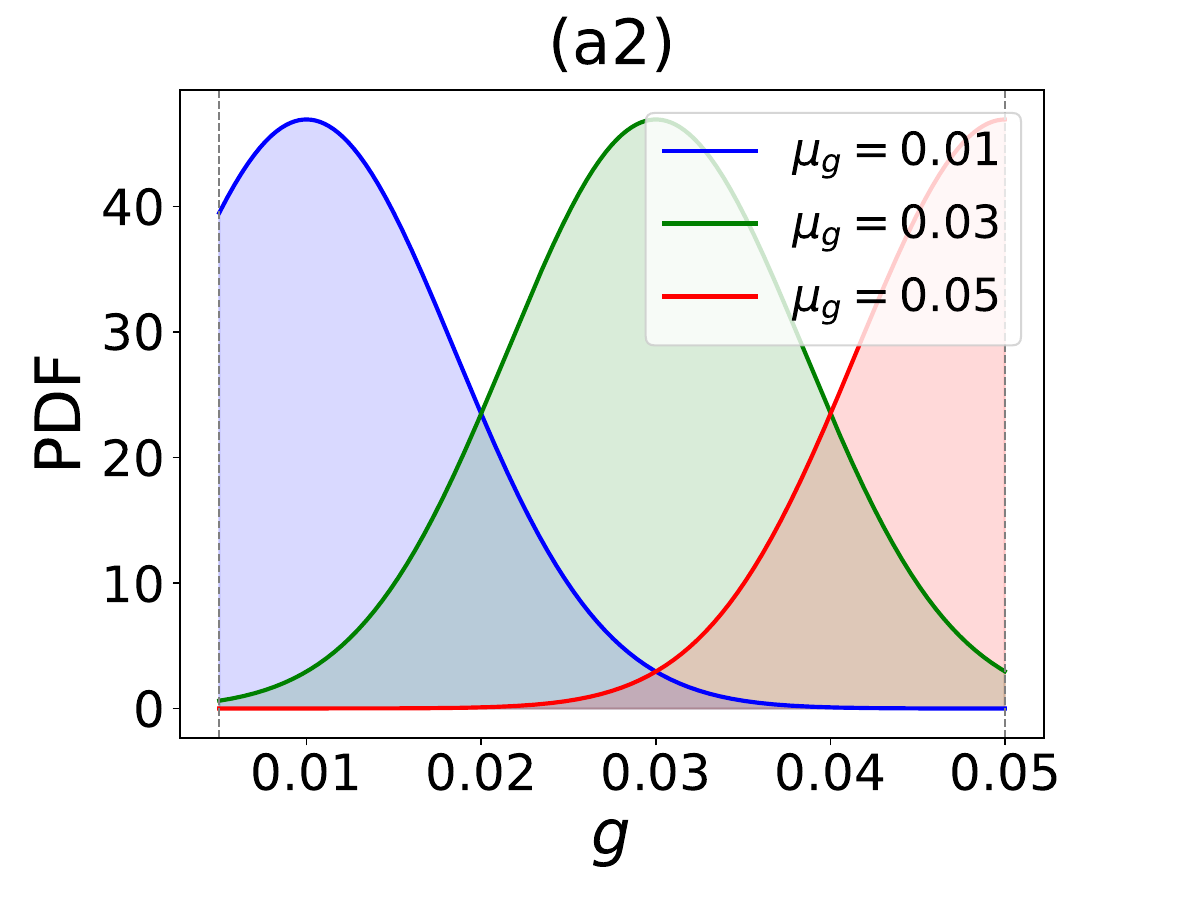}
\includegraphics[width=0.25\textwidth]{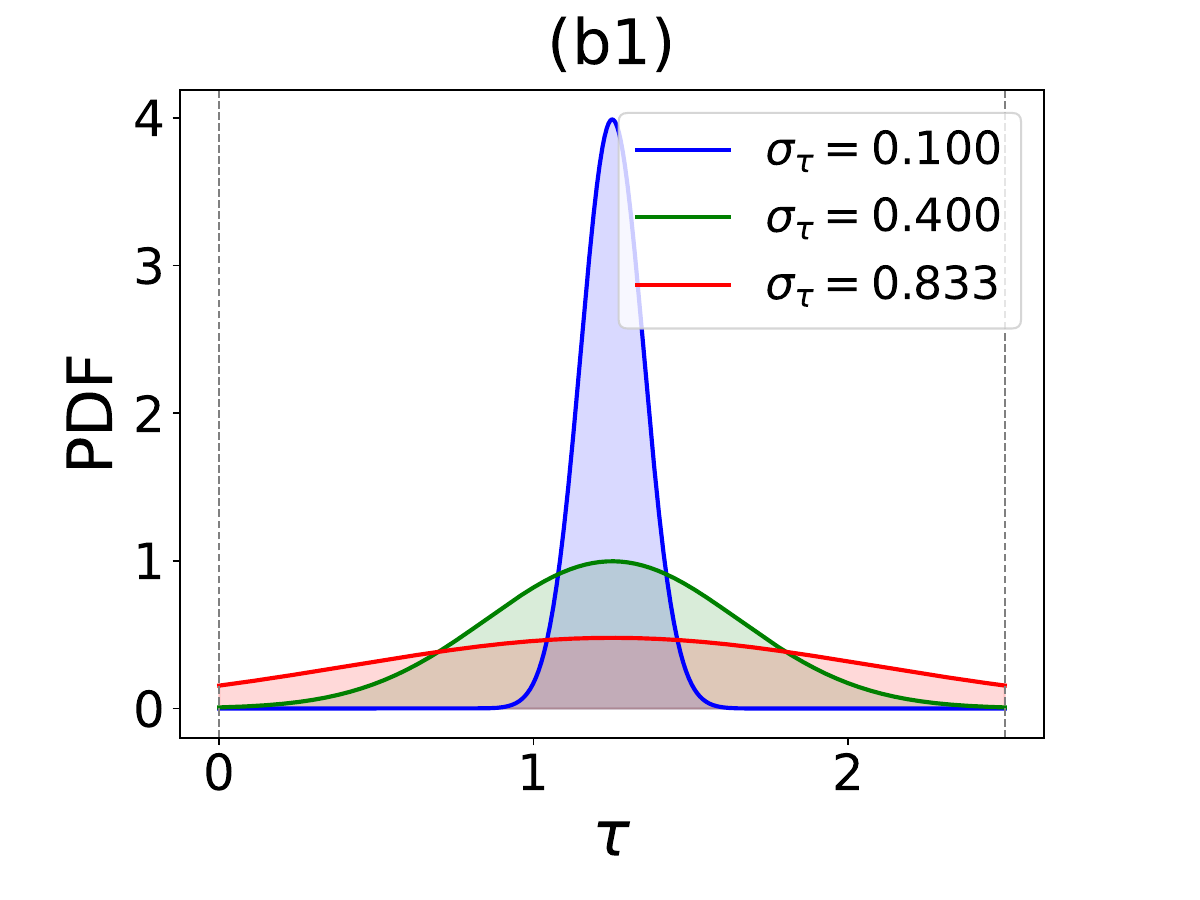}\includegraphics[width=0.25\textwidth]{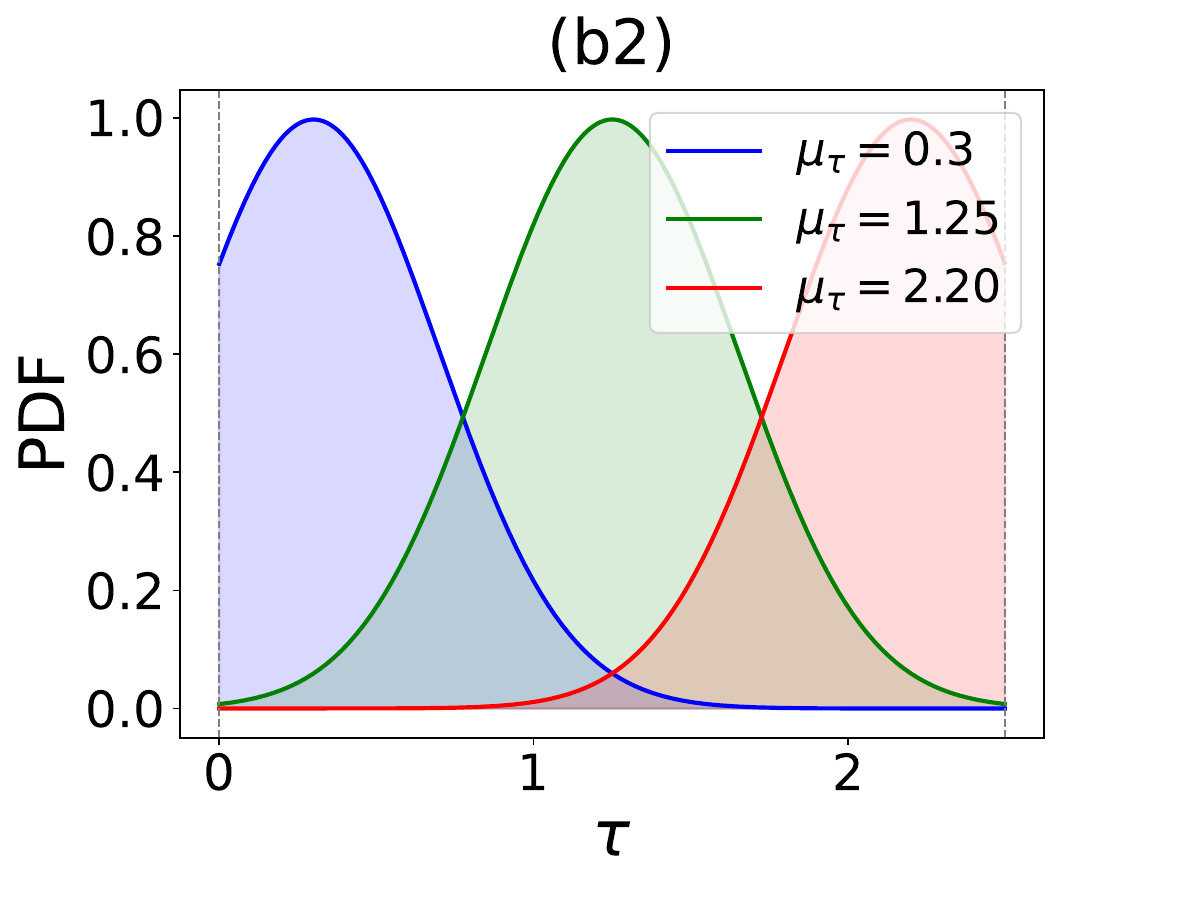}
\caption{\label{fig:g_gauss_pdfs}Representative PDFs of the underlying unprojected Gaussian sampling
distributions for coupling strengths and time delays.
Coupling-strength densities shown over the admissible range
$g\in[0.005,0.05]$: (a1) varying $\sigma_g$ at fixed
$\mu_g=0.0275$; (a2) varying $\mu_g$ at fixed $\sigma_g=0.0085$.
Time-delay densities shown over the admissible range
$\tau\in[0,2.5]$: (b1) varying $\sigma_\tau$ at fixed
$\mu_\tau=1.25$; (b2) varying $\mu_\tau$ at fixed
$\sigma_\tau=0.4$.}
\end{figure}

The parameter ranges are
\begin{eqnarray}
(\mu_g,\sigma_g)&&\in[0.01,0.05]\times[0,0.017]\nonumber,
\\
(\mu_\tau,\sigma_\tau)&&\in[0,2.5]\times[0,0.833].
\end{eqnarray}
The upper dispersions $\sigma_g^{\max}=0.017$ and
$\sigma_\tau^{\max}=0.833$ are chosen such that, at $\mu_g=0.03$ and
$\mu_\tau=1.25$, respectively, the corresponding $3\sigma$ intervals
extend across the admissible ranges. For $\sigma_g=0$ or $\sigma_\tau=0$,
the corresponding edge parameter is homogeneous; increasing the dispersion
introduces progressively stronger heterogeneity. See Fig. \ref{fig:g_gauss_pdfs}.

\subsection{\label{sec:g_2gmm}Bimodal distribution of coupling strengths and time delays}
For each edge $j\to i$, the coupling strength and time delay are
sampled separately, one at a time, from equal-weight Gaussian mixtures,
\begin{equation}
\begin{aligned}
\widetilde g_{ij}
&\sim
\frac{1}{2}\mathcal{N}\!\left(
\mu_g-\frac{\Delta\mu_g}{2},\,s_g^2
\right)
+
\frac{1}{2}\mathcal{N}\!\left(
\mu_g+\frac{\Delta\mu_g}{2},\,s_g^2
\right),\\
\widetilde\tau_{ij}
&\sim
\frac{1}{2}\mathcal{N}\!\left(
\mu_\tau-\frac{\Delta\mu_\tau}{2},\,s_\tau^2
\right)
+
\frac{1}{2}\mathcal{N}\!\left(
\mu_\tau+\frac{\Delta\mu_\tau}{2},\,s_\tau^2
\right),
\end{aligned}
\end{equation}
with fixed component standard deviations
\begin{equation}
s_g=0.002,
\quad
s_\tau=0.1.
\end{equation}
The samples are projected onto their admissible intervals,
\begin{equation}
g_{ij}=\Pi_{[0.005,\,0.05]}(\widetilde g_{ij}),
\quad
\tau_{ij}=\Pi_{[0,\,2.5]}(\widetilde\tau_{ij}).
\end{equation}

\begin{figure}
\centering
\includegraphics[width=0.25\textwidth]{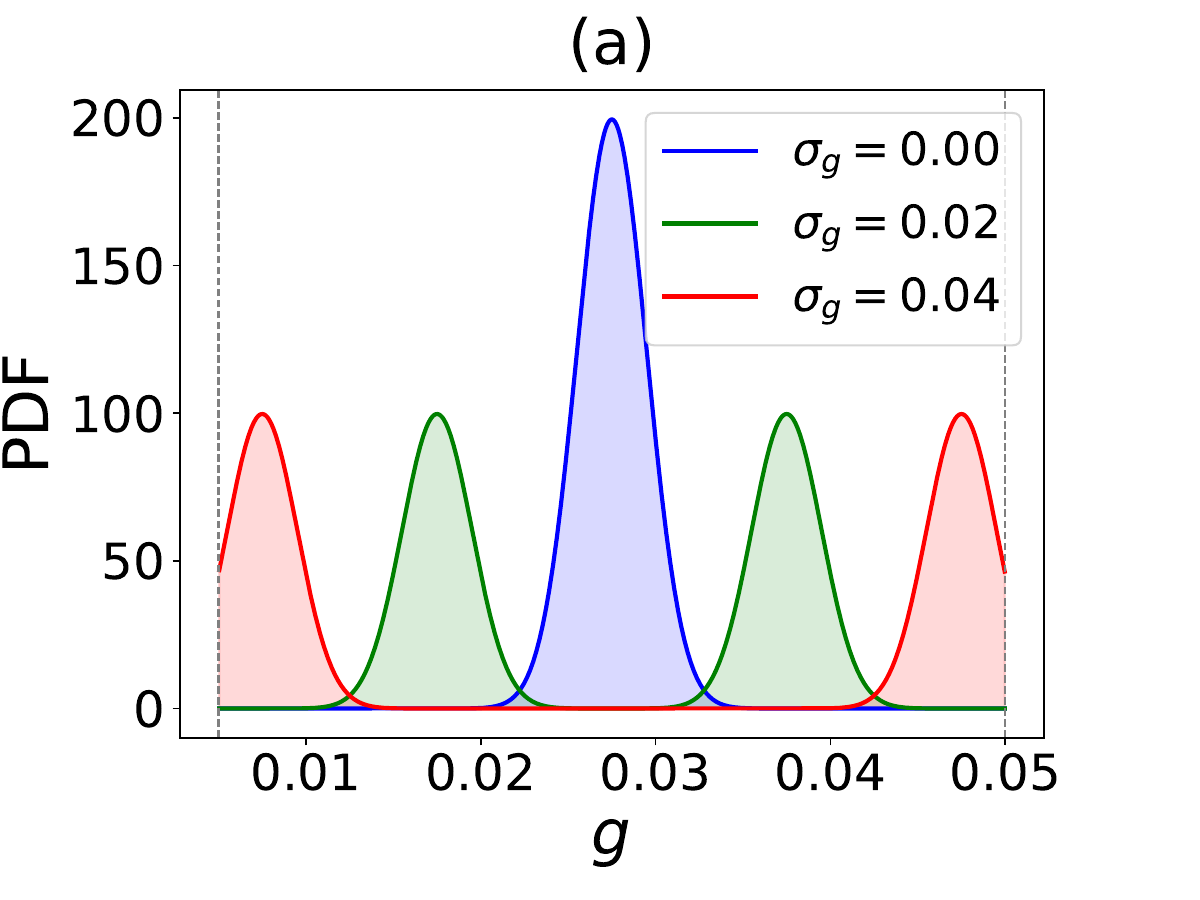}\includegraphics[width=0.25\textwidth]{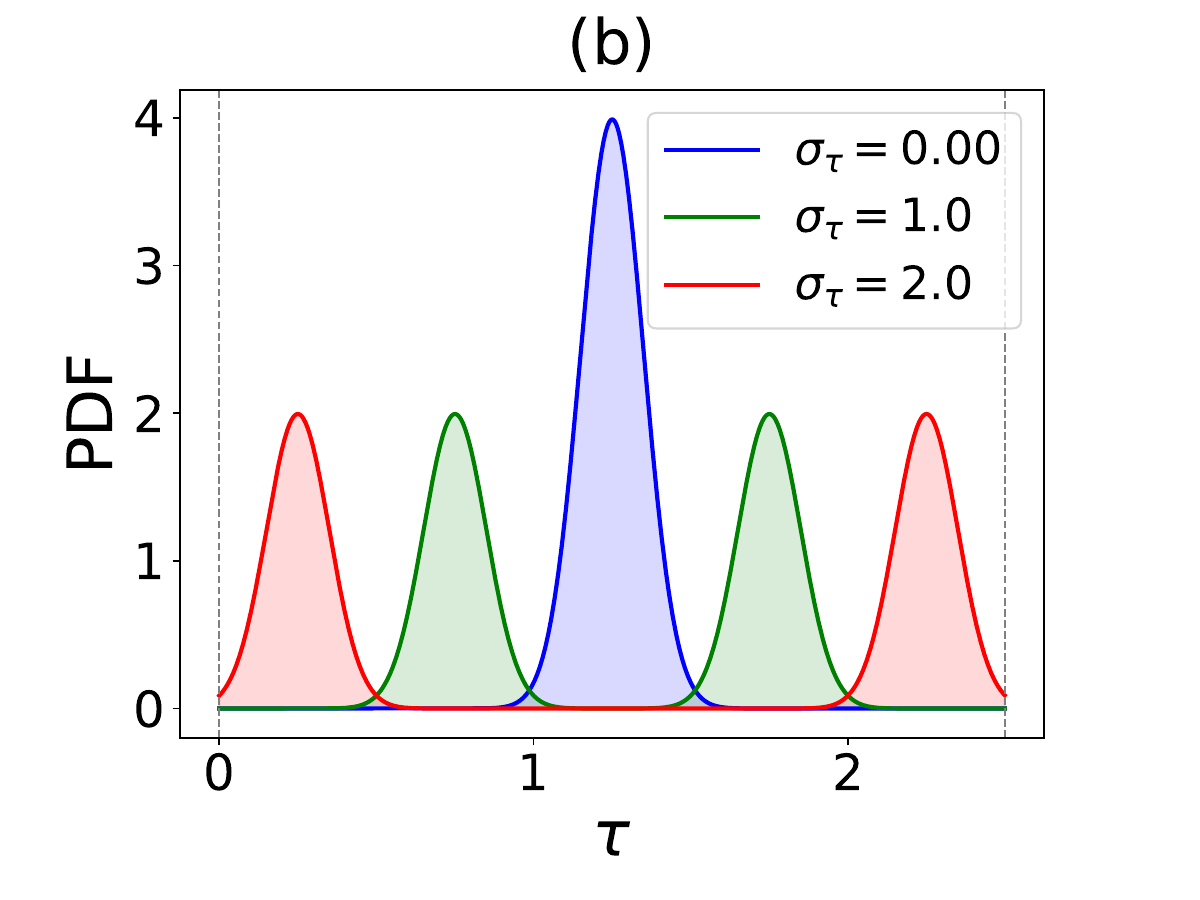}
\caption{\label{fig:g_2gmm_pdfs}Representative PDFs of the underlying unprojected bimodal sampling
distributions for coupling strengths and time delays.
(a) Coupling-strength densities shown over the admissible range
$g\in[0.005,0.05]$: varying $\sigma_g$ at fixed $\mu_g=0.0275$.
(b) Time-delay densities shown over the admissible range
$\tau\in[0,2.5]$: varying $\sigma_\tau$ at fixed $\mu_\tau=1.25$.}
\end{figure}

The component separations define the heterogeneity-control parameters,
\begin{equation}
\sigma_g\equiv\Delta\mu_g,
\quad
\sigma_\tau\equiv\Delta\mu_\tau,
\end{equation}
with parameter ranges
\begin{eqnarray}
(\mu_g,\sigma_g)&&\in[0.01,0.045]\times[0,0.04]\nonumber,
\\
(\mu_\tau,\sigma_\tau)&&\in[0,2.5]\times[0,2.0].
\end{eqnarray}
Thus, $\sigma_g=0$ or $\sigma_\tau=0$ reduces the corresponding mixture
to a single Gaussian, whereas increasing $\sigma_g$ or $\sigma_\tau$
separates the two component means. At $\mu_g=0.03$,
$\sigma_g^{\max}=0.04$ gives component means $0.01$ and $0.05$; at
$\mu_\tau=1.25$, $\sigma_\tau^{\max}=2.0$ gives component means $0.25$
and $2.25$.  See Fig. \ref{fig:g_2gmm_pdfs}.

\subsection{\label{sec:g_exp}Shifted-exponential distribution of coupling strengths and time delays}

For each edge $j\to i$, the coupling strength and time delay are
sampled separately, one at a time, from shifted-exponential distributions,
\begin{equation}
\begin{aligned}
\widetilde g_{ij} &= \mu_g + X_{ij}^{(g)},
& X_{ij}^{(g)} &\sim \operatorname{Exp}(\lambda_g),\\
\widetilde\tau_{ij} &= \mu_\tau + X_{ij}^{(\tau)},
& X_{ij}^{(\tau)} &\sim \operatorname{Exp}(\lambda_\tau),
\end{aligned}
\end{equation}
where
\begin{equation}
\begin{aligned}
\mu_g &= g_0, & \mu_\tau &= \tau_0, \\
\sigma_g &= \lambda_g^{-1}, & \sigma_\tau &= \lambda_\tau^{-1}.
\end{aligned}
\end{equation}
Thus, $\mu_g$ and $\mu_\tau$ determine the lower shifts, whereas
$\sigma_g$ and $\sigma_\tau$ determine the exponential tail scales. The
unprojected means are
\begin{equation}
\mathbb{E}[\widetilde g_{ij}]=\mu_g+\sigma_g,
\quad
\mathbb{E}[\widetilde\tau_{ij}]=\mu_\tau+\sigma_\tau.
\end{equation}

\begin{figure}
\centering
\includegraphics[width=0.24\textwidth]{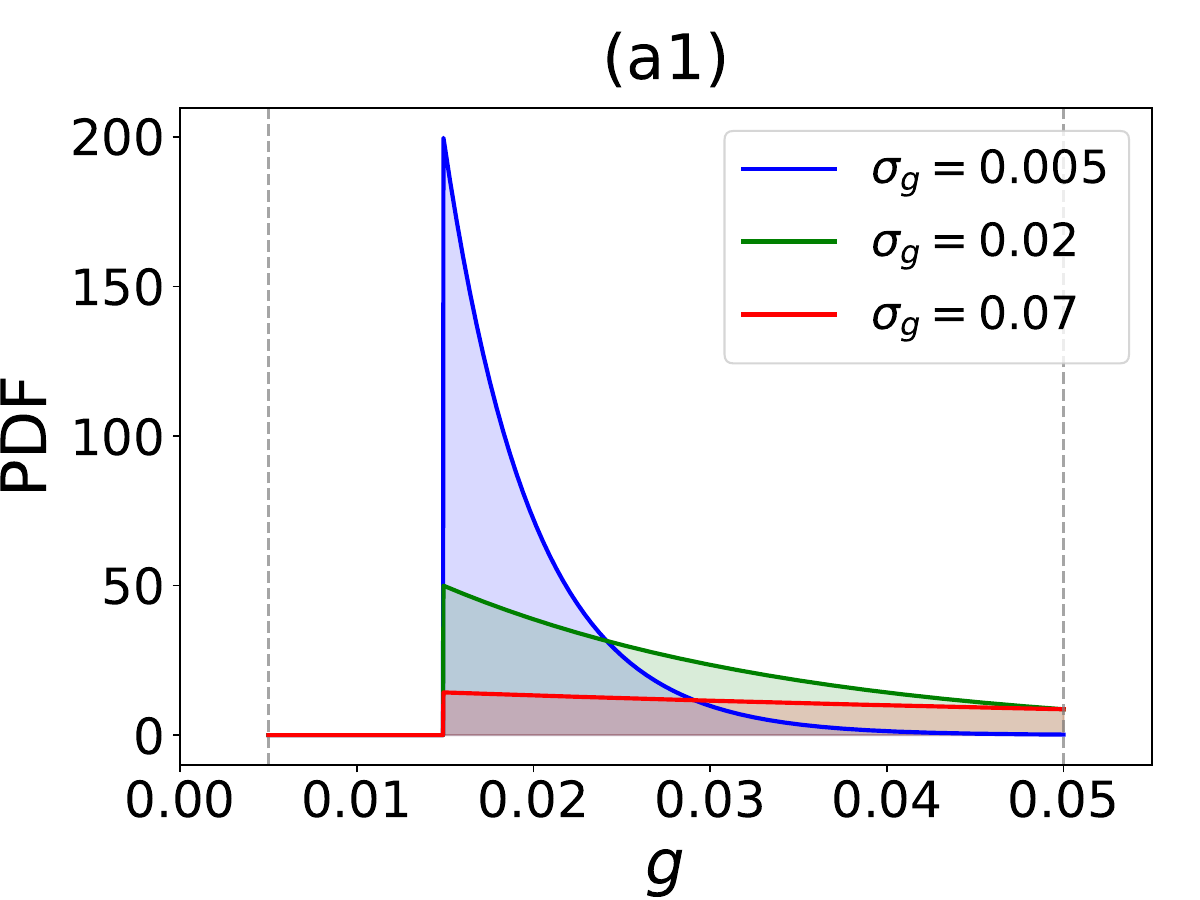}\includegraphics[width=0.24\textwidth]{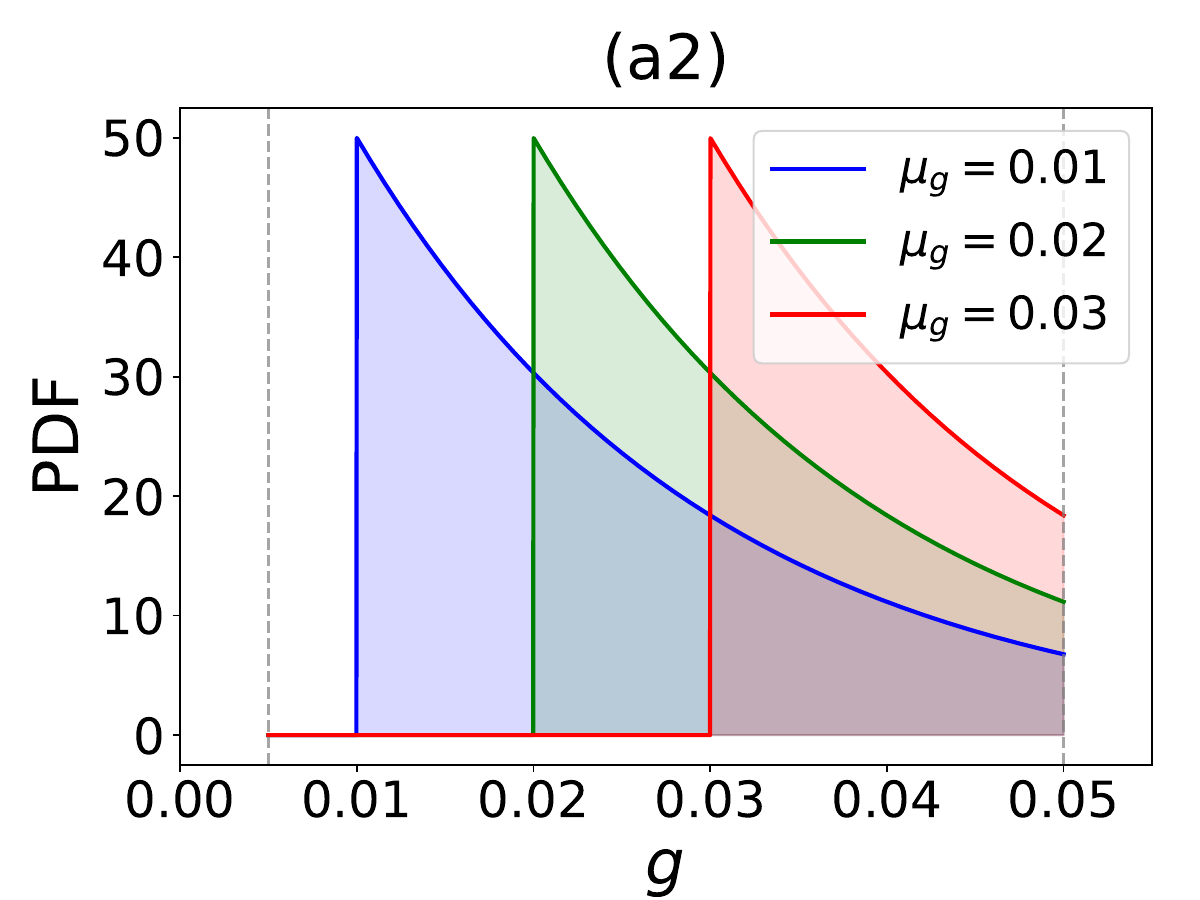}
\includegraphics[width=0.24\textwidth]{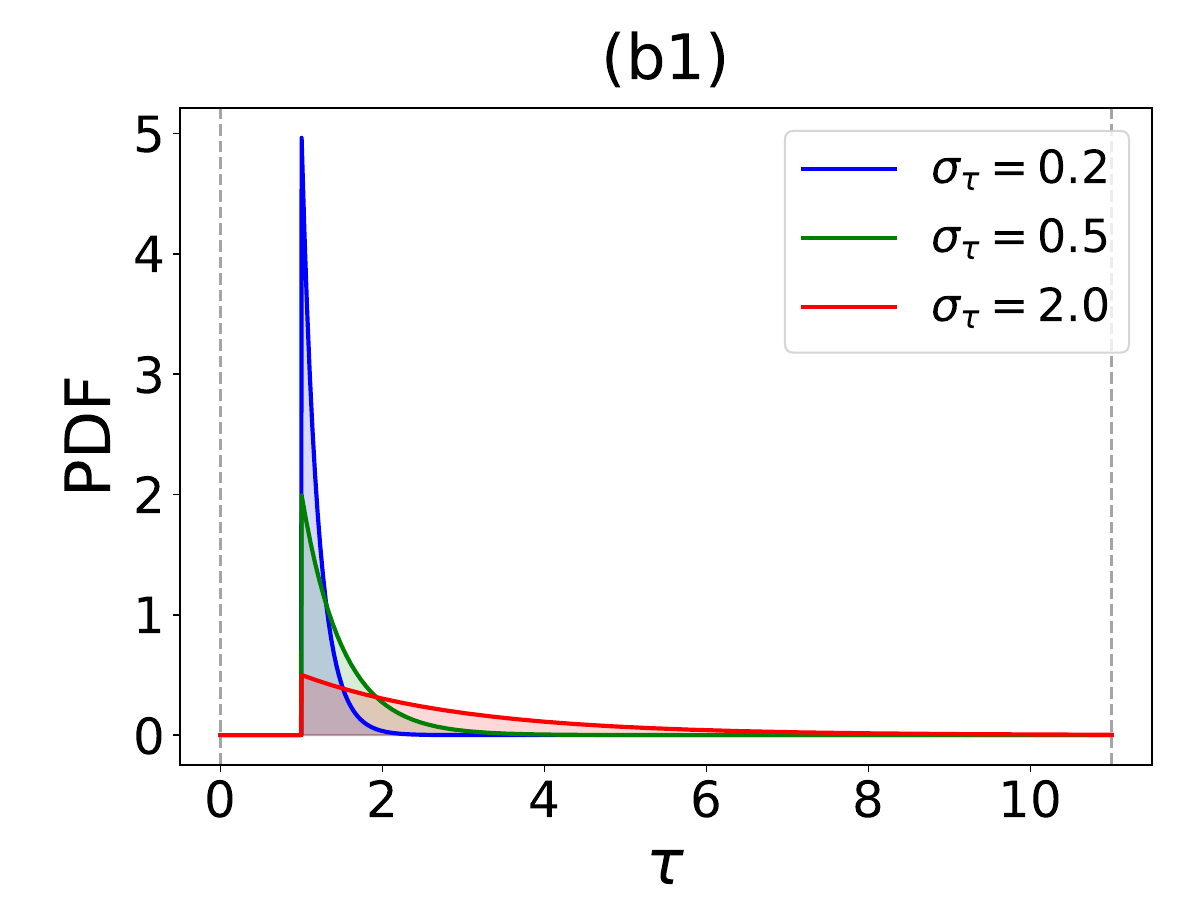}\includegraphics[width=0.24\textwidth]{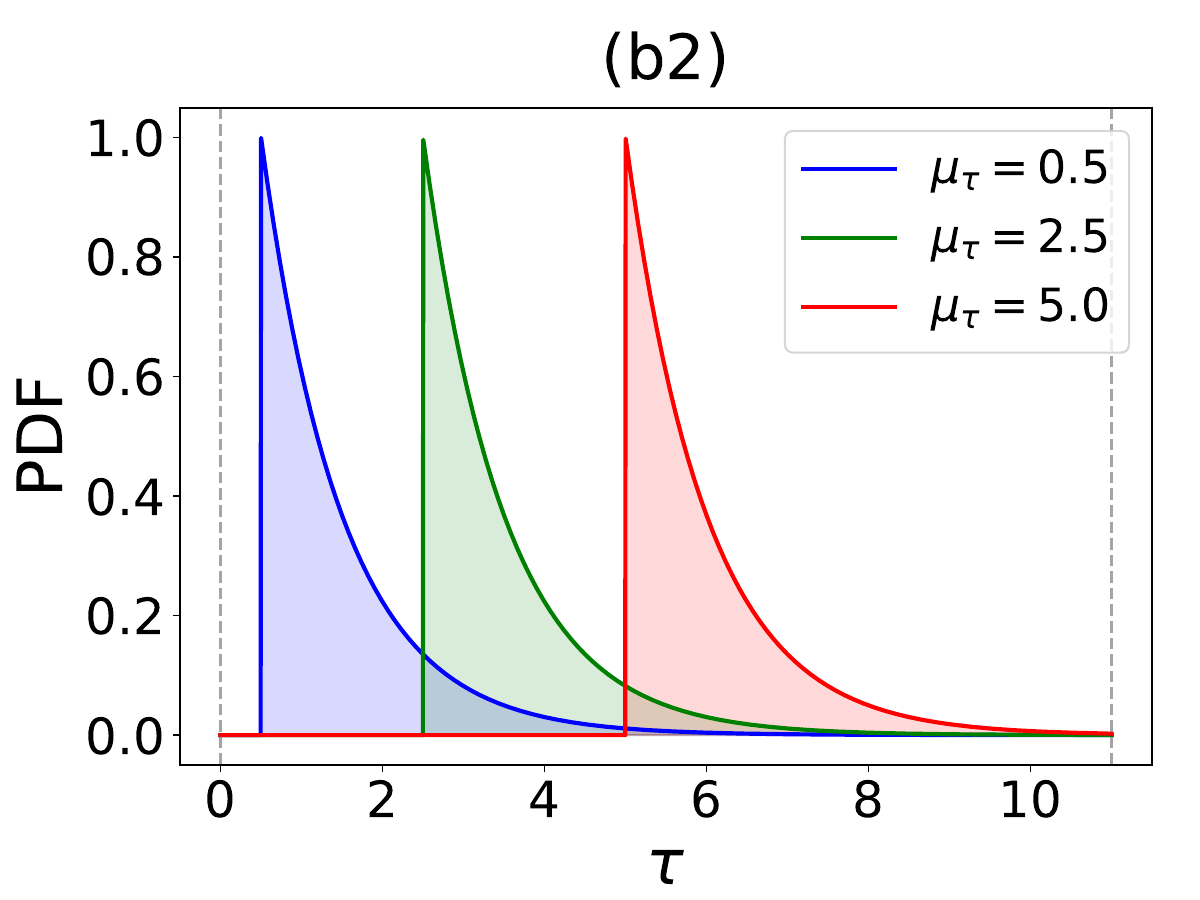}
\caption{\label{fig:g_exp_pdfs}Representative PDFs of the underlying unprojected shifted-exponential
sampling distributions for coupling strengths and time delays.
Coupling-strength densities shown over the admissible range
$g\in[0.005,0.05]$: (a1) varying $\sigma_g$ at fixed
$\mu_g=0.0149$; (a2) varying $\mu_g$ at fixed $\sigma_g=0.02$.
Time-delay densities shown over the admissible range
$\tau\in[0,11]$: (b1) varying $\sigma_\tau$ at fixed
$\mu_\tau=1.0$; (b2) varying $\mu_\tau$ at fixed
$\sigma_\tau=1.0$.}
\end{figure}

The samples are projected onto their admissible intervals,
\begin{equation}
g_{ij}=\Pi_{[0.005,\,0.05]}(\widetilde g_{ij}),
\quad
\tau_{ij}=\Pi_{[0,\,11]}(\widetilde\tau_{ij}).
\end{equation}
The parameter ranges are
\begin{eqnarray}
(\mu_g,\sigma_g)&&\in[0,0.03]\times[0.0005,0.1]\nonumber,\\
(\mu_\tau,\sigma_\tau)&&\in[0,5]\times[0.01,2].
\end{eqnarray}
Increasing $\sigma_g$ or $\sigma_\tau$ broadens the corresponding
right-skewed distribution and increases its unprojected mean.  See Fig. \ref{fig:g_exp_pdfs}.

\section*{Data availability}
The code and data supporting the findings of this study are available from the authors upon reasonable request.


\begin{thebibliography}{31}%
\makeatletter
\providecommand \@ifxundefined [1]{%
 \@ifx{#1\undefined}
}%
\providecommand \@ifnum [1]{%
 \ifnum #1\expandafter \@firstoftwo
 \else \expandafter \@secondoftwo
 \fi
}%
\providecommand \@ifx [1]{%
 \ifx #1\expandafter \@firstoftwo
 \else \expandafter \@secondoftwo
 \fi
}%
\providecommand \natexlab [1]{#1}%
\providecommand \enquote  [1]{``#1''}%
\providecommand \bibnamefont  [1]{#1}%
\providecommand \bibfnamefont [1]{#1}%
\providecommand \citenamefont [1]{#1}%
\providecommand \href@noop [0]{\@secondoftwo}%
\providecommand \href [0]{\begingroup \@sanitize@url \@href}%
\providecommand \@href[1]{\@@startlink{#1}\@@href}%
\providecommand \@@href[1]{\endgroup#1\@@endlink}%
\providecommand \@sanitize@url [0]{\catcode `\\12\catcode `\$12\catcode `\&12\catcode `\#12\catcode `\^12\catcode `\_12\catcode `\%12\relax}%
\providecommand \@@startlink[1]{}%
\providecommand \@@endlink[0]{}%
\providecommand \url  [0]{\begingroup\@sanitize@url \@url }%
\providecommand \@url [1]{\endgroup\@href {#1}{\urlprefix }}%
\providecommand \urlprefix  [0]{URL }%
\providecommand \Eprint [0]{\href }%
\providecommand \doibase [0]{https://doi.org/}%
\providecommand \selectlanguage [0]{\@gobble}%
\providecommand \bibinfo  [0]{\@secondoftwo}%
\providecommand \bibfield  [0]{\@secondoftwo}%
\providecommand \translation [1]{[#1]}%
\providecommand \BibitemOpen [0]{}%
\providecommand \bibitemStop [0]{}%
\providecommand \bibitemNoStop [0]{.\EOS\space}%
\providecommand \EOS [0]{\spacefactor3000\relax}%
\providecommand \BibitemShut  [1]{\csname bibitem#1\endcsname}%
\let\auto@bib@innerbib\@empty
\bibitem [{\citenamefont {Pikovsky}\ and\ \citenamefont {Kurths}(1997)}]{pikovskyCoherenceResonanceNoisedriven1997}%
  \BibitemOpen
  \bibfield  {author} {\bibinfo {author} {\bibfnamefont {A.~S.}\ \bibnamefont {Pikovsky}}\ and\ \bibinfo {author} {\bibfnamefont {J.}~\bibnamefont {Kurths}},\ }\bibfield  {title} {\bibinfo {title} {Coherence resonance in a noise-driven excitable system},\ }\href {https://doi.org/10.1103/PhysRevLett.78.775} {\bibfield  {journal} {\bibinfo  {journal} {Physical Review Letters}\ }\textbf {\bibinfo {volume} {78}},\ \bibinfo {pages} {775} (\bibinfo {year} {1997})}\BibitemShut {NoStop}%
\bibitem [{\citenamefont {Benzi}\ \emph {et~al.}(1981)\citenamefont {Benzi}, \citenamefont {Sutera},\ and\ \citenamefont {Vulpiani}}]{Benzi-1981a}%
  \BibitemOpen
  \bibfield  {author} {\bibinfo {author} {\bibfnamefont {R.}~\bibnamefont {Benzi}}, \bibinfo {author} {\bibfnamefont {A.}~\bibnamefont {Sutera}},\ and\ \bibinfo {author} {\bibfnamefont {A.}~\bibnamefont {Vulpiani}},\ }\bibfield  {title} {\bibinfo {title} {The mechanism of stochastic resonance},\ }\href {https://doi.org/10.1088/0305-4470/14/11/006} {\bibfield  {journal} {\bibinfo  {journal} {Journal of Physics A: Mathematical and General}\ }\textbf {\bibinfo {volume} {14}},\ \bibinfo {pages} {L453} (\bibinfo {year} {1981})}\BibitemShut {NoStop}%
\bibitem [{\citenamefont {Tuckwell}\ and\ \citenamefont {Jost}(2012)}]{tuckwellAnalysisInverseStochastic2012}%
  \BibitemOpen
  \bibfield  {author} {\bibinfo {author} {\bibfnamefont {H.~C.}\ \bibnamefont {Tuckwell}}\ and\ \bibinfo {author} {\bibfnamefont {J.}~\bibnamefont {Jost}},\ }\bibfield  {title} {\bibinfo {title} {Analysis of inverse stochastic resonance and the long-term firing of {{Hodgkin}}--{{Huxley}} neurons with {{Gaussian}} white noise},\ }\href {https://doi.org/10.1016/j.physa.2012.06.019} {\bibfield  {journal} {\bibinfo  {journal} {Physica A: Statistical Mechanics and its Applications}\ }\textbf {\bibinfo {volume} {391}},\ \bibinfo {pages} {5311} (\bibinfo {year} {2012})}\BibitemShut {NoStop}%
\bibitem [{\citenamefont {Longtin}(1993)}]{longtin1993stochastic}%
  \BibitemOpen
  \bibfield  {author} {\bibinfo {author} {\bibfnamefont {A.}~\bibnamefont {Longtin}},\ }\bibfield  {title} {\bibinfo {title} {Stochastic resonance in neuron models},\ }\href {https://doi.org/10.1007/BF01053970} {\bibfield  {journal} {\bibinfo  {journal} {Journal of Statistical Physics}\ }\textbf {\bibinfo {volume} {70}},\ \bibinfo {pages} {309} (\bibinfo {year} {1993})}\BibitemShut {NoStop}%
\bibitem [{\citenamefont {Mori}\ and\ \citenamefont {Kai}(2004)}]{mori2004stochastic}%
  \BibitemOpen
  \bibfield  {author} {\bibinfo {author} {\bibfnamefont {T.}~\bibnamefont {Mori}}\ and\ \bibinfo {author} {\bibfnamefont {S.}~\bibnamefont {Kai}},\ }\bibfield  {title} {\bibinfo {title} {Stochastic resonance in the brain},\ }\href {https://doi.org/10.1002/scj.10398} {\bibfield  {journal} {\bibinfo  {journal} {Systems and Computers in Japan}\ }\textbf {\bibinfo {volume} {35}},\ \bibinfo {pages} {39} (\bibinfo {year} {2004})}\BibitemShut {NoStop}%
\bibitem [{\citenamefont {Li}\ \emph {et~al.}(2024)\citenamefont {Li}, \citenamefont {Yu}, \citenamefont {Wu}, \citenamefont {Ding},\ and\ \citenamefont {Jia}}]{li2024stochastic}%
  \BibitemOpen
  \bibfield  {author} {\bibinfo {author} {\bibfnamefont {T.}~\bibnamefont {Li}}, \bibinfo {author} {\bibfnamefont {D.}~\bibnamefont {Yu}}, \bibinfo {author} {\bibfnamefont {Y.}~\bibnamefont {Wu}}, \bibinfo {author} {\bibfnamefont {Q.}~\bibnamefont {Ding}},\ and\ \bibinfo {author} {\bibfnamefont {Y.}~\bibnamefont {Jia}},\ }\bibfield  {title} {\bibinfo {title} {Stochastic resonance in the small-world networks with higher order neural motifs interactions},\ }\href {https://doi.org/10.1140/epjs/s11734-024-01139-w} {\bibfield  {journal} {\bibinfo  {journal} {The European Physical Journal Special Topics}\ }\textbf {\bibinfo {volume} {233}},\ \bibinfo {pages} {797} (\bibinfo {year} {2024})}\BibitemShut {NoStop}%
\bibitem [{\citenamefont {{Schlabritz}}\ and\ \citenamefont {{Yamakou}}(2026)}]{schlabritz2026interactionordercontrolsstochasticresonance}%
  \BibitemOpen
  \bibfield  {author} {\bibinfo {author} {\bibfnamefont {A.}~\bibnamefont {{Schlabritz}}}\ and\ \bibinfo {author} {\bibfnamefont {M.~E.}\ \bibnamefont {{Yamakou}}},\ }\bibfield  {title} {\bibinfo {title} {Interaction order controls stochastic-resonance enhancement and redistribution in delayed multiplex neural networks},\ }\bibfield  {journal} {\bibinfo  {journal} {arXiv e-prints}\ }\href {https://doi.org/10.48550/arXiv.2608.15421} {10.48550/arXiv.2608.15421} (\bibinfo {year} {2026})\BibitemShut {NoStop}%
\bibitem [{\citenamefont {Marder}\ and\ \citenamefont {Goaillard}(2006)}]{marder2006variability}%
  \BibitemOpen
  \bibfield  {author} {\bibinfo {author} {\bibfnamefont {E.}~\bibnamefont {Marder}}\ and\ \bibinfo {author} {\bibfnamefont {J.-M.}\ \bibnamefont {Goaillard}},\ }\bibfield  {title} {\bibinfo {title} {Variability, compensation and homeostasis in neuron and network function},\ }\href {https://doi.org/10.1038/nrn1949} {\bibfield  {journal} {\bibinfo  {journal} {Nature Reviews Neuroscience}\ }\textbf {\bibinfo {volume} {7}},\ \bibinfo {pages} {563} (\bibinfo {year} {2006})}\BibitemShut {NoStop}%
\bibitem [{\citenamefont {Padmanabhan}\ and\ \citenamefont {Urban}(2010)}]{padmanabhan2010intrinsic}%
  \BibitemOpen
  \bibfield  {author} {\bibinfo {author} {\bibfnamefont {K.}~\bibnamefont {Padmanabhan}}\ and\ \bibinfo {author} {\bibfnamefont {N.~N.}\ \bibnamefont {Urban}},\ }\bibfield  {title} {\bibinfo {title} {Intrinsic biophysical diversity decorrelates neuronal firing while increasing information content},\ }\href {https://doi.org/10.1038/nn.2630} {\bibfield  {journal} {\bibinfo  {journal} {Nature Neuroscience}\ }\textbf {\bibinfo {volume} {13}},\ \bibinfo {pages} {1276} (\bibinfo {year} {2010})}\BibitemShut {NoStop}%
\bibitem [{\citenamefont {Zhou}\ \emph {et~al.}(2001)\citenamefont {Zhou}, \citenamefont {Kurths},\ and\ \citenamefont {Hu}}]{zhouArrayenhancedCoherenceResonance2001}%
  \BibitemOpen
  \bibfield  {author} {\bibinfo {author} {\bibfnamefont {C.}~\bibnamefont {Zhou}}, \bibinfo {author} {\bibfnamefont {J.}~\bibnamefont {Kurths}},\ and\ \bibinfo {author} {\bibfnamefont {B.}~\bibnamefont {Hu}},\ }\bibfield  {title} {\bibinfo {title} {Array-enhanced coherence resonance: Nontrivial effects of heterogeneity and spatial independence of noise},\ }\href {https://doi.org/10.1103/PhysRevLett.87.098101} {\bibfield  {journal} {\bibinfo  {journal} {Physical Review Letters}\ }\textbf {\bibinfo {volume} {87}},\ \bibinfo {pages} {098101} (\bibinfo {year} {2001})}\BibitemShut {NoStop}%
\bibitem [{\citenamefont {Wang}\ and\ \citenamefont {Tu}(2012)}]{wang2012combined}%
  \BibitemOpen
  \bibfield  {author} {\bibinfo {author} {\bibfnamefont {M.}~\bibnamefont {Wang}}\ and\ \bibinfo {author} {\bibfnamefont {Y.}~\bibnamefont {Tu}},\ }\bibfield  {title} {\bibinfo {title} {Combined effects of parameter heterogeneity and noise on spatial coherence in map-based neurons},\ }\href {https://doi.org/10.1142/S0219477512500058} {\bibfield  {journal} {\bibinfo  {journal} {Fluctuation and Noise Letters}\ }\textbf {\bibinfo {volume} {11}},\ \bibinfo {pages} {1250005} (\bibinfo {year} {2012})}\BibitemShut {NoStop}%
\bibitem [{\citenamefont {Kawai}\ \emph {et~al.}(2010)\citenamefont {Kawai}, \citenamefont {Torigoe}, \citenamefont {Yoshida}, \citenamefont {Awazu},\ and\ \citenamefont {Nishimori}}]{kawai2010effective}%
  \BibitemOpen
  \bibfield  {author} {\bibinfo {author} {\bibfnamefont {R.}~\bibnamefont {Kawai}}, \bibinfo {author} {\bibfnamefont {S.}~\bibnamefont {Torigoe}}, \bibinfo {author} {\bibfnamefont {K.}~\bibnamefont {Yoshida}}, \bibinfo {author} {\bibfnamefont {A.}~\bibnamefont {Awazu}},\ and\ \bibinfo {author} {\bibfnamefont {H.}~\bibnamefont {Nishimori}},\ }\bibfield  {title} {\bibinfo {title} {Effective stochastic resonance under noise of heterogeneous amplitude},\ }\href {https://doi.org/10.1103/PhysRevE.82.051122} {\bibfield  {journal} {\bibinfo  {journal} {Physical Review E}\ }\textbf {\bibinfo {volume} {82}},\ \bibinfo {pages} {051122} (\bibinfo {year} {2010})}\BibitemShut {NoStop}%
\bibitem [{\citenamefont {Liu}\ \emph {et~al.}(2022)\citenamefont {Liu}, \citenamefont {Wang}, \citenamefont {Wu}, \citenamefont {Yang},\ and\ \citenamefont {Guan}}]{liu2022diversity}%
  \BibitemOpen
  \bibfield  {author} {\bibinfo {author} {\bibfnamefont {C.}~\bibnamefont {Liu}}, \bibinfo {author} {\bibfnamefont {C.-Y.}\ \bibnamefont {Wang}}, \bibinfo {author} {\bibfnamefont {Z.-X.}\ \bibnamefont {Wu}}, \bibinfo {author} {\bibfnamefont {H.-X.}\ \bibnamefont {Yang}},\ and\ \bibinfo {author} {\bibfnamefont {J.-Y.}\ \bibnamefont {Guan}},\ }\bibfield  {title} {\bibinfo {title} {Diversity-induced resonance in a globally coupled bistable system with diversely distributed heterogeneity},\ }\href {https://doi.org/10.1063/5.0094685} {\bibfield  {journal} {\bibinfo  {journal} {Chaos: An Interdisciplinary Journal of Nonlinear Science}\ }\textbf {\bibinfo {volume} {32}},\ \bibinfo {pages} {083112} (\bibinfo {year} {2022})}\BibitemShut {NoStop}%
\bibitem [{\citenamefont {Scialla}\ \emph {et~al.}(2021)\citenamefont {Scialla}, \citenamefont {Loppini}, \citenamefont {Patriarca},\ and\ \citenamefont {Heinsalu}}]{sciallaHubsDiversitySynchronization2021}%
  \BibitemOpen
  \bibfield  {author} {\bibinfo {author} {\bibfnamefont {S.}~\bibnamefont {Scialla}}, \bibinfo {author} {\bibfnamefont {A.}~\bibnamefont {Loppini}}, \bibinfo {author} {\bibfnamefont {M.}~\bibnamefont {Patriarca}},\ and\ \bibinfo {author} {\bibfnamefont {E.}~\bibnamefont {Heinsalu}},\ }\bibfield  {title} {\bibinfo {title} {Hubs, diversity, and synchronization in {{FitzHugh-Nagumo}} oscillator networks: {{Resonance}} effects and biophysical implications},\ }\href {https://doi.org/10.1103/PhysRevE.103.052211} {\bibfield  {journal} {\bibinfo  {journal} {Physical Review E}\ }\textbf {\bibinfo {volume} {103}},\ \bibinfo {pages} {052211} (\bibinfo {year} {2021})}\BibitemShut {NoStop}%
\bibitem [{\citenamefont {Scialla}\ \emph {et~al.}(2025)\citenamefont {Scialla}, \citenamefont {Patriarca}, \citenamefont {Heinsalu}, \citenamefont {Yamakou},\ and\ \citenamefont {Cartwright}}]{scialla2025effect}%
  \BibitemOpen
  \bibfield  {author} {\bibinfo {author} {\bibfnamefont {S.}~\bibnamefont {Scialla}}, \bibinfo {author} {\bibfnamefont {M.}~\bibnamefont {Patriarca}}, \bibinfo {author} {\bibfnamefont {E.}~\bibnamefont {Heinsalu}}, \bibinfo {author} {\bibfnamefont {M.~E.}\ \bibnamefont {Yamakou}},\ and\ \bibinfo {author} {\bibfnamefont {J.~H.}\ \bibnamefont {Cartwright}},\ }\bibfield  {title} {\bibinfo {title} {Effect of diversity distribution symmetry on global oscillations of networks of excitable units},\ }\href {https://doi.org/10.1103/lvb3-dc11} {\bibfield  {journal} {\bibinfo  {journal} {Physical Review E}\ }\textbf {\bibinfo {volume} {112}},\ \bibinfo {pages} {054201} (\bibinfo {year} {2025})}\BibitemShut {NoStop}%
\bibitem [{\citenamefont {Yamakou}\ \emph {et~al.}(2022)\citenamefont {Yamakou}, \citenamefont {Heinsalu}, \citenamefont {Patriarca},\ and\ \citenamefont {Scialla}}]{yamakou2022diversity}%
  \BibitemOpen
  \bibfield  {author} {\bibinfo {author} {\bibfnamefont {M.~E.}\ \bibnamefont {Yamakou}}, \bibinfo {author} {\bibfnamefont {E.}~\bibnamefont {Heinsalu}}, \bibinfo {author} {\bibfnamefont {M.}~\bibnamefont {Patriarca}},\ and\ \bibinfo {author} {\bibfnamefont {S.}~\bibnamefont {Scialla}},\ }\bibfield  {title} {\bibinfo {title} {Diversity-induced decoherence},\ }\href {https://doi.org/10.1103/PhysRevE.106.L032401} {\bibfield  {journal} {\bibinfo  {journal} {Physical Review E}\ }\textbf {\bibinfo {volume} {106}},\ \bibinfo {pages} {L032401} (\bibinfo {year} {2022})}\BibitemShut {NoStop}%
\bibitem [{\citenamefont {Scialla}\ \emph {et~al.}(2022)\citenamefont {Scialla}, \citenamefont {Patriarca},\ and\ \citenamefont {Heinsalu}}]{sciallaInterplayDiversityNoise2022}%
  \BibitemOpen
  \bibfield  {author} {\bibinfo {author} {\bibfnamefont {S.}~\bibnamefont {Scialla}}, \bibinfo {author} {\bibfnamefont {M.}~\bibnamefont {Patriarca}},\ and\ \bibinfo {author} {\bibfnamefont {E.}~\bibnamefont {Heinsalu}},\ }\bibfield  {title} {\bibinfo {title} {The interplay between diversity and noise in an excitable cell network model},\ }\href {https://doi.org/10.1209/0295-5075/ac5cdb} {\bibfield  {journal} {\bibinfo  {journal} {Europhysics Letters}\ }\textbf {\bibinfo {volume} {137}},\ \bibinfo {pages} {51001} (\bibinfo {year} {2022})}\BibitemShut {NoStop}%
\bibitem [{\citenamefont {Hariharan}\ \emph {et~al.}(2026)\citenamefont {Hariharan}, \citenamefont {Suresh},\ and\ \citenamefont {Chandrasekar}}]{hariharan2026heterogeneous}%
  \BibitemOpen
  \bibfield  {author} {\bibinfo {author} {\bibfnamefont {S.}~\bibnamefont {Hariharan}}, \bibinfo {author} {\bibfnamefont {R.}~\bibnamefont {Suresh}},\ and\ \bibinfo {author} {\bibfnamefont {V.}~\bibnamefont {Chandrasekar}},\ }\bibfield  {title} {\bibinfo {title} {Heterogeneous noise-induced extreme events and synchronization in a globally coupled network of {FitzHugh}--{Nagumo} oscillators},\ }\href {https://doi.org/10.1016/j.chaos.2025.117652} {\bibfield  {journal} {\bibinfo  {journal} {Chaos, Solitons \& Fractals}\ }\textbf {\bibinfo {volume} {203}},\ \bibinfo {pages} {117652} (\bibinfo {year} {2026})}\BibitemShut {NoStop}%
\bibitem [{\citenamefont {Maass}\ \emph {et~al.}(2002)\citenamefont {Maass}, \citenamefont {Natschl{\"a}ger},\ and\ \citenamefont {Markram}}]{maassRealtimeComputingStable2002}%
  \BibitemOpen
  \bibfield  {author} {\bibinfo {author} {\bibfnamefont {W.}~\bibnamefont {Maass}}, \bibinfo {author} {\bibfnamefont {T.}~\bibnamefont {Natschl{\"a}ger}},\ and\ \bibinfo {author} {\bibfnamefont {H.}~\bibnamefont {Markram}},\ }\bibfield  {title} {\bibinfo {title} {Real-time computing without stable states: {{A}} new framework for neural computation based on perturbations},\ }\href {https://doi.org/10.1162/089976602760407955} {\bibfield  {journal} {\bibinfo  {journal} {Neural Computation}\ }\textbf {\bibinfo {volume} {14}},\ \bibinfo {pages} {2531} (\bibinfo {year} {2002})}\BibitemShut {NoStop}%
\bibitem [{\citenamefont {Liao}\ \emph {et~al.}(2021)\citenamefont {Liao}, \citenamefont {Wang}, \citenamefont {Yamahara},\ and\ \citenamefont {Tabata}}]{liao2021echo}%
  \BibitemOpen
  \bibfield  {author} {\bibinfo {author} {\bibfnamefont {Z.}~\bibnamefont {Liao}}, \bibinfo {author} {\bibfnamefont {Z.}~\bibnamefont {Wang}}, \bibinfo {author} {\bibfnamefont {H.}~\bibnamefont {Yamahara}},\ and\ \bibinfo {author} {\bibfnamefont {H.}~\bibnamefont {Tabata}},\ }\bibfield  {title} {\bibinfo {title} {Echo state network activation function based on bistable stochastic resonance},\ }\href {https://doi.org/10.1016/j.chaos.2021.111503} {\bibfield  {journal} {\bibinfo  {journal} {Chaos, Solitons \& Fractals}\ }\textbf {\bibinfo {volume} {153}},\ \bibinfo {pages} {111503} (\bibinfo {year} {2021})}\BibitemShut {NoStop}%
\bibitem [{\citenamefont {Zhai}\ \emph {et~al.}(2023)\citenamefont {Zhai}, \citenamefont {Kong},\ and\ \citenamefont {Lai}}]{zhai2023emergence}%
  \BibitemOpen
  \bibfield  {author} {\bibinfo {author} {\bibfnamefont {Z.-M.}\ \bibnamefont {Zhai}}, \bibinfo {author} {\bibfnamefont {L.-W.}\ \bibnamefont {Kong}},\ and\ \bibinfo {author} {\bibfnamefont {Y.-C.}\ \bibnamefont {Lai}},\ }\bibfield  {title} {\bibinfo {title} {Emergence of a resonance in machine learning},\ }\href {https://doi.org/10.1103/PhysRevResearch.5.033127} {\bibfield  {journal} {\bibinfo  {journal} {Physical Review Research}\ }\textbf {\bibinfo {volume} {5}},\ \bibinfo {pages} {033127} (\bibinfo {year} {2023})}\BibitemShut {NoStop}%
\bibitem [{\citenamefont {FitzHugh}(1955)}]{fitzhughMathematicalModelsThreshold1955}%
  \BibitemOpen
  \bibfield  {author} {\bibinfo {author} {\bibfnamefont {R.}~\bibnamefont {FitzHugh}},\ }\bibfield  {title} {\bibinfo {title} {Mathematical models of threshold phenomena in the nerve membrane},\ }\href {https://doi.org/10.1007/BF02477753} {\bibfield  {journal} {\bibinfo  {journal} {The Bulletin of Mathematical Biophysics}\ }\textbf {\bibinfo {volume} {17}},\ \bibinfo {pages} {257} (\bibinfo {year} {1955})}\BibitemShut {NoStop}%
\bibitem [{\citenamefont {FitzHugh}(1961)}]{fitzhughImpulsesPhysiologicalStates1961}%
  \BibitemOpen
  \bibfield  {author} {\bibinfo {author} {\bibfnamefont {R.}~\bibnamefont {FitzHugh}},\ }\bibfield  {title} {\bibinfo {title} {Impulses and physiological states in theoretical models of nerve membrane},\ }\href {https://doi.org/10.1016/S0006-3495(61)86902-6} {\bibfield  {journal} {\bibinfo  {journal} {Biophysical Journal}\ }\textbf {\bibinfo {volume} {1}},\ \bibinfo {pages} {445} (\bibinfo {year} {1961})}\BibitemShut {NoStop}%
\bibitem [{\citenamefont {Nagumo}\ \emph {et~al.}(1962)\citenamefont {Nagumo}, \citenamefont {Arimoto},\ and\ \citenamefont {Yoshizawa}}]{nagumoActivePulseTransmission1962}%
  \BibitemOpen
  \bibfield  {author} {\bibinfo {author} {\bibfnamefont {J.}~\bibnamefont {Nagumo}}, \bibinfo {author} {\bibfnamefont {S.}~\bibnamefont {Arimoto}},\ and\ \bibinfo {author} {\bibfnamefont {S.}~\bibnamefont {Yoshizawa}},\ }\bibfield  {title} {\bibinfo {title} {An active pulse transmission line simulating nerve axon},\ }\href {https://doi.org/10.1109/JRPROC.1962.288235} {\bibfield  {journal} {\bibinfo  {journal} {Proceedings of the IRE}\ }\textbf {\bibinfo {volume} {50}},\ \bibinfo {pages} {2061} (\bibinfo {year} {1962})}\BibitemShut {NoStop}%
\bibitem [{\citenamefont {Collins}\ \emph {et~al.}(1995)\citenamefont {Collins}, \citenamefont {Chow},\ and\ \citenamefont {Imhoff}}]{collins1995aperiodic}%
  \BibitemOpen
  \bibfield  {author} {\bibinfo {author} {\bibfnamefont {J.}~\bibnamefont {Collins}}, \bibinfo {author} {\bibfnamefont {C.~C.}\ \bibnamefont {Chow}},\ and\ \bibinfo {author} {\bibfnamefont {T.~T.}\ \bibnamefont {Imhoff}},\ }\bibfield  {title} {\bibinfo {title} {Aperiodic stochastic resonance in excitable systems},\ }\href {https://doi.org/10.1103/PhysRevE.52.R3321} {\bibfield  {journal} {\bibinfo  {journal} {Physical Review E}\ }\textbf {\bibinfo {volume} {52}},\ \bibinfo {pages} {R3321} (\bibinfo {year} {1995})}\BibitemShut {NoStop}%
\bibitem [{\citenamefont {Watts}\ and\ \citenamefont {Strogatz}(1998)}]{wattsCollectiveDynamicsSmallworldNetworks1998}%
  \BibitemOpen
  \bibfield  {author} {\bibinfo {author} {\bibfnamefont {D.~J.}\ \bibnamefont {Watts}}\ and\ \bibinfo {author} {\bibfnamefont {S.~H.}\ \bibnamefont {Strogatz}},\ }\bibfield  {title} {\bibinfo {title} {Collective dynamics of `small-world' networks},\ }\href {https://doi.org/10.1038/30918} {\bibfield  {journal} {\bibinfo  {journal} {Nature}\ }\textbf {\bibinfo {volume} {393}},\ \bibinfo {pages} {440} (\bibinfo {year} {1998})}\BibitemShut {NoStop}%
\bibitem [{\citenamefont {Muldoon}\ \emph {et~al.}(2016)\citenamefont {Muldoon}, \citenamefont {Bridgeford},\ and\ \citenamefont {Bassett}}]{muldoonSmallworldPropensityWeighted2016}%
  \BibitemOpen
  \bibfield  {author} {\bibinfo {author} {\bibfnamefont {S.~F.}\ \bibnamefont {Muldoon}}, \bibinfo {author} {\bibfnamefont {E.~W.}\ \bibnamefont {Bridgeford}},\ and\ \bibinfo {author} {\bibfnamefont {D.~S.}\ \bibnamefont {Bassett}},\ }\bibfield  {title} {\bibinfo {title} {Small-world propensity and weighted brain networks},\ }\href {https://doi.org/10.1038/srep22057} {\bibfield  {journal} {\bibinfo  {journal} {Scientific reports}\ }\textbf {\bibinfo {volume} {6}},\ \bibinfo {pages} {22057} (\bibinfo {year} {2016})}\BibitemShut {NoStop}%
\bibitem [{\citenamefont {Higham}(2001)}]{highamAlgorithmicIntroductionNumerical2001}%
  \BibitemOpen
  \bibfield  {author} {\bibinfo {author} {\bibfnamefont {D.~J.}\ \bibnamefont {Higham}},\ }\bibfield  {title} {\bibinfo {title} {An algorithmic introduction to numerical simulation of stochastic differential equations},\ }\href {https://doi.org/10.1137/S0036144500378302} {\bibfield  {journal} {\bibinfo  {journal} {SIAM Review}\ }\textbf {\bibinfo {volume} {43}},\ \bibinfo {pages} {525} (\bibinfo {year} {2001})}\BibitemShut {NoStop}%
\bibitem [{\citenamefont {Ozer}\ \emph {et~al.}(2009)\citenamefont {Ozer}, \citenamefont {Perc},\ and\ \citenamefont {Uzuntarla}}]{ozerStochasticResonanceNewman2009}%
  \BibitemOpen
  \bibfield  {author} {\bibinfo {author} {\bibfnamefont {M.}~\bibnamefont {Ozer}}, \bibinfo {author} {\bibfnamefont {M.}~\bibnamefont {Perc}},\ and\ \bibinfo {author} {\bibfnamefont {M.}~\bibnamefont {Uzuntarla}},\ }\bibfield  {title} {\bibinfo {title} {Stochastic resonance on {{Newman}}--{{Watts}} networks of {{Hodgkin}}--{{Huxley}} neurons with local periodic driving},\ }\href {https://doi.org/10.1016/j.physleta.2009.01.034} {\bibfield  {journal} {\bibinfo  {journal} {Physics Letters A}\ }\textbf {\bibinfo {volume} {373}},\ \bibinfo {pages} {964} (\bibinfo {year} {2009})}\BibitemShut {NoStop}%
\bibitem [{\citenamefont {Hodgkin}\ and\ \citenamefont {Huxley}(1952)}]{hodgkinQuantitativeDescriptionMembrane1952}%
  \BibitemOpen
  \bibfield  {author} {\bibinfo {author} {\bibfnamefont {A.~L.}\ \bibnamefont {Hodgkin}}\ and\ \bibinfo {author} {\bibfnamefont {A.~F.}\ \bibnamefont {Huxley}},\ }\bibfield  {title} {\bibinfo {title} {A quantitative description of membrane current and its application to conduction and excitation in nerve},\ }\href {https://doi.org/10.1113/jphysiol.1952.sp004764} {\bibfield  {journal} {\bibinfo  {journal} {The Journal of Physiology}\ }\textbf {\bibinfo {volume} {117}},\ \bibinfo {pages} {500} (\bibinfo {year} {1952})}\BibitemShut {NoStop}%
\bibitem [{\citenamefont {Yamakou}\ and\ \citenamefont {Kuehn}(2023)}]{yamakou2023combined}%
  \BibitemOpen
  \bibfield  {author} {\bibinfo {author} {\bibfnamefont {M.~E.}\ \bibnamefont {Yamakou}}\ and\ \bibinfo {author} {\bibfnamefont {C.}~\bibnamefont {Kuehn}},\ }\bibfield  {title} {\bibinfo {title} {Combined effects of spike-timing-dependent plasticity and homeostatic structural plasticity on coherence resonance},\ }\href {https://doi.org/10.1103/PhysRevE.107.044302} {\bibfield  {journal} {\bibinfo  {journal} {Physical Review E}\ }\textbf {\bibinfo {volume} {107}},\ \bibinfo {pages} {044302} (\bibinfo {year} {2023})}\BibitemShut {NoStop}%
\end{thebibliography}
%

\end{document}